\documentclass[prd,showpacs,preprintnumbers]{revtex4} 

\usepackage{graphicx}
\usepackage{dcolumn}
\usepackage{bm}
\usepackage{latexsym}
\usepackage{amsfonts}
\usepackage{amssymb}
\usepackage[usenames]{color}
\usepackage{wick}        

\begin{document}
\preprint{KUNS 2267}

\title{Hawking Radiation from Fluctuating Black Holes}

\author{Tomohiro Takahashi}
\author{Jiro Soda}
\affiliation{Department of Physics,  Kyoto University, Kyoto, 606-8501, Japan
}

\date{\today}

\begin{abstract}
Classically, black Holes have the rigid event horizon. 
However, quantum mechanically, 
the event horizon of black holes becomes fuzzy due to quantum fluctuations. 
We study Hawking radiation of a real scalar field from a fluctuating black hole.  
To quantize metric perturbations, we derive the quadratic 
action for those in the black hole background. Then, we 
calculate the cubic interaction terms in the action for the scalar field. 
Using these results, we obtain the spectrum of Hawking radiation in the presence of
interaction between the scalar field and the metric. It turns out that the spectrum deviates
from the Planck spectrum due to quantum fluctuations of the metric.
\end{abstract}

\maketitle

\section{Introduction}

Black holes in general relativity have shown us many fertile aspects of spacetime
such as the event horizon, the singularity, Hawking radiation, and the connection
with thermodynamics. Among these phenomena, Hawking radiation is particularly
interesting because it gives a strong support to the thermodynamical interpretation
of black holes. Moreover, it reveals an aspect of quantum gravity. 
Usually, Hawking radiation is explained by quantizing a free 
field in a fixed black hole background~\cite{Hawking:1974sw}.  
To be specific, we consider a real scalar field $\phi$
in the Schwarzschild black hole background
\begin{eqnarray}
  ds^2 = - \left( 1-\frac{r_H}{r}\right) dt^2
  +\left( 1-\frac{r_H}{r}\right)^{-1} dr^2
  + r^2 \left( d\theta^2 +\sin\theta^2 d\phi^2 \right) \ ,\nonumber
\end{eqnarray}
where $r_H$ is a constant of integration. When the radius coordinate $r$ becomes
$r_H$, the spacetime seems to be singular. Of course, we know it is just a coordinate
singularity and the position $r =r_H$ is the event horizon of the spacetime.
The action for gravity with a scalar field reads
\begin{eqnarray}
 S = \frac{1}{16\pi G}\int d^4 x \sqrt{-g} R 
 - \frac{1}{2} \int d^4 x \sqrt{-g} \partial_\mu \phi \partial^\mu \phi \ ,\nonumber
\end{eqnarray}
where $G$, $g$ and $R$ are the Newton constant, 
the determinant of the metric $g_{\mu\nu}$ and the scalar curvature.
Classically, nothing can come out from the inside of the horizon.  Quantum mechanically,
however, it has been shown that black holes can emit radiation, 
the so-called Hawking radiation~\cite{Hawking:1974sw, Unruh:1976db}.  
Remarkably, it turns out that Hawking radiation has the Planck spectrum. 
This notable feature conforms to the black hole thermodynamics.   
However, taking look at the above action, we notice that the metric 
should be also quantized. Then, the event horizon becomes fuzzy, that is,
 black holes are fluctuating. The purpose of this paper is to study 
 Hawking radiation from fluctuating black holes. Technically, we consider 
 the interaction between the scalar field and the metric fluctuations. 
 In particular, it is interesting to see if the Planck spectrum is
modified by the interaction~\cite{Unruh:1983ac,Birrell:1978ng,Leahy:1983vb}
whatever small it is. 
Apparently, this effect becomes significant if the  horizon radius of black holes $r_h$ 
is close to the Planck length $\ell_p$. 

Recently, various Planck scale black holes and Hawking radiation from them have been 
considered~\cite{Kanti:2008eq,Ida:2002ez,Murata:2007jh,Ishihara:2007ni}.
 These black holes can be created at the LHC or in the early universe. 
Indeed, black holes would be created at the LHC if the dimension of space-time 
is more than six and braneworld picture is correct~\cite{Giddings:2001bu}. 
This is because the higher dimensional Planck mass becomes $\sim 1$TeV which can be reached 
at the LHC. In other words, the mass of a black hole created at the LHC 
is of the order of the higher dimensional Planck mass. 
On the other hand, density fluctuations in the early universe 
could produce black holes with any mass. 
Black holes with $10^{15}$g created in the early universe 
lose their masses by Hawking radiation and become Planck scale just now. 
The spectrum of Hawking radiation from these black holes might be distorted 
due to the nonlinear interaction between quantum fields and the metric. 

Here, we stress that, even for black holes much larger than the Planck length, 
the fluctuations of black holes might be relevant to Hawking radiation. 
The point is that the Planck spectrum of Hawking radiation 
stems from the exponential elongation
of short wavelength quantum fluctuations which can exceed the Planck scale.
Hence, the interaction between the scalar field and quantum mechanical
fluctuations of the metric would be relevant even for the large black holes.

In this paper, we calculate the effect of nonlinear interaction 
on Hawking radiation of a real scalar field from a
 4-dimensional Schwarzschild black hole. 
In order to achieve this aim, the canonically normalized quadratic action for
 metric perturbations and the interaction Hamiltonian are needed. 
 Therefore, we first construct these actions
  by perturbing the Einstein-Hilbert action with a real scalar field. 
After that, we give a formalism to study Hawking radiation from
fluctuating black holes.

The organization of this paper is as follows. 
In section \ref{3}, we construct the quadratic action for metric perturbations by 
perturbing Einstein-Hilbert action in 4-dimensional Schwarzschild background. 
In section \ref{2}, we quantize the field and derive Hawking radiation. 
In section \ref{4}, we present a method for treating interaction
 in spherical symmetric space-time and then 
calculate the cubic interaction Hamiltonian from the action of a real scalar field. 
Using these results, we calculate the effect of nonlinear interaction
 on Hawking radiation of a real scalar field. 
Section \ref{5} is devoted to the conclusion.   
Technical details can be found in Appendixes.

In this paper, we use the convention $16\pi G=1$ and write the Schwarzschild metric as
\begin{eqnarray}
	ds^2=-f(r)dt^2+f^{-1}(r)dr^2+r^2\gamma_{ab}dx^adx^b \ , \qquad
	f(r)=1-\frac{2M}{r}
	\ ,\label{}\nonumber
\end{eqnarray}
where $\gamma_{ab}$ represents the metric of the sphere $S^2$ 
and $M$ corresponds to the mass of the black hole.

\section{Metric Perturbations in Black Hole Background}
\label{3}
In this section, we derive the quadratic action for metric perturbations $h_{\mu\nu}$
in the Schwarzschild black hole background $g^{(0)}_{\mu\nu}$.
 Note that our derivation is slightly different from
the original one by Moncrief~\cite{Moncrief:1974am}.
Now, the metric is given by
\begin{eqnarray}
g_{\mu\nu}=g^{(0)}_{\mu\nu}+h_{\mu\nu}  \ .
\end{eqnarray}
Here, we can use the general coordinate invariance to reduce the number of variables.
Under infinitesimal coordinate transformations $x^{\mu}\rightarrow x^{\mu}+\xi^{\mu}$, 
$h_{\mu\nu}$ transforms as
\begin{eqnarray}
h_{\mu\nu}\rightarrow h_{\mu\nu}+\xi_{\mu;\nu}+\xi_{\nu;\mu}\ ,
\label{gauge_trans}
\end{eqnarray}
which shows that there are 4 gauge degrees of freedom. Hence, we can fix
4 variables. Moreover, we can classify 
the remaining 6 variables into 4 scalar type and 2 vector type ones.
Once the gauge is completely fixed, we can obtain the quadratic action. 
After eliminating unphysical variables using constraint equations, 
we finally obtain the reduced action for physical variables. 

It is easy to extend the analysis in this section to more general cases.
In Appendix \ref{appendixA}, we present the analysis for
 topological Schwarzschild black holes with a cosmological constant. 

\subsection{Scalar Perturbation}

First, we consider the scalar type perturbations 
\begin{eqnarray}
	 h_{\mu\nu}=
	\left(
	\begin{array}{cc|cc}
	f{\bar H}&H_1&\multicolumn{2}{c}{v_{|a}}\\
	sym&H/f&\multicolumn{2}{c}{w_{|a}}\\ \hline
	sym&sym&\multicolumn{2}{c}{r^2K\gamma_{ab}}\\
	sym&sym&\multicolumn{2}{r}{+B_{|ab}}\\
	\end{array}
	\right)\ ,\label{evenmode_pert}
\end{eqnarray}
where $|a$ denotes a covariant derivative with respect to $\gamma_{ab}$.
Under the scalar type gauge transformations $\xi_{\mu}$ with 
\begin{eqnarray}
\xi_{\mu}=
\left(
\xi_0 \ , \xi_r \ , \xi_{L|a}
\right) \ ,
\label{eq:}
\end{eqnarray}
the perturbed variables change as 
\begin{eqnarray}
&& {\bar H}\rightarrow {\bar H}+2\dot{\xi_0}/f-f^{'}\xi_r \ ,\quad
H_1\rightarrow H_1+\xi_0^{'}+\dot{\xi_r}-f^{'}\xi_0/f \ ,\quad
v\rightarrow v+\xi_0 +\dot{\xi_L} \ , \\
&& H\rightarrow H+2f\xi_r^{'}+f^{'}\xi_r \ ,\quad
w\rightarrow w + \xi_r + \xi_L^{'}- 2 \xi_L /r \ ,\quad
K\rightarrow K + 2f\xi_r /r \ ,\quad
B\rightarrow B + 2 \xi_L
\ ,
\label{even_gauge_transport}
\end{eqnarray}
where $'$ and $\cdot$ denote derivatives with respect to $r$ and $t$, respectively.
Then, choosing $\xi_L =-B/2$, $\xi_0 =-v-\dot{B}/2$ and $\xi_r =-rK/(2f)$, we 
get the following metric perturbations
\begin{eqnarray}
	\left(
	\begin{array}{cc|cc}
	f{\bar H}&H_1&\multicolumn{2}{c}{0}\\
	sym&H/f&\multicolumn{2}{c}{w_{|a}}\\ \hline
	sym&sym&0&0\\
	sym&sym&0&0\\
	\end{array}
	\right)\ .
 \label{fs_gauge}
\end{eqnarray} 

Using the above gauge, we obtain the quadratic action 
\begin{eqnarray}
	&\ &\sum_{\ell, m>0}
      \Biggl\{\int dr dt\Biggl[\frac{\gamma_s}{2}|H_{1}{}_{\ell m}|^2
      -\gamma_s{\dot w}_{\ell m}^{*}H_{1}{}_{\ell m}+
	\frac{\gamma_s}{2}|{\dot w}_{\ell m}|^2+2r\dot{H}_{\ell m}^{*}H_1{}_{\ell m} 
      \nonumber\\
	&\ &\hspace{3cm}-\gamma_s\frac{2f+rf^{'}}{2r}w_{\ell m}^{*}H_{\ell m}
      +\frac{(fr)^{'}}{2}|H_{\ell m}|^2+\gamma_s\frac{f}{r^2}|w_{\ell m}|^2
      -\frac{1}{2}{\bar H}_{\ell m}^{*}q_2{}_{\ell m}\Biggr]+(C.C) 
      \Biggr\}\nonumber\\
 & &\hspace{1cm}	+ \sum_{\ell} \int dr dt\Biggl[\frac{\gamma_s}{2}H_{1\ell 0}{}^2
      -\gamma_s{\dot w}_{\ell 0}H_{1\ell 0}+
	\frac{\gamma_s}{2}{\dot w}_{\ell 0}^2+2r\dot{H}_{\ell 0}H_1{}_{\ell 0} \nonumber\\
	&\ &\hspace{4.5cm}-\gamma_s\frac{2f+rf^{'}}{2r}w_{\ell 0}H_{\ell 0}+\frac{(fr)^{'}}{2}H_{\ell 0}^2+\gamma_s\frac{f}{r^2}w_{\ell 0}^2
      -\frac{1}{2}{\bar H}_{\ell 0}q_2{}_{\ell 0}\Biggr] \ ,   
	\label{even_ac}
\end{eqnarray}
where $*$ means complex conjugate and we have defined
\begin{eqnarray}
	q_1{}_{\ell m}=2rfH_{\ell m}-2\gamma_s fw_{\ell m} \ , \quad
	q_2{}_{\ell m}=\gamma_s\left(H_{\ell m}+f^{'}w_{\ell m}
                     -\frac{2fw_{\ell m}}{r}\right)+q_1^{'}{}_{\ell m}
      \ .	 \label{q}
\end{eqnarray}
It is easy to solve as
\begin{eqnarray}
	w_{\ell m}=\frac{r}{\gamma_s T(r)}\left(q_2{}_{\ell m}-q_1^{'}{}_{\ell m}-\frac{\gamma_s}{2rf}q_1{}_{\ell m}\right) \ , \quad
	H_{\ell m}=\frac{1}{T(r)}\left(q_2{}_{\ell m}-q_1^{'}{}_{\ell m}+\frac{rf^{'}-2f}{2rf}q_1{}_{\ell m}\right) \ , 
\label{wH}
\end{eqnarray}
where we have defined $T(r)=rf^{'}-2f+\gamma_s$. 
In Eqs.(\ref{even_ac}), (\ref{q}), and (\ref{wH}), 
we expanded metric perturbations in terms of spherical harmonics $Y_{\ell m}$
 like as $H=\sum H_{\ell m}Y_{\ell m}$.  Here, $\gamma_s\equiv \ell(\ell+1)$. 
 Since spherical harmonics satisfy the relation $Y_{\ell m}=(-1)^mY_{\ell -m}^{*}$, 
 metric perturbations $H_{\ell m}$ satisfy reality conditions
  $H_{\ell m}=(-1)^mH_{\ell -m}^{*}$. 
This allows us to relate negative $m$ modes to positive $m$ modes.
 Hence, we can consider only $m\geq 0$ modes in (\ref{even_ac}). 
Note that $m=0$ mode is real but $m>0$ modes are complex.

From now on, we concentrate on $m=0$ mode and omit the indices. 
We can get the same result for $m>0$ modes with similar calculations.
Taking the variation of the quadratic action (\ref{even_ac}) with respect to $H_1$, 
we obtain the momentum constraint equation
\begin{eqnarray}
	H_1&=&\dot{w}-\frac{2r\dot{H}}{\gamma_s}
      \ .\label{even_constraint}
\end{eqnarray}
This equation enables us to delete $H_1$ from the action (\ref{even_ac}). 
Substituting this result (\ref{even_constraint}) into the action (\ref{even_ac}), 
we obtain the action
\begin{eqnarray}
	\int dr dt\Biggl[-\frac{2r^2}{\gamma_s}\dot{H}^2+2r\dot{H}\dot{w} 
       -\gamma_s\frac{2f+rf^{'}}{2r}wH+\frac{(fr)^{'}}{2}H^2+\gamma_s\frac{f}{r^2}w^2
      -\frac{1}{2}{\bar H}q_2\Biggr]\ .\label{action2}
\end{eqnarray}
Taking the variation of the action (\ref{action2}) 
with respect to the Lagrange multiplier ${\bar H}$,
we can derive the Hamiltonian constraint
\begin{eqnarray}
q_2=0 \ .
\label{}
\end{eqnarray}
Then, substituting $q_2=0$ into the action (\ref{action2})
and using the relations (\ref{wH}), we can express the action by $q_1$. 
With a new variable $\psi^Z$ defined by
\begin{eqnarray}
\psi^Z= \sqrt{\frac{\lambda}{\lambda+1}}\frac{q_1}{T(r)} \ , \quad
\lambda=\frac{\gamma_s-2}{2}
\ , \label{psiz}
\end{eqnarray}
the quadratic action takes the canonical form  
\begin{eqnarray}
	\int dr^{*} dt \left[\frac{1}{2}(\partial_t \psi^Z)^2-\frac{1}{2}(\partial_{r^{*}} \psi^Z)^2-\frac{1}{2}V_{Z}(r) \psi^{Z2}\right] 
\label{}
\end{eqnarray}
where $r^*$ represents a tortoise coordinate and 
we have defined the potential function as
\begin{eqnarray}
	V_{Z}
      &=&-\frac{f}{4\left(r^2f^{'}-2rf+2r\lambda+2r\right)^2}
      \Bigl[8\lambda(1+\lambda)^2+4(1+\lambda)^2rf^{'}+2(2+\lambda)r^2f^{'2}+r^3f^{'3}\nonumber\\
	&\ &   \hspace{0.3cm}             +4f^2(2\lambda+rf^{'})-4f\left\{2\lambda(2+\lambda)+2(1+\lambda)rf^{'}+r^2f^{'2}\right\}\Bigr]\nonumber\\
	&=&\frac{2f}{r^3}\frac{\lambda^2(\lambda+1)r^3+3\lambda^2Mr^2+9\lambda rM^2+9M^3}{(r\lambda+3M)^2} 
\ .\label{zerilli_potential}
\end{eqnarray}
We can also calculate the quadratic actions for $m>0$ modes.
 Thus, we have 
\begin{eqnarray} 
\sum_{m>0}\left\{\int dr^{*} dt \left[|\partial_t \psi^{Z}_{\ell m}|^2-|\partial_{r^{*}} \psi^{Z}_{\ell m}|^2-V_{Z}(r)|\psi^{Z}_{\ell m}|^2\right]\right\}
 + \frac{1}{2}\int dr^{*} dt \left[(\partial_t \psi^{Z}_{\ell 0})^2-(\partial_{r^{*}} \psi^{Z}_{\ell 0})^2-V_{Z}(r) \psi^{Z}_{\ell 0}{}^2\right] 
\label{even_action} \ ,
\end{eqnarray}
which gives the famous Zerilli equation\cite{Zerilli:1970se}
\begin{eqnarray}
-\partial_{t}^{2}\psi^{Z}_{\ell m}+\partial_{r^{*}}^{2}\psi^{Z}_{\ell m}
-V_{Z}\psi^{Z}_{\ell m}=0
\ . \label{eq:Zerilli_equation}
\end{eqnarray}

For later purposes, we deduce the relation between $\psi_{\ell m}^{Z}$ and 
metric perturbations. 
From Eqs.(\ref{q}), (\ref{wH}), (\ref{even_constraint}) and (\ref{psiz}), 
it is easy to get the relation between 
$\psi_{\ell m}^{Z}$ and metric perturbations except for ${\bar H}$.  
The results are 
\begin{eqnarray}
	H_1{}_{\ell m}&=&\dot{w}_{\ell m}-\frac{2r\dot{H}_{\ell m}}{\gamma_s}\ ,\nonumber\\
	w_{\ell m}&=&-\frac{r}{\gamma_s}\sqrt{\frac{\lambda+1}{\lambda}}\psi^{Z}_{\ell m}{}^{'}-\left(\frac{1}{2f}+\frac{rT^{'}}{\gamma_s T}\right)\sqrt{\frac{\lambda+1}{\lambda}}\psi^Z_{\ell m}\ ,\nonumber\\
	H_{\ell m}&=&-\sqrt{\frac{\lambda+1}{\lambda}}\psi^Z_{\ell m}{}^{'}-\left(\frac{T^{'}}{T}-\frac{rf^{'}-2f}{2rf}\right)\sqrt{\frac{\lambda+1}{\lambda}}\psi^Z_{\ell m}
      \ .\label{even_master1}
\end{eqnarray}
In order to get the relation between $\psi^{Z}_{\ell m}$ and ${\bar H}_{\ell m}$, 
we need to look at the equation of motion for metric perturbations. 
Taking the variation of (\ref{action2}) with respect to $H$ and $w$, 
we can get two equations 
\begin{eqnarray}
&& \frac{4r^2}{\gamma_s}\ddot{H}_{\ell m}-2r\ddot{w}_{\ell m}
-\gamma_s\frac{2f+rf^{'}}{2r}w_{\ell m}+(rf)^{'}H_{\ell m}
-\frac{\gamma_s}{2}{\bar H}_{\ell m}+rf{\bar H}_{\ell m}^{'}=0 \ , \\
&& -2r\ddot{H}_{\ell m}-\gamma_s\frac{rf^{'}+2f}{2r}H_{\ell m}
+2\gamma_s\frac{f}{r^2}w_{\ell m}-\frac{\gamma_s}{2}(f^{'}-\frac{2f}{r}){\bar H}_{\ell m}
-\gamma_sf{\bar H}^{'}_{\ell m}=0
 \ .
\end{eqnarray}
Eliminating ${\bar H}^{'}_{\ell m}$ from the above equations, 
we obtain the equation for ${\bar H}_{\ell m}$  
\begin{eqnarray}
	{\bar H}_{\ell m}=\frac{2}{T(r)}\left[\frac{2r^2}{\gamma_s}\ddot{H}_{\ell m}-2r\ddot{w}_{\ell m}+\frac{rf^{'}}{2}H_{\ell m}+\left(\frac{2f}{r}-\frac{\gamma_s}{2}\frac{rf^{'}+2f}{r}\right)w_{\ell m}\right]\ . 
\end{eqnarray}
Using this equation, the relation (\ref{even_master1}) and Zerilli equation, 
we can show that ${\bar H}_{\ell m}$ is related to $H_{\ell m}$ and $w_{\ell m}$ as 
\begin{eqnarray}
	{\bar H}_{\ell m}=H_{\ell m}-\left(2fw_{\ell m}\right)^{'}
      \ . 
\label{even_master2}
\end{eqnarray}
Since $H_{\ell m}$ and $w_{\ell m}$ are related to $\psi_{\ell m}^Z$ 
by Eqs.(\ref{even_master1}), this equation implies a relation between ${\bar H}_{\ell m}$ 
and $\psi_{\ell m}^Z$.  

\subsection{Vector Perturbation}

Next, we consider the quadratic action for vector type perturbations 
\begin{eqnarray}
	h_{\mu \nu}=
	\left(
	\begin{array}{cc|cc}
	0&0&\multicolumn{2}{c}{v_a}\\
	0&0&\multicolumn{2}{c}{w_a}\\\hline
	{\bf sym}&{\bf sym}&\multicolumn{2}{c}{\frac{1}{2}\left[C_{a|b}+C_{b|a}\right]}\\
	\end{array}
	\right)
      \ , \label{odd_pert}
\end{eqnarray}
where $v_{a}$, $w_a$ and $C_a$ satisfy $v_a{}^{|a}=w_a{}^{|a}=C_a{}^{|a}=0$. 
For vector type perturbations (\ref{odd_pert}), the gauge transformations with 
\begin{eqnarray}
\xi_{\mu}= \left( 0 \, 0 \ , \xi_{a} \right)  
\ , 
\end{eqnarray}
read
\begin{eqnarray}
v_a \rightarrow v_a+\dot{\xi}_a \ , \quad
w_a \rightarrow w_a+\xi_a^{'}-2\xi_a/r \ , \quad
C_a \rightarrow C_a+ 2 \xi_a
\ ,
\end{eqnarray}
where $\xi_a$ satisfies the transverse condition.
Then, we can completely fix the gauge by choosing $\xi_a=-C_a/2$,
 which is called Regge-Wheeler gauge. In the Regge-Wheeler gauge, metric perturbations 
are written as 
\begin{eqnarray}
	h_{\mu \nu}=
	\left(
	\begin{array}{cccc}
	0&0&\multicolumn{2}{c}{v_a}\\
	0&0&\multicolumn{2}{c}{w_a}\\
	sym&sym&0&0\\
	sym&sym&0&0
	\end{array}
	\right)
      \ .\label{odd_p}
\end{eqnarray}
Using the metric perturbations (\ref{odd_p}), we can calculate the quadratic action 
\begin{eqnarray}
&\ &(2\lambda+2)
\sum_{m>0}\left\{\int drdt \Bigl[\frac{1}{2}\left|\dot{w}_{\ell m}+\frac{2}{r}v_{\ell m}-v_{\ell m}^{'}\right|^2+\frac{\lambda}{fr^2}|v_{\ell m}|^2-\frac{\lambda f}{r^2}|w_{\ell m}|^2\Bigr]+(C.C)\right\}\nonumber\\
&\ &\hspace{1cm}+(2\lambda+2)
	 \int drdt  \Bigl[\frac{1}{2}\left(\dot{w}_{\ell 0}+\frac{2}{r}v_{\ell 0}-v^{'}_{\ell 0}\right)^2+\frac{\lambda}{fr^2}v^2_{\ell 0}-\frac{\lambda f}{r^2}w^2_{\ell 0}\Bigr]\ ,
\label{odd_ac}
\end{eqnarray}
where  metric perturbations are expanded by vector harmonics 
$V^a_{\ell m}$  like as $w^a=\sum w_{\ell m}V^a_{\ell m}$. 
Since vector harmonics satisfy relations $V^a_{\ell m}=(-1)^mV^{a*}_{\ell -m}$,
  $w_{\ell m}$ satisfy reality conditions $w_{\ell m}=(-1)^mw_{\ell -m}$.
Hence,  we can consider only $m\geq 0$ modes in (\ref{odd_ac}). 

For a while, we concentrate on $m=0$ mode and omit the indices.
Taking the variation of the action (\ref{odd_ac}) with respect to $v$, 
we can derive momentum constraint equation 
\begin{eqnarray}
	\left[r^2(\dot{w}+\frac{2}{r}v-v^{'})\right]^{'}+\frac{2\lambda}{f}v=0 
\ .\label{odd_const}
\end{eqnarray}
We have to solve this constraint equation. However, it is not easy to do so. 
In order to circumvent this difficulty, we move on to Hamiltonian formalism.
From the quadratic action (\ref{odd_ac}), we can read off the Lagrangian  
\begin{eqnarray} 
	 L=(2\lambda+2)\int dr  \Bigl[\frac{1}{2}\left(\dot{w}+\frac{2}{r}v-v^{'}\right)^2+\frac{\lambda}{fr^2}v^2-\frac{\lambda f}{r^2}w^2\Bigr]\ ,
\label{}
\end{eqnarray}
and also define the momentum conjugate to $w$ as 
\begin{eqnarray}
	p=\frac{\delta L}{\delta \dot{w}}=(2\lambda+2)\left(\dot{w}+\frac{2}{r}v-v^{'}\right) 
\ .
\label{}
\end{eqnarray}
Then, the Hamiltonian $H$ is given by
\begin{eqnarray}
H&=&    \int dr p{\dot w}-L    \nonumber\\
 &=&(2\lambda+2)\int dr \left[\frac{1}{2}\left(\frac{p}{2\lambda+2}\right)^2+\frac{\lambda f}{r^2}w^2-\left(\frac{2}{r}\frac{p}{2\lambda+2}+
           \left(\frac{p}{2\lambda+2}\right)^{'}+\frac{\lambda v}{fr^2}\right)v\right]
\ .
\label{odd_ham}
\end{eqnarray}
Now, the  constraint equation  
\begin{eqnarray}
	 &\ &\frac{2}{r}p+p^{'}+\frac{4\lambda(\lambda+1) v}{fr^2}=0\nonumber
\end{eqnarray}
gives the relation
\begin{eqnarray} 
v=-\frac{fr^2}{4\lambda(\lambda+1)}\left(\frac{2}{r}p+p^{'}\right)
\ .\label{sol_con}
\end{eqnarray}
Substituting this relation into the Hamiltonian (\ref{odd_ham}), 
we obtain the reduced Hamiltonian
\begin{eqnarray}
	H=(2\lambda+2)\int dr\left[\left(\frac{1}{2}-\frac{1}{2\lambda}+\frac{f}{\lambda}\right)\left(\frac{p}{2\lambda+2}\right)^2+\frac{fr^2}{4\lambda}\left(\frac{p}{2\lambda+2}\right)^{'}{}^{2}+\frac{\lambda f}{r^2}w^2\right]
      \ . \label{hamiltonian}
\end{eqnarray}
We further perform the canonical transformation $(w,p) \rightarrow (Q,P)$
to interchange the role of $w$ and $p$
\begin{eqnarray}
Q=-p \ ,\quad w=P \  \ ,
\label{QP}
\end{eqnarray}
which leads to a new Hamiltonian 
\begin{eqnarray}
K(Q,P)&=&H(P,-Q)\nonumber\\
&=&(2\lambda+2)\int dr\left[\left(\frac{1}{2}-\frac{(fr)^{'}}{2\lambda}+\frac{f}{\lambda}\right)\left(\frac{Q}{2\lambda+2}\right)^2+
\frac{fr^2}{4\lambda}\left(\frac{Q}{2\lambda+2}\right)^{'}{}^{2}+\frac{\lambda f}{r^2}P^2\right]
\ . \nonumber
\end{eqnarray}
Using the Hamilton equation for ${\dot Q}$
\begin{eqnarray} 
\dot{Q}=\frac{\delta K}{\delta P}=\frac{4\lambda(\lambda+1) f}{r^2}P
        \ , \label{dotQ}
\end{eqnarray}
we can go back to the Lagrangian formalism 
\begin{eqnarray}
  \int dt\left[ \int dr P\dot{Q}  - K  \right]
  =\frac{1}{2\lambda+2}\int dr dt\Biggl[\frac{r^2}{4\lambda f}\dot{Q}^2-\left(\frac{1}{2}-\frac{(fr)^{'}}{2\lambda}+\frac{f}{\lambda}\right)Q^2-\frac{fr^2}{4\lambda}Q^{'}{}^{2}\Biggr] 
\ .\label{odd_act}
\end{eqnarray}
With a change of normalization  
\begin{eqnarray}
\psi^{RW}\equiv \frac{r}{2\sqrt{\lambda(\lambda+1)}}Q
\ ,\label{psiRW}
\end{eqnarray}
the quadratic action (\ref{odd_act}) becomes 
\begin{eqnarray}
\int dr^{*}dt\left[\frac{1}{2}\left(\partial_t \psi^{RW}\right)^2-\frac{1}{2}\left(\partial_{r^*} \psi^{RW}\right)^2-\frac{1}{2}V_{RW}(r)\psi^{RW}{}^2\right] 
\ ,
\end{eqnarray}
where $r^*$ represents a tortoise coordinate and we have defined the potential function
\begin{eqnarray}
V_{RW}=\frac{f}{r^2}\left(-rf^{'}+2f+2\lambda\right)
\ .
\end{eqnarray}

So far we have only considered the $m=0$ mode in (\ref{odd_ac}). 
Of course, we can also calculate the action for $m>0$ modes. 
The result reads 
\begin{eqnarray}
&& \sum_{m>0}\Biggl\{
\int dr^{*}dt\left[\left|\partial_t \psi^{RW}_{\ell m}\right|^2-\left|\partial_{r^*} \psi^{RW}_{\ell m}\right|^2-V_{RW}(r)|\psi^{RW}_{\ell m}|^2\right]\Biggr\} \nonumber\\
&& \qquad + \frac{1}{2}\int dr^{*}dt\left[\left(\partial_t \psi^{RW}_{\ell 0}\right)^2-\left(\partial_{r^*} \psi^{RW}_{\ell 0}\right)^2-V_{RW}(r)\psi^{RW}_{\ell 0}{}^2\right]
         \ , \label{eq:RWaction}
\end{eqnarray}
which gives the  Regge-Wheeler equation~\cite{Regge:1957td}
\begin{eqnarray}
-\partial_t ^2\psi^{RW}_{\ell m}+\partial_{r^{*}}^2\psi^{RW}_{\ell m}-V_{RW}(r)\psi^{RW}_{\ell m}=0
            \ . \label{eq:RWe}
\end{eqnarray} 

Finally, we clarify the relation between $\psi^{RW}_{\ell m}$ and metric perturbations. 
From Eqs.(\ref{sol_con}), (\ref{QP}), (\ref{dotQ}) and (\ref{psiRW}), 
the relations are given by
\begin{eqnarray}
v_{\ell m}=\frac{f}{2\sqrt{\lambda(\lambda+1)}}\left(r\psi^{RW}_{\ell m}\right)^{'}
           \ , \quad
w_{\ell m}=\frac{r}{2\sqrt{\lambda(\lambda+1)}f}{\dot \psi}^{RW}_{\ell m}
           \ . \label{odd_master}
\end{eqnarray}

\section{Hawking Radiation}
\label{2}
In this section, we review Hawking radiation using a spherically symmetric
 collapsing model\cite{Hawking:1974sw,BD}. 
In this model, it is assumed that outside of the collapsing matter can be 
described by the Schwarzschild metric 
\begin{eqnarray}
ds^2=-f(r)dt^2+1/f(r)dr^2+r^2 \gamma_{ab}dx^a dx^b\ ,
\end{eqnarray}
where $f(r)=1-2M/r$. Here, $M$ and $\gamma_{ab}$ are the ADM mass and 
the metric of $S^2$, respectively.
The action for the real massless scalar field
\begin{eqnarray}
-\frac{1}{2}\int d^4x \sqrt{-g} g^{\mu\nu} \varphi_{,\mu}\varphi_{,\nu}
\label{scalar_action}
\end{eqnarray}
can be written as
\begin{eqnarray}
	&\ & \sum_{m>0}\int  dr^* dt \left[ \left| \partial_t \phi_{\ell m} \right|^2-\left|\partial_{r^*}\phi_{\ell m}\right|^{2}-f\left(\frac{f^{'}}{r}+\frac{\gamma_s}{r^2}\right)\left|\phi_{\ell m}\right|^2\right]\nonumber\\
	&\ &+\frac{1}{2}\int dr^*dt\left[ \left(\partial_t \phi_{\ell 0}\right)^2 - \left(\partial_{r^*}\phi_{\ell 0}\right)^{2}-f\left(\frac{f^{'}}{r}+\frac{\gamma_s}{r^2}\right)\phi_{\ell 0}^2\right]\ ,
	\label{action_phi}
\end{eqnarray} 
where $r^*$ is a tortoise coordinate.
Here, we used the harmonic expansion $\varphi=\sum \phi_{\ell m} Y_{\ell m}/r$. 
Note that $\varphi$ is real. Hence, complex coefficients $\phi_{\ell m}$
 satisfy reality conditions $\phi_{\ell m}=(-1)^m\phi_{\ell -m}^{*}$.  
Then, the equation of motion outside the matter
$
	\varphi^{;\mu}{}_{;\mu}=0  	
$
can be written as
\begin{eqnarray}
	\partial_{t}^2\phi_{\ell m}-\partial_{r^{*}}^2\phi_{\ell m}+V_{\phi}\phi_{\ell m}=0
      \ , \quad 
	V_{\phi}=f(r)\left(\frac{f^{'}}{r}+\frac{\ell(\ell+1)}{r^2}\right)
	\ . \label{Klein}
\end{eqnarray}
It should be noted that the quantization of metric perturbations can be
performed in the same way as that of the scalar field explained below.

\subsection{Quantization }

Let us quantize the scalar field $\varphi$ 
by promoting $\varphi$ to the operator:
\begin{eqnarray}
\varphi=\sum_{\ell m}\int d\omega \frac{1}{\sqrt{2\omega}}\left[a_{\omega \ell m}\frac{1}{r}Y_{\ell m}u_{\omega \ell}(t,r^{*})+a_{\omega \ell m}^{\dagger}\frac{1}{r}Y_{\ell m}^{*}u_{\omega \ell}^{*}(t,r^{*})\right]
          \ . \label{decompose_var}
\end{eqnarray}
In the above formula, $\frac{1}{\sqrt{2\omega}}\frac{1}{r}Y_{\ell m}u_{\omega \ell}$
 are complete orthonormal basis satisfying Klein-Gordon equation 
\begin{eqnarray}
-\partial_t^2u_{\omega\ell}+\partial_{r^{*}}^2u_{\omega \ell}-f(r)\left(\frac{f^{'}}{r}+\frac{\ell(\ell+1)}{r^2}\right)u_{\omega \ell}=0 \ .
\end{eqnarray}
Note that this equation is independent of $m$, so we used $u_{\omega \ell}$
 instead of $u_{\omega\ell m}$. 
Here, the norm is  defined by the inner product 
\begin{eqnarray}
(\psi_1,\psi_2)
=-i\int_{\Sigma} \sqrt{-g_{\Sigma}}d\Sigma^{\mu}
\left( \psi_{1} \partial_{\mu} \psi_2^{*} - \psi_{2}^{*} \partial_{\mu} \psi_1 \right)
 \ ,
\label{inner}
\end{eqnarray} 
where $\Sigma$ is a Cauchy surface and $d\Sigma^{\mu}$ means area element times
a unit normal vector. 
Now, the canonical commutation relation implies the  following commutation relations;
\begin{eqnarray}
	[a_{\omega \ell m},a_{\omega^{'} \ell^{'} m^{'}}^{\dagger}]
      = \delta (\omega - \omega^{'} ) \ \delta_{\ell,\ell^{'}}\delta_{m,m^{'}}
      \ . \label{com}
\end{eqnarray} 
 Using the property of spherical harmonics $Y_{\ell m}=(-1)^mY_{\ell -m}^{*}$, 
we can rewrite (\ref{decompose_var})  as 
\begin{eqnarray}
	 \varphi=\sum_{\ell m}\int d\omega \frac{1}{\sqrt{2\omega}}\left[a_{\omega \ell m}u_{\omega \ell}(t,r^{*})+(-1)^{m}a_{\omega \ell -m}^{\dagger}u_{\omega \ell}^{*}(t,r^{*})\right]\frac{Y_{\ell m}}{r}
       \ .\label{decompose2}
\end{eqnarray} 
Let us define $\phi_{\ell m}$ by $\varphi=\sum \phi_{\ell m}Y_{\ell m}/r$.
Then, we have 
\begin{eqnarray}
	\phi_{\ell m}=\int d\omega\frac{1}{\sqrt{2\omega}} \left[b_{\omega \ell m}u_{\omega \ell}(t,r^{*})+d_{\omega \ell m}^{\dagger}u_{\omega \ell}^{*}(t,r^{*})\right] 
      \ .
\end{eqnarray} 
From Eq.(\ref{decompose2}), we see the following relations 
\begin{eqnarray}
	b_{\omega \ell m}=a_{\omega \ell m},\ d_{\omega \ell m}=(-1)^{m}a_{\omega \ell -m}\ . \label{abd}
\end{eqnarray}
These relations are equivalent to 
\begin{eqnarray}
	\phi_{\ell m}^{\dagger}=(-1)^{m}\phi_{\ell -m}\ , \label{}
\end{eqnarray}
which is nothing but the reality condition for $\varphi$. 
And this reality condition suggests that we can only consider $m\geq 0$ modes. 
From (\ref{com}) and (\ref{abd}), we can deduce the commutation relations for $b_{\omega \ell m}$ and $d_{\omega \ell m}$ for $m\geq 0$ modes.
The results are as follows.

\begin{itemize}
\item $m=0$ modes

We can define $b_{\omega \ell 0}=a_{\omega \ell 0}=d_{\omega \ell 0}(\equiv c_{\omega \ell})$ from the relations (\ref{abd}).
 Using the commutation relations (\ref{com}), we can show that 
\begin{eqnarray} 
	[c_{\omega \ell},c_{\omega^{'} \ell^{'}}^{\dagger}]
      =\delta (\omega - \omega^{'} ) \ \delta_{\ell,\ell^{'}}
      \ . 
\end{eqnarray}
The commutation relations show that $\phi_{\ell 0}$ are real scalar fields whose annihilation operators are $c_{\omega \ell}$.
Note that these scalar fields $\phi_{\ell 0}$ have the action (\ref{action_phi}). 

\item $m>0$ modes

From the commutation relations (\ref{com}), it is easy to show
\begin{eqnarray}
[b_{\omega \ell m},b_{\omega^{'} \ell^{'} m^{'}}^{\dagger}]
=\delta (\omega - \omega^{'})\ 
\delta_{\ell,\ell^{'}}\delta_{m,m^{'}} \ , \quad 
[d_{\omega \ell m},d_{\omega^{'} \ell^{'} m^{'}}^{\dagger}]
=\delta ( \omega - \omega^{'}) \ \delta_{\ell,\ell^{'}}\delta_{m,m^{'}}
      \ . 
\end{eqnarray}
Thus, $\phi_{\ell m}$ $(m>0)$ are complex scalar fields whose 
annihilation operators are $b_{\omega\ell m}$ and $d_{\omega\ell m}$. 
Note that complex scalar fields $\phi_{\ell m}$ have the action (\ref{action_phi}).
\end{itemize}

\subsection{Bogolubov Transformations }

To introduce the Bogolubov transformation, we define complete orthonormal basis  
\begin{eqnarray}
	v_{\omega \ell m}=\frac{1}{\sqrt{2\omega}}\frac{1}{r}Y_{\ell m}u_{\omega \ell}\ . \label{}
\end{eqnarray}
Suppose that there are two kind of complete orthonormal basis of $\varphi$,
 $\{v_{\omega \ell m}\}$ and $\{{\bar v}_{\omega \ell m}\}$.
 As both $\{v_{\omega\ell m}\}$ and $\{{\bar v}_{\omega\ell m}\}$ 
are complete basis, we can expand $\varphi$ in two ways 
\begin{eqnarray}
	\varphi 
  = \sum_{\ell m} \int d\omega \left( a_{\omega \ell m}  
  v_{\omega\ell m}+a_{\omega\ell m}^{\dagger}v_{\omega\ell m}^{*} \right)
  = \sum_{\ell m} \int d\omega \left({\bar a}_{\omega\ell m}{\bar v}_{\omega\ell m}
  +{\bar a}_{\omega\ell m}^{\dagger}{\bar v}_{\omega\ell m}^{*} \right)
                         \ ,\label{tennkai}
\end{eqnarray}
where $a_{\omega \ell m}$ and ${\bar a}_{\omega\ell m}$ are annihilation operators
 for $v_{\omega \ell m}$ and ${\bar v}_{\omega \ell m}$, respectively. 
Using these annihilation operators, we can define two kind of vacua as 
\begin{eqnarray}
	a_{\omega\ell m}|0\rangle=0\ ,\quad
	{\bar a}_{\omega\ell m}|{\bar 0}\rangle=0  
      \ .\label{}
\end{eqnarray}
Due to the completeness of $\{v_{\omega\ell m}\}$ and $\{{\bar v}_{\omega\ell m}\}$, 
we can expand $v_{\omega\ell m}$ as
\begin{eqnarray}
	v_{\omega\ell m}
     =\sum_{\ell' m'}  \int d\omega' \left(
   \alpha_{\omega^{'}\ell^{'}m^{'}, \omega\ell m}^{*}{\bar v}_{\omega^{'} \ell^{'} m^{'}}
-\beta_{\omega^{'}\ell^{'}m^{'}, \omega\ell m}{\bar v}_{\omega^{'}\ell^{'} m^{'}}^{*}
  \right)  \ , \label{v_barv}
\end{eqnarray}
or conversely 
\begin{eqnarray}
	{\bar v}_{\omega\ell m}
  =\sum_{\ell' m'} \int d\omega' \left(
  \alpha_{\omega\ell m, \omega^{'}\ell^{'} m^{'}}  v_{\omega^{'} \ell^{'} m^{'}}
+\beta_{\omega\ell m, \omega^{'}\ell^{'} m^{'}} v_{\omega^{'}\ell^{'} m^{'}}^{*}
  \right)  \ , \label{barv_v}
\end{eqnarray}
where $\alpha_{\omega\ell m, \omega^{'}\ell^{'} m^{'}}$ and 
$\beta_{\omega\ell m, \omega^{'}\ell^{'} m^{'}}$ are called  Bogolubov coefficients.
From the above formula, we can deduce the Bogolubov coefficients as 
\begin{eqnarray}
\alpha_{\omega\ell m,\omega^{'}\ell^{'} m^{'}}
  = ({\bar v}_{\omega\ell m},v_{\omega^{'}\ell^{'} m^{'}}) \ , \quad
\beta_{\omega\ell m, \omega^{'}\ell^{'} m^{'}}
  = -({\bar v}_{\omega\ell m},v_{\omega^{'}\ell^{'} m^{'}}^{*})
	\ . \label{beta}
\end{eqnarray}
Since both complete basis are proportional to the spherical harmonics $Y_{\ell m}$,  
 Bogolubov coefficients are proportional to $\delta_{\ell, \ell^{'}}$ and 
$\delta_{m, m^{'}}$, that is 
\begin{eqnarray}
	\alpha_{\omega \ell m,\omega^{'} \ell^{'}m^{'}}
      =\alpha_{\omega\ell,\omega^{'}\ell}\delta_{\ell,\ell^{'}}\delta_{m,m^{'}}
                                   \ , \quad
	\beta_{\omega \ell m,\omega^{'} \ell^{'}m^{'}}
      =(-1)^{m}\beta_{\omega\ell,\omega^{'}\ell}\delta_{\ell,\ell^{'}}\delta_{m,-m^{'}}
	 \label{angular}
	\ .
\end{eqnarray}
Substituting (\ref{v_barv}) and (\ref{barv_v}) into (\ref{tennkai}) and using (\ref{angular}), we can get Bogolubov transformations between $a_{\omega \ell m}$ and 
${\bar a}_{\omega\ell m}$ as follows;
\begin{eqnarray}
	a_{\omega \ell m}
  &=&\sum_{\ell^{'},m^{'}} \int d\omega' \left(
  \alpha_{\omega^{'} \ell^{'}m^{'},\omega \ell m}{\bar a}_{\omega^{'} \ell^{'}m^{'}}+\beta^{*}_{\omega^{'} \ell^{'}m^{'},\omega \ell m}{\bar a}^{\dagger}_{\omega^{'} \ell^{'}m^{'}}
  \right) \nonumber\\
 &=& \int d\omega' \left(
 \alpha_{\omega^{'}\ell,\omega\ell}{\bar a}_{\omega^{'} \ell m}+(-1)^{m}\beta^{*}_{\omega^{'}\ell,\omega \ell}{\bar a}^{\dagger}_{\omega^{'} \ell -m}
 \right)  \ ,\nonumber\\
	{\bar a}_{\omega\ell m}
   &=& \sum_{\ell^{'},m^{'}} \int d\omega' \left(
   \alpha^{*}_{\omega\ell m, \omega^{'}\ell^{'} m^{'}} a_{\omega^{'}\ell^{'} m^{'}}-\beta_{\omega\ell m,\omega^{'}\ell^{'} m^{'}}^{*} a_{\omega^{'}\ell^{'} m^{'}}^{\dagger}
   \right)  \nonumber\\
  &=& \int d\omega' \left(
  \alpha^{*}_{\omega\ell, \omega^{'}\ell} a_{\omega^{'}\ell m}-(-1)^m\beta_{\omega\ell,\omega^{'}\ell}^{*} a_{\omega^{'}\ell -m}^{\dagger}
        \right)  \label{a_bara}
\ .
\end{eqnarray}
Furthermore, substituting the second relation in (\ref{a_bara}) into 
the first relation in (\ref{a_bara}), we see that Bogolubov coefficients satisfy 
\begin{eqnarray}
\left\{
\begin{array}{l}
\int d\omega'' \left(
\alpha_{\omega\ell, \omega^{''}\ell}\alpha^{*}_{\omega^{'}\ell,\omega^{''}\ell}-\beta_{\omega \ell,\omega^{''}\ell}\beta^{*}_{\omega^{'}\ell, \omega^{''}\ell} \right)
=\delta_{\omega,\omega^{'}}\\
\int d\omega'' \left(
\alpha_{\omega\ell, \omega^{''}\ell}\beta_{\omega^{'}\ell, \omega^{''}\ell}-\beta_{\omega\ell, \omega^{''}\ell}\alpha_{\omega^{'}\ell,\omega^{''}\ell}  \right)
=0
\end{array}
\right.\ .
\label{property}
\end{eqnarray}
So far, we have considered Bogolubov transformation for $\varphi$. 
We can also describe Bogolubov transformation in terms of $\phi_{\ell m}\ (m\geq 0)$.
In order to get Bogolubov transformation for $c_{\omega\ell}$ 
which is  defined by $c_{\omega\ell}=a_{\omega\ell 0}$, 
we set $m=0$ in (\ref{a_bara}). Then, we get 
\begin{eqnarray}
	c_{\omega \ell}
      =\int d\omega' \left(
      \alpha_{\omega^{'}\ell,\omega\ell}{\bar c}_{\omega^{'} l}+\beta^{*}_{\omega^{'}\ell,\omega \ell}{\bar c}^{\dagger}_{\omega^{'} \ell} \right)
      \ . 
\end{eqnarray}
Similarly, setting $m>0$ in (\ref{a_bara}),
 we have Bogolubov transformation for $b_{\omega\ell m}$ 
which is defined in (\ref{abd});
\begin{eqnarray}
	b_{\omega \ell m}
      =\int d\omega' \left(
      \alpha_{\omega^{'}\ell,\omega \ell}{\bar b}_{\omega^{'} \ell m}+\beta^{*}_{\omega^{'}\ell,\omega \ell}{\bar d}^{\dagger}_{\omega^{'} \ell m} \right)
      \ . \label{b_bogo}
\end{eqnarray}
And we obtain  Bogolubov transformation for $d_{\omega\ell m}$
 by considering $m<0$ in (\ref{a_bara}) 
\begin{eqnarray}
	d_{\omega \ell m}
      =\int d\omega' \left(
      \alpha_{\omega^{'}\ell,\omega\ell}{\bar d}_{\omega^{'} \ell m}+\beta^{*}_{\omega^{'}\ell,\omega \ell}{\bar b}^{\dagger}_{\omega^{'} \ell m} \right)
      \ . 
\end{eqnarray}

Finally, we calculate particles created due to the time evolution of space-time.  
It is assumed that $\{{\bar v}_{\omega \ell m}\}$ is a natural base at the initial time
 ( ``in region") and $\{v_{\omega\ell m}\}$ is natural at the final time ( ``out region"). 
Furthermore, we assume that the initial state is $|{\bar 0}\rangle$. Note that this state does not change because we use a Heisenberg picture. 
Under these assumptions, the number spectrum of $\varphi$ particles 
at the final time can be expressed by
\begin{eqnarray}
	\sum_{\ell m}\langle {\bar 0}|N_{\omega \ell m}|{\bar 0}\rangle 
      &=&\sum_{\ell m}\langle {\bar 0}|a_{\omega \ell m}^{\dagger}a_{\omega \ell m}|{\bar 0}\rangle \nonumber\\
	&=&\sum_{\ell}\Bigl[\langle {\bar 0}|c_{\omega \ell}^{\dagger}c_{\omega \ell}|{\bar 0}\rangle 
      + \sum_{0< m\leq \ell}\left\{\langle {\bar 0}|b_{\omega \ell m}^{\dagger}b_{\omega \ell m}|{\bar 0}\rangle +\langle {\bar 0}|d_{\omega \ell m}^{\dagger}d_{\omega \ell m}|{\bar 0}\rangle \right\}\Bigr]
      \ . 
\end{eqnarray}
Then, using Bogolubov transformations for $b_{\omega\ell m}$, $c_{\omega\ell}$
 and $d_{\omega\ell m}$, we can calculate the number spectrum 
\begin{eqnarray}
	\sum_{\ell m}\langle {\bar 0}|N_{\omega \ell m}|{\bar 0}\rangle =\sum_{\ell}\left[(2\ell+1)\int d\omega' \left|\beta_{\omega\ell,\omega^{'}\ell}\right|^2\right]
	\ . \label{harm_beta}
\end{eqnarray}
This shows that we need $\beta_{\omega\ell,\omega^{'}\ell}$ 
 defined in (\ref{beta}) and (\ref{angular}) 
in order to know the number spectrum.

\subsection{Hawking Radiation}

\begin{figure}[tb]
 \begin{center}
  \includegraphics[width=60mm]{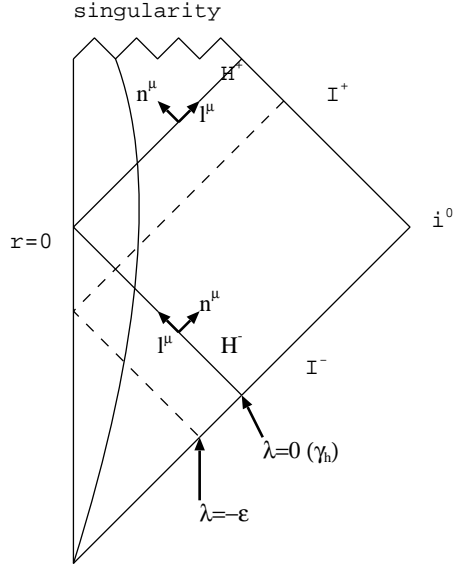}
 \end{center}
 \caption{Penrose diagram for a collapsing model. In this figure, $\lambda$ means 
 an affine parameter}
 \label{fig1}
\end{figure}

Let us consider Hawking radiation by using spherically symmetric collapsing model
described by the Penrose diagram in FIG.\ref{fig1}. In this figure, $I^{-}$ 
corresponds to ``in region" and $I^{+}$ represents ``out region". 

In order to calculate $\beta_{\omega \ell,\omega^{'}\ell}$, we must define natural basis in ``in region" and 
``out region". We assume that outside of the collapsing matter 
is described by the Schwarzschild metric. Then, $\phi_{\ell m}$ satisfy
\begin{eqnarray}
	\partial_t^2\phi_{\ell m}-\partial_{r^{*}}^2\phi_{\ell m}=0 \label{asymptotic}
\end{eqnarray}
in this region. Here, $V_{\phi}$ is ignored because $V_{\phi}$ dumps as $O(1/r^2)$. 
Noticing (\ref{asymptotic}), we choose the natural basis 
$\{u_{\omega\ell}^{in}\}$ and $\{u_{\omega\ell}^{out}\}$ defined by
\begin{eqnarray}
u_{\omega\ell}^{in}\rightarrow 
e^{-i\omega(t+r^{*})}\ ({\rm near}\ I^{-})\ ,\quad
u_{\omega\ell}^{out}\rightarrow 
e^{-i\omega(t-r^{*})}\ ({\rm near}\ I^{+}) \ .
\label{in_condition}
\end{eqnarray} 
%
Given the complete basis, we can calculate $\beta_{\omega\ell,\omega^{'}\ell}$ 
defined in (\ref{beta}) and (\ref{angular}). 

We have to choose $\Sigma$ to calculate the inner product in the formula (\ref{beta}).
 Here, we take $I^{-}$. Hence, it is necessary to know the behavior of 
 $u_{\omega\ell}^{out}$ near $I^{-}$. For that purpose, 
 we use geometric optics approximation where the worldline of a particle
 is a null ray of a constant phase and trace this null line backward to $I^{-}$. 
This approximation is quite good near the horizon $H^{+}$ because the 
frequency effectively becomes very large in this region. 
Hereafter, we write $u=t-r^*$ and $v=t+r^*$. 
Furthermore, we define $\lambda$ as an affine parameter with respect to $n^{\mu}$ which satisfies $\lambda=0$ on $\gamma_h$.
Near $H^{\pm}$, this parameter can be expressed as 
\begin{eqnarray}
\left\{
	\begin{array}{l}
	\lambda=-Ce^{-\kappa u} \quad ({\rm near\ horizon})\nonumber\\
	\lambda=Dv \quad ({\rm near\ H^{-}}) 
\end{array}
\right.\ ,
\end{eqnarray}
where $\kappa=1/4M $ is the surface gravity and $C$ and $D$ are some constants. 
Because $\lambda={\rm const.}$ line is also null line, geometric optics approximation suggests that 
$\lambda={\rm const.}$ surface is equivalent to a constant phase surface. 
Then, the constant phase surface 
$\lambda=-\epsilon$ satisfies
\begin{eqnarray}
	-\epsilon&=&-Ce^{-\kappa u} \quad ({\rm near\ horizon})\ ,\nonumber\\
	-\epsilon&=&Dv \quad ({\rm near\ H^{-}})
      \ ,
\end{eqnarray}
and so 
\begin{eqnarray}
	u=-\frac{1}{\kappa}\log\left(-\frac{D}{C}v\right)\ . \label{}
\end{eqnarray}
This means that $u_{\omega\ell}^{out}$ approximately behaves as
\begin{eqnarray}
	u_{\omega\ell}^{out}\simeq\left\{
	\begin{array}{l}
	0 \qquad ({\rm for}\ v>0)    \\
	\frac{1}{\sqrt{2\omega}}e^{\frac{i\omega}{\kappa}\log\left(-\frac{D}{C}v\right)}
      \qquad ({\rm for}\ v<0\ {\rm and}\ |v|\ll 1)
	\end{array}
	\right. 
      \label{uout}
\end{eqnarray}
%
near $I^{-}$. Then, we can calculate the inner products in (\ref{beta}) using (\ref{uout}). Note that 
(\ref{uout}) is a good approximation in $|v|\ll 1$ but the effect from this region is dominant because 
the phase of $u_{\omega\ell}^{out}$ diverges. Therefore it is also a good approximation 
to use the expression (\ref{uout}) on $I^{-}$ in calculating inner products. 
The results of calculations are 
\begin{eqnarray}
\alpha_{\omega\ell,\omega^{'}\ell}&=&\sqrt{\frac{\omega^{'}}{\omega}}\int_{-\infty}^{0}e^{i\omega^{'}v}\exp\left(\frac{i\omega}{\kappa}\ln(-Dv/C)\right)dv\nonumber\\
&=&\frac{1}{\sqrt{\omega\omega^{'}}}\left(\frac{D}{C}\right)^{i\omega/\kappa}e^{\pi\omega/2\kappa}(\omega^{'})^{-i\omega/\kappa}\Gamma\left(1+\frac{i\omega}{\kappa}\right)
\end{eqnarray}
and
\begin{eqnarray}
\beta_{\omega\ell,\omega^{'}\ell}&=&\sqrt{\frac{\omega^{'}}{\omega}}\int_{-\infty}^{0}e^{-i\omega^{'}v}\exp\left(\frac{i\omega}{\kappa}\ln(-Dv/C)\right)dv\nonumber\\
&=&-\frac{1}{\sqrt{\omega\omega^{'}}}\left(\frac{D}{C}\right)^{i\omega/\kappa}e^{-\pi\omega/2\kappa}(\omega^{'})^{-i\omega/\kappa}\Gamma\left(1+\frac{i\omega}{\kappa}\right)
   \ .  \label{Hawking_beta}
\end{eqnarray}
This gives 
\begin{eqnarray}
|\alpha_{\omega\ell,\omega^{'}\ell}|^2=e^{2\pi\omega/\kappa}|\beta_{\omega\ell,\omega^{'}\ell}|^2  \ .
\end{eqnarray}
Using the relation
\begin{eqnarray}
\int d\omega' \left(|\alpha_{\omega\ell,\omega^{'}\ell}|^2-|\beta_{\omega\ell,\omega^{'}\ell}|^2\right)=1
\end{eqnarray}
derived from (\ref{property}), we obtain Bogolubov coefficients $\beta$  
\begin{eqnarray}
	\int d\omega' |\beta_{\omega\ell,\omega^{'}\ell}|^2
      =\frac{1}{e^{2\pi\omega/\kappa}-1} \ .
      \label{mondai}
\end{eqnarray}
Substituting this result into (\ref{harm_beta}),
 we find that the observer in the asymptotic region 
see the thermal spectrum of $\varphi$ particles.

Actually, Eq.(\ref{mondai}) diverges and is proportional to the delta
function $\delta (0)$. We interpret this as
\begin{eqnarray}
  \delta (0) = \lim_{\omega_1 \rightarrow \omega, L\rightarrow \infty}
  \frac{1}{2\pi} \int^{\pi L}_{-\pi L} \exp i(\omega_1 -\omega )t
\end{eqnarray}
and introduce the regularization $\delta (0) =L$ hereafter.
Namely, what we obtained is the number of particles created
per frequency and  per unit time. In the above formula, we have discarded
this factor. However, we should keep in mind this fact in the 
following calculations.

\section{Hawking Radiation from Fluctuating Black Holes}
\label{4}

In this section, we consider interaction between the scalar field $\phi$ and 
gravitons $\psi$. First, we introduce a method for treating interaction 
in the spherically symmetric space-time~\cite{BD}. 
Next, we calculate cubic interaction terms from the action (\ref{scalar_action}). 
Finally, we evaluate the effect of interaction on Hawking radiation. 

Here, we follow the notations in Sec.\ref{2}, namely, we use $\phi_{\ell m}$
 defined by $\varphi=\sum \phi_{\ell m}Y_{\ell m}/r$ 
and annihilation operators $b_{\omega\ell m}$,
 $c_{\omega\ell}$, $d_{\omega\ell m}$.
 
\subsection{Interacting Quantum Fields in Curved Spacetime}
\label{effect}

Let us consider the Lagrangian 
\begin{eqnarray}
	{\cal L}={\cal L}_0+{\cal L}_{int}\ ,\nonumber \label{}
\end{eqnarray}
where ${\cal L}_0$ represents the free field part of the action
 and ${\cal L}_{int}$ is the interaction part. 
Furthermore, we assume that ${\cal L}_{int}$ vanish both in the remote past 
(or ``in region") and  in the remote future (or ``out region").
Under these assumptions, we can define asymptotic fields 
\begin{eqnarray}
\varphi\rightarrow \varphi^{in} \quad {\rm (for\ in\ region)} \ ,\qquad
\varphi\rightarrow\varphi^{out} \quad {\rm (for\ out\ region)}
\ , 
\end{eqnarray}
where $\varphi^{in}$ and $\varphi^{out}$ satisfy equation of motion 
derived from ${\cal L}_{0}$. 
Therefore, using basis $\{\frac{1}{\sqrt{2\omega}}\frac{Y_{\ell m}}{r}u^{in}_{\omega \ell }(t,r^{*})\}$  in ``in region" 
and $\{\frac{1}{\sqrt{2\omega}}Y_{\ell m}u_{\omega \ell }^{out}(t,r^{*})\}$ 
 in ``out region", 
we can expand asymptotic fields $\varphi^{in}$  as 
\begin{eqnarray}
	\varphi^{in}&=&
 \sum_{ \ell m}\int d\omega \left(
 a_{\omega \ell m}^{in}\frac{1}{\sqrt{2\omega}}\frac{Y_{\ell m}}{r}u_{\omega \ell }^{in}+a_{\omega \ell m}^{in}{}^{\dagger}\frac{1}{\sqrt{2\omega}}\frac{Y_{\ell m}^{*}}{r}u_{\omega \ell }^{in}{}^{*}
 \right) \nonumber\\
 &=&\sum_{ \ell m} \int d\omega \left(
 {\bar a_{\omega \ell m}^{in}}\frac{1}{\sqrt{2\omega}}\frac{Y_{\ell m}}{r}u_{\omega \ell }^{out}+{\bar a_{\omega \ell m}^{in}{}^{\dagger}}\frac{1}{\sqrt{2\omega}}\frac{Y_{\ell m}^{*}}{r}u_{\omega \ell }^{out}{}^{*}
 \right) \ ,
\end{eqnarray}
and  $\varphi^{out}$ as
\begin{eqnarray} 
	\varphi^{out}&=&
  \sum_{ \ell m}\int d\omega \left(
  a_{\omega \ell m}^{out}\frac{1}{\sqrt{2\omega}}\frac{Y_{\ell m}}{r}u_{\omega \ell }^{in}+a_{\omega \ell m}^{out}{}^{\dagger}\frac{1}{\sqrt{2\omega}}\frac{Y_{\ell m}^{*}}{r}u_{\omega \ell }^{in}{}^{*}
  \right) \nonumber\\
  &=& \sum_{ \ell m} \int d\omega \left(
  {\bar a_{\omega \ell m}^{out}}\frac{1}{\sqrt{2\omega}}\frac{Y_{\ell m}}{r}u_{\omega \ell }^{out}+{\bar a_{\omega \ell m}^{out}{}^{\dagger}}\frac{1}{\sqrt{2\omega}}\frac{Y_{\ell m}^{*}}{r}u_{\omega \ell }^{out}{}^{*} \right)
                  \ .
\end{eqnarray}
From the above expansion, we can define four kind of vacua 
\begin{eqnarray}
	a_{\omega \ell m}^{in}|0\ {\rm in}\rangle = 0\ ,\quad
	{\bar a_{\omega \ell m}^{in}}|{\bar 0}\ {\rm in}\rangle = 0\ ,\quad
	a_{\omega \ell m}^{out}|0\ {\rm out}\rangle = 0\ ,\quad
	{\bar a_{\omega \ell m}^{out}}|{\bar 0}\ {\rm out}\rangle = 0
      \ .
\end{eqnarray}
Here, $|0\ {\rm in}\rangle$ is a natural vacuum in ``in region", while  
$|{\bar 0}\ {\rm out}\rangle$ is natural in ``out region".
 Other vacua are introduced merely for the computational purpose. 

Now we calculate the number of particles  created due to the time evolution of 
space-time in the presence of interaction. 
Given the natural vacuum $|0\ {\rm in}\rangle$,
 the number spectrum of $\varphi$ becomes 
\begin{eqnarray}
	\sum_{\ell m}\langle 0\ {\rm in}|{\bar N}_{\omega \ell m}^{out}|0\ {\rm in}\rangle &=&\sum_{\ell m}\langle 0\ {\rm in}|{\bar a}_{\omega \ell m}^{out\dagger}{\bar a}_{\omega \ell m}^{out}|0\ {\rm in}\rangle \nonumber\\
	&=&\sum_{\ell }\Bigl[\langle 0\ {\rm in}|{\bar c}_{\omega \ell }^{out\dagger}{\bar c}^{out}_{\omega \ell }|0\ {\rm in}\rangle 
  +\sum_{0<m\leq \ell }\left\{\langle 0\ {\rm in}|{\bar b}_{\omega \ell m}^{out\dagger}{\bar b}_{\omega \ell m}^{out}|0\ {\rm in}\rangle 
	+\langle 0\ {\rm in}|{\bar d}_{\omega \ell m}^{out\dagger}{\bar d}_{\omega \ell m}^{out}|0\ {\rm in}\rangle \right\}\Bigr]\ , \label{keisanshitai}
\end{eqnarray}

First, we evaluate $\langle 0\ {\rm in}|{\bar b}_{\omega \ell m}^{out\dagger}{\bar b}_{\omega \ell m}^{out}|0\ {\rm in}\rangle$. 
 The Bogolubov transformations (\ref{b_bogo}) for ${\bar b}_{\omega\ell m}^{out}$ 
 read
\begin{eqnarray}
	{\bar b}_{\omega \ell m}^{out}
   =\int d\omega' \left(
   \alpha_{\omega\ell,\omega^{'}\ell}^{*}b_{\omega^{'}\ell m}^{out}-\beta_{\omega\ell,\omega^{'}\ell}^{*}d_{\omega^{'}\ell m}^{out}{}^{\dagger} \right)  \ . 
\end{eqnarray}
Then, ${\bar b}_{\omega \ell m}^{out\dagger}{\bar b}_{\omega \ell m}^{out}$
 can be expressed by
\begin{eqnarray}
{\bar b}_{\omega \ell m}^{out\dagger}{\bar b}_{\omega \ell m}^{out}
      &=&
      \int d\omega' 
      \left|\beta_{\omega\ell,\omega^{'}\ell}\right|^2 \nonumber\\
  &&    +\int d\omega' \int d\omega''
  \Biggl[\alpha_{\omega\ell,\omega^{''}\ell}\alpha_{\omega\ell,\omega^{'}\ell}^{*}b_{\omega^{''}\ell m}^{out\dagger} b_{\omega^{'}\ell m}^{out}+\beta_{\omega\ell,\omega^{'}\ell}\beta_{\omega\ell,\omega^{''}\ell}^{*}d_{\omega^{''}\ell m}^{out\dagger} d_{\omega^{'}\ell m}^{out}\nonumber\\
	&\ &\hspace{4cm}-(\alpha_{\omega\ell,\omega^{''}\ell}\beta_{\omega\ell,\omega^{'}\ell}^{*}b_{\omega^{''}\ell m}^{out\dagger} d_{\omega^{'}\ell m}^{out\dagger}+h.c)\Biggr]
      \ . 
\end{eqnarray} 
The expectation value of this number operator in the state $|0\ {\rm in}\rangle$
is given by 
\begin{eqnarray}
 \langle 0\ {\rm in}|{\bar b}_{\omega \ell m}^{out\dagger}{\bar b}_{\omega \ell m}^{out}|0\ {\rm in}\rangle
&=&
\int d\omega' 
\left|\beta_{\omega\ell,\omega^{'}\ell}\right|^2 \nonumber\\
&&      +\int d\omega' \int d\omega''
\Biggl[\alpha_{\omega\ell,\omega^{''}\ell}\alpha_{\omega\ell,\omega^{'}\ell}^{*}\langle 0\ {\rm in}|b_{\omega^{''}\ell m}^{out\dagger} b_{\omega^{'}\ell m}^{out}|0\ {\rm in}\rangle \nonumber\\
	&\ &\hspace{2cm}+\beta_{\omega\ell,\omega^{'}\ell}\beta_{\omega\ell,\omega^{''}\ell}^{*}\langle 0\ {\rm in}|d_{\omega^{''}\ell m}^{out\dagger} d_{\omega^{'}\ell m}^{out}|0\ {\rm in}\rangle \nonumber\\
	&\ &\hspace{2.5cm}-2\Re\left\{\alpha_{\omega\ell,\omega^{''}\ell}\beta_{\omega\ell,\omega^{'}\ell}^{*}\langle 0\ {\rm in}|b_{\omega^{''}\ell m}^{out\dagger} d_{\omega^{'}\ell m}^{out\dagger}|0\ {\rm in}\rangle \right\}\Biggr]
      \ .\label{effect_from_b}
\end{eqnarray}
The first term of the right hand side of this equation is the
 same as (\ref{harm_beta}) and so this term represents the conventional
Hawking radiation. Hence, other terms in (\ref{effect_from_b}) are induced by 
interaction. In fact, these terms vanish if there is no interaction. 
Then, in order to know the effect of interaction on 
Hawking radiation, what we have to do is to calculate these terms. 
For later purpose, we decompose these terms into $\omega^{'}=\omega^{''}$ cases
 and $\omega^{'}\neq\omega^{''}$ cases. 
Using $L=\delta(\omega=0)$, for example, $\int d\omega' \int d\omega''
\left[\alpha_{\omega\ell,\omega^{''}\ell}\alpha_{\omega\ell,\omega^{'}\ell}^{*}\langle 0\ {\rm in}|b_{\omega^{''}\ell m}^{out\dagger} b_{\omega^{'}\ell m}^{out}|0\ {\rm in}\rangle\right]$ in (\ref{effect_from_b}) can be written as 
\begin{eqnarray}
&\ &\int d\omega^{'} d\omega^{''}\alpha_{\omega\ell,\omega^{''}\ell}\alpha_{\omega\ell,\omega^{'}\ell}^{*}\langle 0\ {\rm in}|b_{\omega^{''}\ell m}^{out\dagger} b_{\omega^{'}\ell m}^{out}|0\ {\rm in}\rangle  \nonumber\\
&=&\int d\omega^{'} d\omega^{''}\alpha_{\omega\ell,\omega^{''}\ell}\alpha_{\omega\ell,\omega^{'}\ell}^{*}\Biggl[\langle 0\ {\rm in}|b_{\omega^{''}\ell m}^{out\dagger} b_{\omega^{'}\ell m}^{out}|0\ {\rm in}\rangle-\frac{\delta (\omega^{'}-\omega^{''})}{L}\langle 0\ {\rm in}|b_{\omega^{''}\ell m}^{out\dagger} b_{\omega^{'}\ell m}^{out}|0\ {\rm in}\rangle\nonumber\\
&\ &\hspace{5cm}
+\frac{\delta (\omega^{'}-\omega^{''})}{L}\langle 0\ {\rm in}|b_{\omega^{''}\ell m}^{out\dagger} b_{\omega^{'}\ell m}^{out}|0\ {\rm in}\rangle\Biggr]\nonumber\\
&=&\frac{1}{L}\int d\omega^{'}\left[\left|\alpha_{\omega\ell,\omega^{'}\ell}\right|^2\langle 0\ {\rm in}|b_{\omega^{'}\ell m}^{out\dagger} b_{\omega^{'}\ell m}^{out}|0\ {\rm in}\rangle \right]\nonumber\\
&\ &\  +\int d\omega^{'} d\omega^{''}\alpha_{\omega\ell,\omega^{''}\ell}\alpha_{\omega\ell,\omega^{'}\ell}^{*}\Biggl[\langle 0\ {\rm in}|b_{\omega^{''}\ell m}^{out\dagger} b_{\omega^{'}\ell m}^{out}|0\ {\rm in}\rangle-\frac{\delta (\omega^{'}-\omega^{''})}{L}\langle 0\ {\rm in}|b_{\omega^{'}\ell m}^{out\dagger} b_{\omega^{'}\ell m}^{out}|0\ {\rm in}\rangle\Biggr]\ .
\label{dec_equal_neq}
\end{eqnarray}
The second term in the above equation reads
\begin{eqnarray}
\langle 0\ {\rm in}|b_{\omega^{''}\ell m}^{out\dagger} b_{\omega^{'}\ell m}^{out}|0\ {\rm in}\rangle-\frac{\delta (\omega^{'}-\omega^{''})}{L}\langle 0\ {\rm in}|b_{\omega^{'}\ell m}^{out\dagger} b_{\omega^{'}\ell m}^{out}|0\ {\rm in}\rangle =
\left\{
\begin{array}{l}
\langle 0\ {\rm in}|b_{\omega^{''}\ell m}^{out\dagger} b_{\omega^{'}\ell m}^{out}|0\ {\rm in}\rangle \quad ({\rm for\ }\omega^{'}\neq\omega^{''})\\
0\hspace{3.6cm}({\rm for\ }\omega^{'}=\omega^{''})
\end{array}
\right. \ .
\label{}
\end{eqnarray}
Therefore, Eq.(\ref{dec_equal_neq}) can be rewritten as 
\begin{eqnarray}
\frac{1}{L}\int d\omega^{'}\left[\left|\alpha_{\omega\ell,\omega^{'}\ell}\right|^2\langle 0\ {\rm in}|b_{\omega^{'}\ell m}^{out\dagger} b_{\omega^{'}\ell m}^{out}|0\ {\rm in}\rangle \right] +\int_{\omega^{'}\neq\omega^{''}} d\omega^{'} d\omega^{''}\left[\alpha_{\omega\ell,\omega^{''}\ell}\alpha_{\omega\ell,\omega^{'}\ell}^{*}\langle 0\ {\rm in}|b_{\omega^{''}\ell m}^{out\dagger} b_{\omega^{'}\ell m}^{out}|0\ {\rm in}\rangle\right]\ .
\label{effect_b_int}
\end{eqnarray}
Similarly, we can evaluate expectation values
$\langle 0\ {\rm in}|{\bar d}_{\omega \ell m}^{out\dagger}{\bar d}_{\omega \ell m}^{out}|0\ {\rm in}\rangle$ and 
$\langle 0\ {\rm in}|{\bar c}_{\omega \ell }^{out\dagger}{\bar c}^{out}_{\omega \ell }|0\ {\rm in}\rangle$ in (\ref{keisanshitai}) 
 as 
\begin{eqnarray}
\langle 0\ {\rm in}|{\bar d}_{\omega \ell m}^{out\dagger}{\bar d}_{\omega \ell m}^{out}|0\ {\rm in}\rangle
      &=&\int d\omega' 
      \left|\beta_{\omega\ell,\omega^{'}\ell}\right|^2
      + \int d\omega' \int d\omega''
      \Biggl[\alpha_{\omega\ell,\omega^{''}\ell}\alpha_{\omega\ell,\omega^{'}\ell}^{*}\langle 0\ {\rm in}|d_{\omega^{''}\ell m}^{out\dagger} d_{\omega^{'}\ell m}^{out}|0\ {\rm in}\rangle \nonumber\\
	&\ &\hspace{4cm}+\beta_{\omega\ell,\omega^{'}\ell}\beta_{\omega\ell,\omega^{''}\ell}^{*}\langle 0\ {\rm in}|b_{\omega^{''}\ell m}^{out\dagger} b_{\omega^{'}\ell m}^{out}|0\ {\rm in}\rangle \nonumber\\
	&\ &\hspace{4cm}-2\Re\left\{\alpha_{\omega\ell,\omega^{''}\ell}\beta_{\omega\ell,\omega^{'}\ell}^{*}\langle 0\ {\rm in}|d_{\omega^{''}\ell m}^{out\dagger} b_{\omega^{'}\ell m}^{out\dagger}|0\ {\rm in}\rangle \right\}\Biggr]
      \ ,\nonumber\\ 
 \langle 0\ {\rm in}|{\bar c}_{\omega \ell }^{out\dagger}{\bar c}^{out}_{\omega \ell }|0\ {\rm in}\rangle
      &=&\int d\omega' 
      \left|\beta_{\omega\ell,\omega^{'}\ell}\right|^2
      +\int d\omega' \int d\omega''
      \Biggl[(\alpha_{\omega\ell,\omega^{''}\ell}\alpha_{\omega\ell,\omega^{'}\ell}^{*}+\beta_{\omega\ell,\omega^{'}\ell}\beta_{\omega\ell,\omega^{''}\ell}^{*})\langle 0\ {\rm in}|c_{\omega^{''} \ell}^{out\dagger} c_{\omega^{'} \ell}^{out}|0\ {\rm in}\rangle \nonumber\\
	&\ &\hspace{4cm}-2\Re\left\{\alpha_{\omega\ell,\omega^{''}\ell}\beta_{\omega\ell,\omega^{'}\ell}^{*}\langle 0\ {\rm in}|c_{\omega^{''}\ell }^{out\dagger} c_{\omega^{'}\ell }^{out\dagger}|0\ {\rm in}\rangle \right\}\Biggr]
      \ .  \label{jyuuyou}
\end{eqnarray}
These results are almost the same as $\langle 0\ {\rm in}|{\bar b}_{\omega \ell m}^{out\dagger}{\bar b}_{\omega \ell m}^{out}|0\ {\rm in}\rangle$ 
and the terms other than the first term of the above formula have information of
 interaction. 
Substituting (\ref{effect_from_b}) and (\ref{jyuuyou}) into (\ref{keisanshitai}), we 
 find the deviation from the Planck spectrum  due to the interaction as 
\begin{eqnarray}
&\ &\sum_{\ell}\int d\omega' \int d\omega''
\Biggl[(\alpha_{\omega\ell,\omega^{''}\ell}\alpha_{\omega\ell,\omega^{'}\ell}^{*}+\beta_{\omega\ell,\omega^{'}\ell}\beta_{\omega\ell,\omega^{''}\ell}^{*})
 \Biggl\{\langle 0\ {\rm in}|c_{\omega^{''} \ell}^{out\dagger} c_{\omega^{'} \ell}^{out}|0\ {\rm in}\rangle\nonumber\\
&\ &\hspace{4cm}+\sum_{0<m\leq \ell}\left(
\langle 0\ {\rm in}|b_{\omega^{''}\ell m}^{out\dagger} b_{\omega^{'}\ell m}^{out}|0\ {\rm in}\rangle+
\langle 0\ {\rm in}|d_{\omega^{''}\ell m}^{out\dagger} d_{\omega^{'}\ell m}^{out}|0\ {\rm in}\rangle\right)\Biggr\}\nonumber\\
&\ &\hspace{1cm}
-2\Re\Biggl\{(\alpha_{\omega\ell,\omega^{''}\ell}\beta_{\omega\ell,\omega^{'}\ell}^{*})
\langle 0\ {\rm in}|c_{\omega^{''}\ell }^{out\dagger} c_{\omega^{'}\ell }^{out\dagger}|0\ {\rm in}\rangle
\nonumber\\
&\ &\hspace{3cm}
+(\alpha_{\omega\ell,\omega^{''}\ell}\beta_{\omega\ell,\omega^{'}\ell}^{*})
\sum_{0<m\leq \ell}\biggl(
\langle 0\ {\rm in}|b_{\omega^{''}\ell m}^{out\dagger} d_{\omega^{'}\ell m}^{out\dagger}|0\ {\rm in}\rangle
+\langle 0\ {\rm in}|d_{\omega^{''}\ell m}^{out\dagger} b_{\omega^{'}\ell m}^{out\dagger}|0\ {\rm in}\rangle\biggr)
\Biggl\}
\Biggr] \ .
\label{keisanshitai2}
\end{eqnarray}

In order to calculate $\langle 0\ {\rm in}|b_{\omega^{''}\ell m}^{out\dagger} b_{\omega^{'}\ell m}^{out}|0\ {\rm in}\rangle $ etc., it is useful to define the basis
\begin{eqnarray}
	|n_b\cdots; n_c\cdots; n_d\cdots {\rm out}\rangle=\frac{1}{\sqrt{n_b!\cdots n_c! \cdots n_d!\cdots}}\left(b^{out\dagger}\right)^{n_b}\cdots\left(c^{out\dagger}\right)^{n_c}\cdots\left(d^{out\dagger}\right)^{n_d}\cdots |0\ {\rm out}\rangle 
      \ , \nonumber 
\end{eqnarray}
which satisfy the completeness relation 
\begin{eqnarray}
1=\sum \frac{|n_b\cdots;n_c\cdots;n_d\cdots\ {\rm out}\rangle\langle n_b\cdots;n_c\cdots;n_d\cdots\ {\rm out}|}{\langle n_b\cdots;n_c\cdots;n_d\cdots\ {\rm out}| n_b\cdots;n_c\cdots;n_d\cdots\ {\rm out}\rangle}
\ .
\label{completenes}
\end{eqnarray}
Substituting (\ref{completenes}) into just after $\langle 0\ {\rm in}|$ 
and just before $|0\ {\rm in}\rangle$, for example, we can calculate
$\langle 0\ {\rm in}|b_{\omega^{''}\ell m}^{out\dagger} b_{\omega^{'}\ell m}^{out}|0\ {\rm in}\rangle $ as follows:
\begin{itemize}
\item If $\omega^{''}=\omega^{'}$, 
\begin{eqnarray}
\langle 0\ {\rm in}|b_{\omega^{'}\ell m}^{out\dagger} b_{\omega^{'}\ell m}^{out}|0\ {\rm in}\rangle 
=\sum_{n_{b\omega^{'}\ell m}}\sum_ {\{others\}}n_{b\omega^{'}\ell m}\frac{L\left|\langle 0\ {\rm in}|n_{b\omega^{'}\ell m};\{others\}\ {\rm out}\rangle \right|^2}{\langle n_{b\omega^{'}\ell m};\{others\}\ {\rm out}|n_{b\omega^{'}\ell m};\{others\}\ {\rm out}\rangle }\ , \label{hituyou1}
\end{eqnarray}
where $L=\delta(\omega=0)$.
\item If $\omega^{'}\neq\omega^{''}$, 
\begin{eqnarray}
\langle 0\ {\rm in}|b_{\omega^{''}\ell m}^{out\dagger} b_{\omega^{'}\ell m}^{out}|0\ {\rm in}\rangle
&=&\sum_{n^{'}_{b\omega^{'}\ell m}}\sum_{n^{''}_{b\omega^{''}\ell m}}\sum_{ \{others\}}\frac{L}{\langle n^{'}_{b\omega^{'}\ell m},n^{''}_{b\omega^{''}\ell m};\{others\}\ {\rm out}|n^{'}_{b\omega^{'}\ell m},n^{''}_{b\omega^{''}\ell m};\{others\}\ {\rm out}\rangle} \nonumber\\ 
&\ &\times\Biggl [\sqrt{n^{''}_{b\omega^{''}\ell m}(n^{'}_{b\omega^{'}\ell m}+1)}
\langle 0\ {\rm in}|n^{'}_{b\omega^{'}\ell m},n^{''}_{b\omega^{''}\ell m};\{others\}\ {\rm out}\rangle\nonumber\\
&\ &\hspace{3.5cm} \times\langle n^{'}_{b\omega^{'}\ell m}+1,n^{''}_{b\omega^{''}\ell m}-1;\{others\}\ {\rm out}|0\ {\rm in}\rangle \Biggr]\ . \label{hituyou2}
\end{eqnarray}
\end{itemize} 
Similar rules apply to $\langle 0\ {\rm in}|c_{\omega^{''}\ell }^{out\dagger} c_{\omega^{'}\ell }^{out}|0\ {\rm in}\rangle $ and $\langle 0\ {\rm in}|d_{\omega^{''}\ell m}^{out\dagger} d_{\omega^{'}\ell m}^{out}|0\ {\rm in}\rangle $.
In the same way, $\langle 0\ {\rm in}|c_{\omega^{''}\ell }^{out\dagger} c_{\omega^{'}\ell }^{out\dagger}|0\ {\rm in}\rangle$
becomes
\begin{itemize}
\item If $\omega^{'}=\omega^{''}$,
\begin{eqnarray}
\langle 0\ {\rm in}|c_{\omega^{'}\ell }^{out\dagger} c_{\omega^{'}\ell }^{out\dagger}|0\ {\rm in}\rangle
&=&\sum_{n_{c\omega^{'}\ell }}\sum_{\{others\}}
L  \sqrt{(n_{c\omega^{'}\ell }+1)(n_{c\omega^{'}\ell }+2)}\nonumber\\
&\ &\times\frac{\langle 0\ {\rm in}|n_{c\omega^{'}\ell }+2;\{others\}\ {\rm out}\rangle \langle n_{c\omega^{'}\ell };\{others\}\ {\rm out}|0 {\rm\ in}\rangle}{\langle n_{c\omega^{'}\ell };\{others\}\ {\rm out}|n_{c\omega^{'}\ell };\{others\}\ {\rm out}\rangle}
\ .  \label{hituyou3}
\end{eqnarray}

\item If $\omega^{'}\neq\omega^{''}$, 
\begin{eqnarray}
\langle 0\ {\rm in}|c_{\omega^{''}\ell }^{out\dagger} c_{\omega^{'}\ell }^{out\dagger}|0\ {\rm in}\rangle
      &=&\sum_{n^{'}_{c\omega^{'}\ell }}\sum_{ n^{''}_{c\omega^{''}\ell }}\sum_{\{others\}}
      \frac{L}{\langle n^{'}_{c\omega^{'}\ell },n^{''}_{c\omega^{''}\ell };\{others\}\ {\rm out}|n^{'}_{c\omega^{'}\ell },n^{''}_{c\omega^{''}\ell };\{others\}\ {\rm out}\rangle}\nonumber\\
      &\ &\times\Biggl[\sqrt{(n^{'}_{c\omega^{'}\ell }+1)(n^{''}_{c\omega^{''}\ell }+1)}
      \langle 0\ {\rm in}|n^{'}_{c\omega^{'}\ell }+1,n^{''}_{c\omega^{''}\ell }+1;\{others\}\ {\rm out}\rangle \nonumber\\
	&\ &\hspace{3.5cm}\times\langle n^{'}_{c\omega^{'}\ell },n^{''}_{c\omega^{''}\ell };\{others\}\ {\rm out}|0\ {\rm in}\rangle \Biggr]\ . \label{hituyou4}
\end{eqnarray}
\end{itemize}
And also $\langle 0\ {\rm in}|d_{\omega^{'}\ell  m}^{out\dagger}b_{\omega^{''}\ell  m}^{out\dagger}|0\ {\rm in}\rangle $ is given by
\begin{eqnarray}
\langle 0\ {\rm in}|d_{\omega^{'}\ell  m}^{out\dagger}b_{\omega^{''}\ell  m}^{out\dagger}|0\ {\rm in}\rangle 
 &=&\sum_{n^{'}_{d\omega^{'}\ell m}}\sum_{ n^{''}_{b\omega^{''}\ell m}}\sum_{\{others\}}
 \frac{L}{\langle n^{'}_{d\omega^{'}\ell m},n^{''}_{b\omega^{''}\ell m};\{others\}\ {\rm out}|n^{'}_{d\omega^{'}\ell m},n^{''}_{b\omega^{''}\ell m};\{others\}\ {\rm out}\rangle}\nonumber\\
 &\ &\times\Biggl[\sqrt{(n^{'}_{d\omega^{'}\ell m}+1)(n^{''}_{b\omega^{''}\ell m}+1)}
 \langle 0\ {\rm in}|n^{'}_{d\omega^{'}\ell m}+1,n^{''}_{b\omega^{''}\ell m}+1;\{others\}\ {\rm out}\rangle \nonumber\\
	&\ &\hspace{3.5cm}\times\langle n^{'}_{d\omega^{'}\ell m},n^{''}_{b\omega^{''}\ell m};\{others\}\ {\rm out}|0\ {\rm in}\rangle \Biggr]
      \ . \label{hituyou5}
\end{eqnarray}
These results will be used in the calculation of S-matrix. 
Before moving on to calculations of S-matrix elements
and clarify the effect of interaction on Hawking radiation,
we need to calculate the interaction Hamiltonian. 

\subsection{Cubic Interaction terms}
\label{calculation}

In this subsection, we derive cubic interaction terms in the following action 
\begin{eqnarray}
-\frac{1}{2}\int d^4 x \sqrt{-g}g^{\mu\nu}\varphi_{,\mu}\varphi_{,\nu}\ . 
\label{scalar-action}
\end{eqnarray}
We put the metric 
\begin{eqnarray}
g_{\mu\nu}=g^{(0)}_{\mu\nu}+h_{\mu\nu}\ ,
\end{eqnarray}
where $g^{(0)}_{\mu\nu}$ is the background Schwarzschild metric. 
In the Regge-Wheeler and the convenient gauge, $h_{\mu\nu}$ reads 
\begin{eqnarray}
	 h_{\mu\nu}=
	\left(
	\begin{array}{cc|cc}
	f{\bar H}&H_1&\multicolumn{2}{c}{v_{a}}\\
	sym&H/f&\multicolumn{2}{c}{w_{a}+w_{|a}}\\ \hline
	sym&sym&0&0\\
	sym&sym&0&0
	\end{array} \right)
      \ .\label{total-pert-metric}
\end{eqnarray}  
Using this gauge, we obtain the cubic action 
\begin{eqnarray}
& &
-\frac{1}{2}\int r^2\sqrt{\gamma}\frac{1}{2}\left(H-{\bar H}\right)\left[-\frac{1}{f}{\dot \varphi}^2+f\varphi^{'2}+\frac{\gamma^{ab}}{r^2}\varphi_{|a}\varphi_{|b}\right]\nonumber\\
& & \qquad
- \frac{1}{2}\int r^2\sqrt{\gamma}\left[\frac{{\bar H}}{f}{\dot \varphi}^2+2H_1\varphi^{'}{\dot \varphi}+\frac{2}{r^2f}v^{a}{\dot \varphi}\varphi_{|a}-fH\varphi^{'2}-\frac{2f}{r^2}\left(w^{|a}+w^{a}\right)\varphi^{'}\varphi_{|a}\right]
\label{all_interaction}
\end{eqnarray}
We see that scalar and vector perturbations are decoupled, so 
we will calculate them separately. 

\subsubsection{ Vector perturbations}

In the action (\ref{all_interaction}), terms including vector perturbations are
\begin{eqnarray}
\int d^4x \sqrt{\gamma}\left[-\frac{1}{f}v^{a}{\dot \varphi}\varphi_{|a}+fw^a\varphi^{'}\varphi_{|a}\right] \ .
\end{eqnarray}
We expand these variables by harmonics; for example,
\begin{eqnarray}
w^a=\sum_{\ell m}w_{\ell m}V^a_{\ell m}  \ ,\quad 
 \varphi=\sum_{\ell m}\varphi_{\ell m}Y_{\ell m}\ , \nonumber
\end{eqnarray}
where $V^a_{\ell m}$ are vector harmonics and $Y_{\ell m}$ are spherical harmonics. 
The result is 
\begin{eqnarray}
D^{\ell m}_{\ell^{'}m^{'};\ell^{''}m^{''}}\int dr dt\left[-\frac{1}{f}v_{\ell m}^{*}{\dot \varphi}_{\ell^{'} m^{'}}\varphi_{\ell^{''}m^{''}}+fw^{*}_{\ell m}\varphi_{\ell^{'}m^{'}}^{'}\varphi_{\ell^{''}m^{''}}\right]\ ,\label{odd_D}
\end{eqnarray}
where 
\begin{eqnarray}
D^{\ell m}_{\ell^{'}m^{'};\ell^{''}m^{''}}\equiv \int d\Theta d\Phi\sqrt{\gamma}V^{a*}_{\ell m}Y_{\ell^{'} m^{'}}Y_{\ell^{''}m^{''}|a}\ .
\end{eqnarray}
It is easy to show the relation
 $D^{\ell m}_{\ell^{'}m^{'};\ell^{''}m^{''}}=-D^{\ell m}_{\ell^{''}m^{''};\ell^{'}m^{'}}$ 
 using the property $V^{a}_{\ell m}{}_{|a}=0$. 
We also see that $D^{\ell m}_{\ell^{'}m^{'};\ell^{''}m^{''}}$ are pure imaginary numbers
 proportional to $\delta_{m,m^{'}+m^{''}}$ because both vector harmonics and spherical harmonics are proportional to $e^{im\Phi}$. 
Furthermore, because $D^{\ell m}_{\ell^{'}m^{'};\ell^{''}m^{''}}$ are pure imaginary and both harmonics have property such as 
$Y_{\ell m}=(-1)^mY_{\ell m}^{*}$,  $D^{\ell m}_{\ell^{'}m^{'};\ell^{''}m^{''}}$, 
we can derive the following formula, 
$D^{\ell m}_{\ell ^{'}m^{'};\ell ^{''}m^{''}}=-D^{\ell m}_{\ell ^{'}m^{'};\ell ^{''}m^{''}}{}^{*}=(-1)^{m+m^{'}+m^{''}+1}D^{\ell -m}_{\ell ^{'}-m^{'};\ell ^{''}-m^{''}}$.  
Note that we omitted $\sum_{\ell m}\sum_{\ell^{'}m^{'}}\sum_{\ell^{''}m^{''}}$
 in (\ref{odd_D}). 
Thus, using (\ref{odd_master}) and $\varphi_{\ell m}=\phi_{\ell m}/r$, 
we can get cubic interaction terms  
\begin{eqnarray}
D^{\ell m}_{\ell^{'}m^{'};\ell^{''}m^{''}}\int dr^{*}dt\frac{1}{\sqrt{\lambda(\lambda+1)}}\left[-\frac{f}{r^2}\psi^{RW*}_{\ell m}{\dot \phi}_{\ell^{'}m^{'}}\phi_{\ell^{''}m^{''}}+\frac{1}{r}\psi^{RW*}_{\ell m}{\dot \phi}_{\ell^{'}m^{'}}\partial_{r^{*}}\phi_{\ell^{''}m^{''}}\right]\ .
\label{odd_interaction}
\end{eqnarray} 

\subsubsection{Scalar perturbations}

Terms including scalar perturbations in the action (\ref{all_interaction}) read
\begin{eqnarray}
&\ &-\frac{1}{2}\int r^2\sqrt{\gamma}\frac{1}{2}\left(H-{\bar H}\right)\left[-\frac{1}{f}{\dot \varphi}^2+f\varphi^{'2}+\frac{\gamma^{ab}}{r^2}\varphi_{|a}\varphi_{|b}\right]\nonumber\\
&\ &\hspace{0.5cm}-\frac{1}{2}\int r^2\sqrt{\gamma}\left[\frac{{\bar H}}{f}{\dot \varphi}^2+2H_1\varphi^{'}{\dot \varphi}-fH\varphi^{'2}-\frac{2f}{r^2}w^{|a}\varphi^{'}\varphi_{|a}\right]\ .
\end{eqnarray}
Using expansion by spherical harmonics like
\begin{eqnarray}
H=\sum H_{\ell m}Y_{\ell m} \ , \quad
\varphi=\sum \varphi_{\ell m}Y_{\ell m} \ ,
\nonumber
\end{eqnarray}
we can write the action in the following form 
\begin{eqnarray}
&\ &-\frac{1}{2}C^{\ell m}_{\ell^{'}m^{'};\ell^{''}m^{''}}\int dr dt r^{2}\Biggl[
\frac{1}{f}\left((fw_{\ell m}^{*})^{'}-H_{\ell m}^{*}\right){\dot \varphi}_{\ell^{'} m^{'}}{\dot \varphi_{\ell^{''}m^{''}}}
+f\left((fw_{\ell m}^{*})^{'}-H_{\ell m}^{*}\right)\varphi_{\ell^{'} m^{'}}^{'}\varphi_{\ell^{''}m^{''}}^{'}\nonumber\\
&\ &\hspace{5cm}2\left({\dot w}_{\ell m}^{*}-\frac{2r{\dot H}_{\ell m}^{*}}{\gamma_s}\right)
+2fw_{\ell m}^{*}\varphi_{\ell^{'}m^{'}}^{'}\varphi_{\ell m}^{|a}{}_{|a}
\Biggr]\ ,
\label{even_interaction1}
\end{eqnarray}
where we have defined
\begin{eqnarray}
C^{\ell m}_{\ell^{'}m^{'};\ell^{''}m^{''}}\equiv\int \sqrt{\gamma}Y_{\ell m}^{*}Y_{\ell^{'}m^{'}}Y_{\ell^{''}m^{''}}\ .
\label{def_of_C}
\end{eqnarray}
These coefficients $C^{\ell m}_{\ell^{'}m^{'};\ell^{''}m^{''}}$ have almost the 
same property as $D^{\ell m}_{\ell^{'}m^{'};\ell^{''}m^{''}}$. 
It is easy to derive the relation
 $C^{\ell m}_{\ell^{'}m^{'};\ell^{''}m^{''}}=C^{\ell m}_{\ell^{''}m^{''};\ell^{'}m^{'}}$. 
Because of $Y_{\ell m}\propto e^{im\Phi}$, $C^{\ell m}_{\ell^{'}m^{'};\ell^{''}m^{''}}$
 are real numbers proportional to $\delta_{m,m^{'}+m^{''}}$. 
The reality of $C^{\ell m}_{\ell^{'}m^{'};\ell^{''}m^{''}}$  implies that 
$C^{\ell m}_{\ell ^{'}m^{'};\ell ^{''}m^{''}}
=C^{\ell m}_{\ell ^{'}m^{'};\ell ^{''}m^{''}}{}^{*}
=(-1)^{m+m^{'}+m^{''}}C^{\ell -m}_{\ell ^{'}-m^{'};\ell ^{''}-m^{''}}$.
Using (\ref{even_master1}), (\ref{even_master2}) and $\varphi_{\ell m}=\phi_{\ell m}/r$,
we can derive cubic interaction terms for scalar perturbations 
\begin{eqnarray}
-\frac{1}{2}C^{\ell m}_{\ell^{'}m^{'};\ell^{''}m^{''}}\int dr^{*}dt\sqrt{\frac{\lambda+1}{\lambda}}\Biggl[&&
\left(-\frac{2r}{\gamma_sf}\partial_{r^{*}}^2\psi_{\ell m}^{Z*}+\left(\frac{rf^{'}}{\gamma_s f}+A(r)\right)\partial_{r^{*}}\psi_{\ell m}^{Z*}+B(r)\psi^{Z*}_{\ell m}\right){\dot \phi}_{\ell^{'} m^{'}}{\dot \phi}_{\ell^{''} m^{''}}\nonumber\\
&\ &+\left(-\frac{2r}{\gamma_s f}\partial_{r^{*}}^{2}\psi_{\ell m}^{Z*}+\left(\frac{rf^{'}}{\gamma_s f}+C(r)\right)\partial_{r^{*}}\psi_{\ell m}^{Z*}+D(r)\psi^{Z*}_{\ell m}\right)\nonumber\\
&\ &\hspace{0.5cm}\times\left(\frac{f^2}{r^2}\phi_{\ell^{'}m^{'}}\phi_{\ell^{''}m^{''}}-\frac{2f}{r}\phi_{\ell^{'}m^{'}}\partial_{r^{*}}\phi_{\ell^{''}m^{''}}+\partial_{r^{*}} \phi_{\ell^{'}m^{'}} \partial_{r^{*}}\phi_{\ell^{''}m^{''}}\right)\nonumber\\
&\ &+\left(\frac{4r}{\gamma_s f}\partial_{r^{*}}{\dot \psi}_{\ell m}^{Z*}-\left(\frac{2rf^{'}}{\gamma_s f}-E(r)\right){\dot \psi}_{\ell m}^{Z*}\right){\dot \phi}_{\ell^{'}m^{'}}\left(-\frac{f}{r}\phi_{\ell^{''}m^{''}}+\partial_{r^{*}}\phi_{\ell^{''}m^{''}}\right)
\Biggr]\ ,\nonumber\\
\label{even_interaction}
\end{eqnarray}
where 
\begin{eqnarray}
&& A(r)=-\frac{2}{\gamma_s}-\frac{2(\gamma_s-2)}{\gamma_sT} \ , \quad
B(r)=\frac{T{'}}{\gamma_s}-2\left(\frac{frT{'}}{\gamma_s T}\right)^{'}\ ,\quad
C(r)=\frac{2}{\gamma_s}-\frac{2(\gamma_s-2)}{\gamma_s}\frac{1}{T}\ ,\nonumber\\
&&
D(r)=\frac{T^{'}}{T}+\frac{2}{r}-\frac{rf^{'}T^{'}}{\gamma_sT}-\frac{2rf}{\gamma_s}\left(\frac{T^{'}}{T}\right)^{'}\ ,\quad
E(r)=\frac{4(\gamma_s-2)}{\gamma_s T}\ .
\end{eqnarray}
These function $A(r)$ etc. are regular outside the horizon. However, the other coupling functions in (\ref{even_interaction}) are 
singular at the horizon because these terms are proportional to $1/f(r)$. 
We also have a relation
\begin{eqnarray}
A(r)+C(r)+E(r)=0\ .
\label{ACE}
\end{eqnarray}
Note that we ignored the terms which is related to equation of motion for
 $\varphi$ in (\ref{even_interaction}). This is because 
these terms do not affect  our calculation. 

\subsubsection{Total cubic interaction}
 
We notice that the cubic interaction Hamiltonian have the following structure  
\begin{eqnarray}
H_{int}&=&C^{\ell m}_{\ell^{'}m^{'};\ell^{''}m^{''}}\sum_{int}\int dr^{*}dt h_{int}(r)\psi^{Z*}_{\ell m}\phi_{\ell^{'}m^{'}}{\bar \phi}_{\ell^{''}m^{''}}
\nonumber\\&\ &+D^{\ell m}_{\ell^{'}m^{'};\ell^{''}m^{''}}\sum_{int}\int dr^{*}dt w_{int}(r)\psi^{RW*}_{\ell m}\phi_{\ell^{'}m^{'}}{\bar \phi}_{\ell^{''}m^{''}}\ .
\label{7_interaction}
\end{eqnarray}
From the interaction Hamiltonian (\ref{7_interaction}), 
the first term of (\ref{odd_interaction}) can be reproduced by the replacement 
\begin{eqnarray}
w_{int}(r) \Rightarrow -\frac{1}{\sqrt{\lambda(\lambda+1)}}\frac{f}{r^2} \ ,\quad
\psi^{RW}_{\ell m} \Rightarrow \psi^{RW}_{\ell m}  \ , \quad
\phi_{\ell^{'}m^{'}}\Rightarrow  {\dot \phi}_{\ell^{'}m^{'}} \ , \quad
{\bar \phi}_{\ell^{''}m^{''}}\Rightarrow \phi_{\ell^{''}m^{''}}  
\ .
\end{eqnarray}
Note that $m$ takes the integer in the range $-\ell\leq m\leq \ell$; 
$m^{'}$ and $m^{''}$ take the value in the same range. As is shown in Sec.\ref{2}, 
 $\phi_{\ell m}$ ($m>0$) are complex scalar fields and $\phi_{\ell 0}$ are 
real scalar field. By using reality conditions, we can rewrite the interaction 
Hamiltonian (\ref{7_interaction}) using only $m\geq 0$ modes. 
From now on, $m$ moves in the range $0<m\leq \ell$ unless we specify its range. 
With this notation, each term for scalar perturbation in (\ref{7_interaction}) can be rewritten as
\begin{eqnarray}
& &\sum_{\ell ,\ell ^{'},\ell ^{''}}\sum_{\ell \leq m\leq \ell}\sum_{\ell^{'} \leq m^{'}\leq \ell^{'}}\sum_{\ell^{''} \leq m^{''}\leq \ell^{''}}
C^{\ell m}_{\ell ^{'}m^{'};\ell ^{''}m^{''}}\psi^{\dagger}_{\ell m}\phi_{\ell ^{'}m^{'}}{\bar \phi}_{\ell ^{''}m^{''}}\nonumber\\
&& = C^{\ell m}_{\ell ^{'}m^{'};\ell ^{''}m^{''}}\psi^{\dagger}_{\ell m}\phi_{\ell ^{'}m^{'}}{\bar \phi}_{\ell ^{''}m^{''}}
+C^{\ell 0}_{\ell ^{'}0;\ell ^{''}0}\psi_{\ell 0}\phi_{\ell ^{'}0}{\bar\phi}_{\ell ^{''}0}
+C^{\ell m}_{\ell ^{'}m^{'};\ell ^{''}m^{''}}\psi_{\ell m}\phi^{\dagger}_{\ell ^{'}m^{'}}{\bar\phi}_{\ell ^{''}m^{''}}^{\dagger}\nonumber\\
& & \qquad +C^{\ell m}_{\ell ^{'}m^{'};\ell ^{''}0}(\psi^{\dagger}_{\ell m}\phi_{\ell ^{'}m^{'}}{\bar \phi}_{\ell ^{''}0}+\psi^{\dagger}_{\ell m}{\bar \phi}_{\ell ^{'}m^{'}}\phi_{\ell ^{''}0})+
C^{\ell m}_{\ell ^{'}m^{'};\ell ^{''}0}(\psi_{\ell m}\phi_{\ell ^{'}m^{'}}^{\dagger}{\bar \phi}_{\ell ^{''}0}+
\psi_{\ell m}{\bar \phi}_{\ell ^{'}m^{'}}^{\dagger}\phi_{\ell ^{''}0})\nonumber\\
&& \qquad+ (-1)^{m^{''}}C^{\ell m}_{\ell ^{'}m^{'};\ell ^{''}-m^{''}}(\psi^{\dagger}_{\ell m}\phi_{\ell ^{'}m^{'}}{\bar \phi}^{\dagger}_{\ell ^{''}m^{''}}+
\psi^{\dagger}_{\ell m}{\bar \phi}_{\ell ^{'}m^{'}} \phi^{\dagger}_{\ell ^{''}m^{''}})\nonumber\\
& & \hspace{1cm}+(-1)^{m^{''}}C^{\ell 0}_{\ell ^{'}m^{'};\ell ^{''}-m^{''}}(\psi_{\ell 0}\phi_{\ell ^{'}m^{'}}{\bar \phi}_{\ell ^{''}m^{''}}^{\dagger}+
\psi_{\ell 0}{\bar \phi}_{\ell ^{'}m^{'}}\phi_{\ell ^{''}m^{''}}^{\dagger})
\nonumber\\&\ &\hspace{2cm}+(-1)^{m^{''}}C^{\ell m}_{\ell^{'}m^{'};\ell^{''}-m^{''}}(\psi_{\ell m}\phi_{\ell ^{'}m^{'}}^{\dagger}{\bar \phi}_{\ell ^{''}m^{''}}+
\psi_{\ell m}{\bar \phi}_{\ell ^{'}m^{'}}^{\dagger} \phi_{\ell ^{''}m^{''}})\ .
\label{C_expand}
\end{eqnarray}
In this decomposition, we use the property
 $C^{\ell m}_{\ell^{'}m^{'};\ell^{''}m^{''}}$ and reality conditions for 
 $\psi_{\ell m}^{Z}$ and $\phi_{\ell m}$.  
Note that there exist other terms in (\ref{C_expand}) such as
 $C^{\ell 0}_{\ell ^{'}m^{'};\ell ^{''}m^{''}}\psi_{\ell 0}^{Z}\phi_{\ell^{'}m^{'}}
 \phi_{\ell^{''}m^{''}}$. 
However, such terms must vanish due to the fact
 $C^{\ell m}_{\ell ^{'}m^{'};\ell ^{''}m^{''}}\propto \delta_{m,m^{'}+m^{''}}$. 

Similarly, the terms related to vector perturbations read 
\begin{eqnarray}
& &\sum_{\ell ,\ell ^{'},\ell ^{''}}\sum_{m,m^{'},m^{''}}D^{\ell m}_{\ell ^{'}m^{'};\ell ^{''}m^{''}}\psi^{\dagger}_{\ell m}\phi_{\ell ^{'}m^{'}}{\bar \phi}_{\ell ^{''}m^{''}}\nonumber\\
&&=D^{\ell m}_{\ell ^{'}m^{'};\ell ^{''}m^{''}}\psi^{\dagger}_{\ell m}\phi_{\ell ^{'}m^{'}}{\bar \phi}_{\ell ^{''}m^{''}}
+D^{\ell 0}_{\ell ^{'}0;\ell ^{''}0}\psi_{\ell 0}\phi_{\ell ^{'}0}{\bar\phi}_{\ell ^{''}0}
-D^{\ell m}_{\ell ^{'}m^{'};\ell ^{''}m^{''}}\psi_{\ell m}\phi^{\dagger}_{\ell ^{'}m^{'}}{\bar\phi}_{\ell ^{''}m^{''}}^{\dagger}\nonumber\\
&&\qquad +D^{\ell m}_{\ell ^{'}m^{'};\ell ^{''}0}(\psi^{\dagger}_{\ell m}\phi_{\ell ^{'}m^{'}}{\bar \phi}_{\ell ^{''}0}-
\psi^{\dagger}_{\ell m}{\bar \phi}_{\ell ^{'}m^{'}}\phi_{\ell ^{''}0})-
D^{\ell m}_{\ell ^{'}m^{'};\ell ^{''}0}(\psi_{\ell m}\phi_{\ell ^{'}m^{'}}^{\dagger}{\bar \phi}_{\ell ^{''}0}-
\psi_{\ell m}{\bar \phi}_{\ell ^{'}m^{'}}^{\dagger}\phi_{\ell ^{''}0})\nonumber\\
&&\qquad +(-1)^{m^{''}}D^{\ell m}_{\ell ^{'}m^{'};\ell ^{''}-m^{''}}(\psi^{\dagger}_{\ell m}\phi_{\ell ^{'}m^{'}}{\bar \phi}^{\dagger}_{\ell ^{''}m^{''}}-
\psi^{\dagger}_{\ell m}{\bar \phi}_{\ell ^{'}m^{'}} \phi^{\dagger}_{\ell ^{''}m^{''}})\nonumber\\
& & \hspace{1cm}+(-1)^{m^{''}}D^{\ell 0}_{\ell ^{'}m^{'};\ell ^{''}-m^{''}}(\psi_{\ell 0}\phi_{\ell ^{'}m^{'}}{\bar \phi}_{\ell ^{''}m^{''}}^{\dagger}-
\psi_{\ell 0}{\bar \phi}_{\ell ^{'}m^{'}}\phi_{\ell ^{''}m^{''}}^{\dagger})
\nonumber\\&\ &\hspace{2cm}-(-1)^{m^{''}}D^{\ell m}_{\ell^{'}m^{'};\ell^{''}-m^{''}}(\psi_{\ell m}\phi_{\ell ^{'}m^{'}}^{\dagger}{\bar \phi}_{\ell ^{''}m^{''}}-
\psi_{\ell m}{\bar \phi}_{\ell ^{'}m^{'}}^{\dagger} \phi_{\ell ^{''}m^{''}})\ .
\label{D_expand}
\end{eqnarray} 

\subsection{The spectrum of Hawking Radiation}
\label{tree}

In this subsection, we calculate the S-matrix elements which constitute the expectation 
values (\ref{hituyou1}), (\ref{hituyou2}), (\ref{hituyou3}), (\ref{hituyou4})
 and (\ref{hituyou5}) by using the cubic interaction (\ref{7_interaction}). 
We here only consider the first order interaction and 
do not deal with loop diagram like FIG.\ref{FIG3}. 
Under these assumptions, we consider  the S-matrix elements like 
\begin{eqnarray}
\langle \phi\ {\rm particle}\times 2;{\rm graviton}\times 1\ {\rm out}|0\ {\rm in}\rangle \ .
\end{eqnarray}
Then, we only consider (\ref{hituyou1}) and (\ref{hituyou2}) because 
the others do not include such S-matrix elements.  

\begin{figure}[tbp]
 \begin{center}
  \includegraphics[width=50mm]{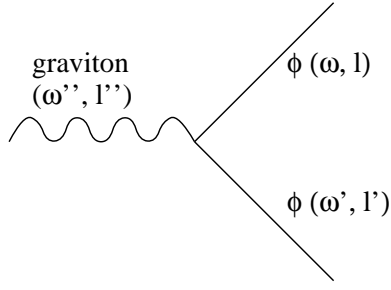}
 \end{center}
 \caption{This diagram affects S-matrix we consider. This diagram corresponds to $K^{\omega^{''}\ell ^{''}}_{\omega \ell ;\omega^{'}\ell ^{'}}$.}
 \label{FIG2}
\end{figure}
\begin{figure}[tbp]
 \begin{center}
  \includegraphics[width=50mm]{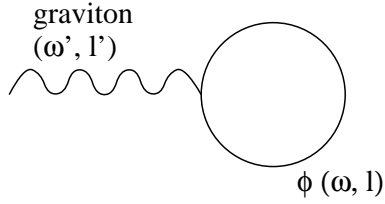}
 \end{center}
 \caption{Loop diagram which appear in the first order calculation.}
 \label{FIG3}
 \end{figure} 
 
Now, we derive the Feynman rules. Recall that
$\phi_{\ell m}$ and $\phi_{\ell 0}$ are written as 
\begin{eqnarray}
\phi_{\ell m} =
\int d\omega
\frac{1}{\sqrt{2\omega}}\left(u_{\omega \ell}^{\phi}b_{\omega\ell m}^{\phi}
+u^{\phi*}_{\omega\ell }d_{\omega\ell m}^{\phi\dagger}\right)\ ,\quad
\phi_{\ell 0} =
\int d\omega
\frac{1}{\sqrt{2\omega}}\left(u^{\phi}_{\omega \ell}c^{\phi}_{\omega \ell}
+u^{\phi*}_{\omega\ell}c^{\phi\dagger}_{\omega \ell}\right)  \ ,
\end{eqnarray}
where $u^{\phi}_{\omega\ell}$ satisfy Eq.(\ref{Klein}) and the boundary condition 
in ``in-region" (\ref{in_condition}). 
Then, we can deduce the Feynman rules as follows:
%
\begin{eqnarray}
\wick{1}{<1 \phi_{\ell^{'} m^{'}}|>1b^{\phi}_{\omega \ell m}}\rangle 
=\frac{1}{\sqrt{2\omega}}u^{\phi}_{\omega\ell}\delta_{\ell \ell^{'}}\delta_{mm^{'}}
\ ,\quad
\wick{1}{<1 \phi_{\ell^{'} m^{'}}^{\dagger}|>1d^{\phi}_{\omega \ell m}}\rangle 
=\frac{1}{\sqrt{2\omega}}u^{\phi}_{\omega\ell}\delta_{\ell \ell^{'}}\delta_{mm^{'}}
\ ,\quad
\wick{1}{<1 \phi_{\ell^{'} 0}|>1c^{\phi}_{\omega \ell }}\rangle 
=\frac{1}{\sqrt{2\omega}}u^{\phi}_{\omega\ell}\delta_{\ell \ell^{'}}
                         \ .
\end{eqnarray}
We can also obtain the Feynman rules for $\psi^{Z}$ and $\psi^{RW}$. Note that 
$\psi^{Z}_{\ell m}$ and $\psi^{Z}_{\ell 0}$ can be expanded as  
\begin{eqnarray}
\psi^Z_{\ell m}=
\int d\omega
\frac{1}{\sqrt{2\omega}}\left(u_{\omega \ell}^{Z}b_{\omega\ell m}^{Z}
+u^{Z*}_{\omega\ell }d_{\omega\ell m}^{Z\dagger}\right)\ ,\quad
\psi^Z_{\ell 0}=
\int d\omega
\frac{1}{\sqrt{2\omega}}\left(u^{Z}_{\omega \ell}c^{Z}_{\omega \ell}
+u^{Z*}_{\omega\ell}c^{Z\dagger}_{\omega \ell}\right)  \ ,
\end{eqnarray}
where $u^{Z}_{\omega\ell}$ satisfy the Zerilli equation and the boundary condition 
in ``in-region" (\ref{in_condition}). Then, the Feynman rules for $\psi^{Z}$ become 
\begin{eqnarray}
\wick{1}{<1 \psi_{\ell^{'} m^{'}}^{Z}|>1b^{Z}_{\omega \ell m}}\rangle 
=\frac{1}{\sqrt{2\omega}}u^{Z}_{\omega\ell}\delta_{\ell \ell^{'}}\delta_{mm^{'}}
\ ,\quad
\wick{1}{<1 \psi_{\ell^{'} m^{'}}^{Z\dagger}|>1d^{Z}_{\omega \ell m}}\rangle 
=\frac{1}{\sqrt{2\omega}}u^{Z}_{\omega\ell}\delta_{\ell \ell^{'}}\delta_{mm^{'}}
\ ,\quad
\wick{1}{<1 \psi_{\ell^{'} 0}^{Z}|>1c^{\phi}_{\omega \ell }}\rangle 
=\frac{1}{\sqrt{2\omega}}u^{Z}_{\omega\ell}\delta_{\ell \ell^{'}}
                        \ .
\end{eqnarray}
The same rules apply to $\psi^{RW}$.
Using such rules and Eq.(\ref{C_expand}), we can perform the following calculations
\begin{eqnarray}
H_{int}|d^{Z}_{\omega\ell m},b^{\phi}_{\omega^{'}\ell^{'}m^{'}},b^{\phi}_{\omega^{''}\ell^{''}m^{''}}\rangle 
&=& \sum_{int}\int dr^*dt h_{int} C^{\ell_1 m_1}_{\ell_2m_2;\ell_3m_3}
\wick{123}{<1 \psi^{Z\dagger}_{\ell_1m_1}<2\phi_{\ell_2m_2}<3 {\bar \phi}_{\ell_3m_3}|>1d^{Z}_{\omega\ell m},>2 b^{\phi}_{\omega\ell^{'}m^{'}},>3 b^{\phi}_{\omega\ell^{''}m^{''}}}\rangle \nonumber\\
& & \qquad +\sum_{int}\int dr^*dt h_{int} C^{\ell_1 m_1}_{\ell_2m_2;\ell_3m_3}
\wick{123}{<1 \psi^{Z\dagger}_{\ell_1m_1}<2\phi_{\ell_2m_2}<3{\bar \phi}_{\ell_3m_3}|>1d^{Z}_{\omega\ell m},>3 b^{\phi}_{\omega\ell^{'}m^{'}},>2 b^{\phi}_{\omega\ell^{''}m^{''}}}\rangle \nonumber\\
&=&  C^{\ell m}_{\ell^{'}m^{'};\ell^{''}m^{''}}\sum_{int}\int dr^*dt \frac{h_{int}}{\sqrt{8\omega\omega^{'}\omega^{''}}}u^{Z}_{\omega\ell}\left(u^{\phi}_{\omega^{'}\ell^{'}}
{\bar u}^{\phi}_{\omega^{''}\ell^{''}}
+{\bar u}^{\phi}_{\ell^{'}m^{'}}u^{\phi}_{\ell^{''}m^{''}}
\right)\ .
\end{eqnarray}
Note that  only one term in (\ref{C_expand}) is related to this calculation. 
Consider another state $|d^{Z}_{\omega\ell m},b^{\phi}_{\omega^{'}\ell^{'}m^{'}},c^{\phi}_{\omega^{''}\ell^{''}}\rangle $. It is easy to obtain
\begin{eqnarray}
H_{int}|d^{Z}_{\omega\ell m},b^{\phi}_{\omega^{'}\ell^{'}m^{'}},c^{\phi}_{\omega^{''}\ell^{''}}\rangle 
&=& \sum_{int}\int dr^*dt h_{int} C^{\ell_1 m_1}_{\ell_2m_2;\ell_3 0}
\wick{123}{<1 \psi^{Z\dagger}_{\ell_1m_1}<2\phi_{\ell_2m_2}<3{\bar \phi}_{\ell_3 0}|>1d^{Z}_{\omega\ell m},>2 b^{\phi}_{\omega\ell^{'}m^{'}},>3 c^{\phi}_{\omega\ell^{''}0}}\rangle \nonumber\\
& &\qquad +\sum_{int}\int dr^*dt h_{int} C^{\ell_1 m_1}_{\ell_2m_2;\ell_30}
\wick{123}{<1 \psi^{Z\dagger}_{\ell_1m_1}<2{\bar \phi}_{\ell_2m_2}<3 \phi_{\ell_3 0}|>1d^{Z}_{\omega\ell m},>2 b^{\phi}_{\omega\ell^{'}m^{'}},>3 c^{\phi}_{\omega\ell^{''}0}}\rangle \nonumber\\
&=& C^{\ell m}_{\ell^{'}m^{'};\ell^{''}0}\sum_{int}\int dr^*dt \frac{h_{int}}{\sqrt{8\omega\omega^{'}\omega^{''}}}u^{Z}_{\omega\ell}\left(u^{\phi}_{\omega^{'}\ell^{'}}
{\bar u}^{\phi}_{\omega^{''}\ell^{''}}
+{\bar u}^{\phi}_{\ell^{'}m^{'}}u^{\phi}_{\ell^{''}m^{''}}
\right)  \ ,
\end{eqnarray}
where the two terms in (\ref{C_expand}) are relevant. 
In the same manner, we can consider the terms related to $\psi^{RW}$. 
For the state $|d^{RW}_{\omega\ell m},b^{\phi}_{\omega^{'}\ell^{'}m^{'}},
b^{\phi}_{\omega^{''}\ell^{''}m^{''}}\rangle $, we have 
\begin{eqnarray}
H_{int}|d^{RW}_{\omega\ell m},b^{\phi}_{\omega^{'}\ell^{'}m^{'}},b^{\phi}_{\omega^{''}\ell^{''}m^{''}}\rangle 
&=&\sum_{int}\int dr^*dt h_{int} D^{\ell_1 m_1}_{\ell_2m_2;\ell_3m_3}
\wick{123}{<1 \psi^{RW\dagger}_{\ell_1m_1}<2\phi_{\ell_2m_2}<3{\bar \phi}_{\ell_3m_3}|>1d^{RW}_{\omega\ell m},>2 b^{\phi}_{\omega\ell^{'}m^{'}},>3 b^{\phi}_{\omega\ell^{''}m^{''}}}\rangle \nonumber\\
& &\qquad+\sum_{int}\int dr^*dt h_{int} D^{\ell_1 m_1}_{\ell_2m_2;\ell_3m_3}
\wick{123}{<1 \psi^{RW\dagger}_{\ell_1m_1}<2\phi_{\ell_2m_2}<3{\bar \phi}_{\ell_3m_3}|>1d^{RW}_{\omega\ell m},>3 b^{\phi}_{\omega\ell^{'}m^{'}},>2 b^{\phi}_{\omega\ell^{''}m^{''}}}\rangle \nonumber\\
&=& D^{\ell m}_{\ell^{'}m^{'};\ell^{''}m^{''}}\sum_{int}\int dr^*dt \frac{w_{int}}{\sqrt{8\omega\omega^{'}\omega^{''}}}u^{RW}_{\omega\ell}\left(u^{\phi}_{\omega^{'}\ell^{'}}
{\bar u}^{\phi}_{\omega^{''}\ell^{''}}
-{\bar u}^{\phi}_{\ell^{'}m^{'}}u^{\phi}_{\ell^{''}m^{''}}
\right)\ .
\end{eqnarray}
Note that additional sign comes from the antisymmetric property $D^{\ell m}_{\ell^{'}m^{'};\ell^{''}m^{''}}=-D^{\ell m}_{\ell^{''}m^{''};\ell^{'}m^{'}}$. 
These results suggest that it is useful to define 
\begin{eqnarray}
K^{\omega\ell}_{\omega^{'}\ell^{'};\omega^{''}\ell^{''}}&\equiv&\sum_{int}\int dr^*dt\frac{h_{int}}{\sqrt{8\omega\omega^{'}\omega^{''}}}
u^{Z*}_{\omega\ell}\left(u^{\phi*}_{\omega^{'}\ell^{'}}{\bar u}^{\phi *}_{\omega^{''}\ell^{''}}+{\bar u}^{\phi *}_{\omega^{'}\ell^{'}}u_{\omega^{''}\ell^{''}}^{\phi *}\right)\nonumber\\
H^{\omega\ell}_{\omega^{'}\ell^{'};\omega^{''}\ell^{''}}&\equiv&\sum_{int}\int dr^*dt\frac{w_{int}}{\sqrt{8\omega\omega^{'}\omega^{''}}}
u^{RW*}_{\omega\ell}\left(u^{\phi*}_{\omega^{'}\ell^{'}}{\bar u}^{\phi *}_{\omega^{''}\ell^{''}}-{\bar u}^{\phi *}_{\omega^{'}\ell^{'}}u_{\omega^{''}\ell^{''}}^{\phi *}\right)
\ .\label{def_KH}
\end{eqnarray}
In fact, using these notations, all S-matrix elements can be written as  
(coefficient of (\ref{C_expand}))$\times K^{\omega\ell}_{\omega^{'}\ell^{'};\omega^{''}\ell^{''}}{}^{*}$ 
or (coefficient of (\ref{D_expand}))$\times H^{\omega\ell}_{\omega^{'}\ell^{'};\omega^{''}\ell^{''}}{}^{*}$. 
Noticing that the effect of the first term of (\ref{odd_interaction})
on $H^{\omega\ell}_{\omega^{'}\ell^{'};\omega^{''}\ell^{''}}$ is given by
\begin{eqnarray}
- \int dr^*dt 
\frac{1}{\sqrt{8\omega\omega^{'}\omega^{''}}}\frac{1}{\sqrt{\lambda(\lambda+1)}}\frac{f}{r^2}
u^{RW*}_{\omega\ell}\left({\dot u}^{\phi*}_{\omega^{'}\ell^{'}} u^{\phi *}_{\omega^{''}\ell^{''}}-u^{\phi *}_{\omega^{'}\ell^{'}}{\dot u}^{\phi *}_{\omega^{''}\ell^{''}}\right)
\ .
\end{eqnarray}
Using these notations, we calculate (\ref{hituyou1}) and (\ref{hituyou2})
in the following.

\subsubsection{ Calculation of (\ref{hituyou2})}

First, we consider $\psi^{Z}_{\ell m}$. 
In the formula (\ref{hituyou2}), we only need the cases
 ``$n^{'}_{b\omega^{'}\ell m}=n^{''}_{b\omega^{''}\ell m}=1$ and one graviton " or 
``$n^{'}_{b\omega^{'}\ell m}=0,\ n^{''}_{b\omega^{''}\ell m}=2$ and one graviton". 
From (\ref{C_expand}), for both cases, this graviton 
should be ``d-particle". In the former case, (\ref{hituyou2}) becomes 
\begin{eqnarray}
&\ &\frac{1}{L^2}\sum_{\omega^{'''}}\sum_{\ell ^{'''}}\sum_{0<m^{'''}\leq \ell ^{'''}}\sqrt{2}\langle 0\ {\rm in}|b^{\phi}_{\omega^{'}\ell m},b^{\phi}_{\omega^{''}\ell m},d^{Z}_{\omega^{'''}\ell ^{'''}m^{'''}}\ {\rm out }\rangle 
\langle \left(b^{\phi}_{\omega^{'}\ell m}\right)^2,d^{Z}_{\omega^{'''}\ell ^{'''}m^{'''}}\ {\rm out}|0\ {\rm in}\rangle \nonumber\\
&=&\frac{1}{L} \int d\omega'''
\sum_{\ell ^{'''}}\sum_{m^{'''}}\sqrt{2}\left[C^{\ell ^{'''}m^{'''}}_{\ell m;\ell m}K^{\omega^{'''}\ell ^{'''}}_{\omega^{'}\ell ;\omega^{''}\ell }\right]^{*}\left[\frac{1}{\sqrt{2}}C^{\ell ^{'''}m^{'''}}_{\ell m;\ell m}K^{\omega^{'''}\ell ^{'''}}_{\omega^{'}\ell ;\omega^{'}\ell }\right]\nonumber\\
&=&\frac{1}{L}\int d\omega'''
\sum_{\ell ^{'''}}\left|C^{\ell ^{'''}2m}_{\ell m;\ell m}\right|^2K^{\omega^{'''}\ell ^{'''}}_{\omega^{'}\ell ;\omega^{''}\ell }{}^{*}K^{\omega^{'''}\ell ^{'''}}_{\omega^{'}\ell ;\omega^{'}\ell }  \ ,
\end{eqnarray} 
where we have introduced the regularization
$\langle b^{\phi}_{\omega^{'}\ell m},b^{\phi}_{\omega^{''}\ell m},d^{Z}_{\omega^{'''}\ell ^{'''}m^{'''}}\ {\rm out }|b^{\phi}_{\omega^{'}\ell m},b^{\phi}_{\omega^{''}\ell m},d^{Z}_{\omega^{'''}\ell ^{'''}m^{'''}}\ {\rm out }\rangle=(\delta(\omega=0))^3\equiv L^3$.
Here, $L$ has the dimension of the length.
Note that we used $C^{\ell m}_{\ell ^{'}m^{'};\ell ^{''}m^{''}}\propto \delta_{m,m^{'}+m^{''}}$ in the last equality. 
In the above calculation, we used the formula $\sum_{\omega^{'''}}=L\int d\omega^{'''} $. 
We can get the same result as this for the latter case and adding these results gives 
\begin{eqnarray}
(\ref{hituyou2})=
\frac{1}{L}\int d\omega'''
\sum_{\ell ^{'''}}\left|C^{\ell ^{'''}2m}_{\ell m;\ell m}\right|^2
\left\{K^{\omega^{'''}\ell ^{'''}}_{\omega^{'}\ell ;\omega^{''}\ell }{}^{*}K^{\omega^{'''}\ell ^{'''}}_{\omega^{'}\ell ;\omega^{'}\ell}+
K^{\omega^{'''}\ell ^{'''}}_{\omega^{''}\ell ;\omega^{''}\ell }{}^{*}K^{\omega^{'''}\ell ^{'''}}_{\omega^{'}\ell ;\omega^{''}\ell}
\right\}  \ .
\end{eqnarray}
 We only calculated the case that $\phi_{\ell m}$ is ``b-particle" so far.
  The same calculation leads to the result for ``c-particle" 
\begin{eqnarray}
\frac{1}{L}\int d\omega'''
\sum_{\ell ^{'''}}\left|C^{\ell ^{'''}0}_{\ell 0;\ell 0}\right|^2
\left\{K^{\omega^{'''}\ell ^{'''}}_{\omega^{'}\ell ;\omega^{''}\ell }{}^{*}K^{\omega^{'''}\ell ^{'''}}_{\omega^{'}\ell ;\omega^{'}\ell}+
K^{\omega^{'''}\ell ^{'''}}_{\omega^{''}\ell ;\omega^{''}\ell }{}^{*}K^{\omega^{'''}\ell ^{'''}}_{\omega^{'}\ell ;\omega^{''}\ell}
\right\}\ ,
\end{eqnarray}
and that for ``d-particle" 
\begin{eqnarray}
&\ &\frac{1}{L}\int d\omega'''
\sum_{\ell ^{'''}}\left|C^{\ell ^{'''}2m}_{\ell m;\ell m}\right|^2
\left\{K^{\omega^{'''}\ell ^{'''}}_{\omega^{'}\ell ;\omega^{''}\ell }{}^{*}K^{\omega^{'''}\ell ^{'''}}_{\omega^{'}\ell ;\omega^{'}\ell}+
K^{\omega^{'''}\ell ^{'''}}_{\omega^{''}\ell ;\omega^{''}\ell }{}^{*}K^{\omega^{'''}\ell ^{'''}}_{\omega^{'}\ell ;\omega^{''}\ell}
\right\}\nonumber\\
&=&\frac{1}{L}\int d\omega'''
\sum_{\ell ^{'''}}\left|C^{\ell ^{'''}-2m}_{\ell -m;\ell -m}\right|^2
\left\{K^{\omega^{'''}\ell ^{'''}}_{\omega^{'}\ell ;\omega^{''}\ell }{}^{*}K^{\omega^{'''}\ell ^{'''}}_{\omega^{'}\ell ;\omega^{'}\ell}+
K^{\omega^{'''}\ell ^{'''}}_{\omega^{''}\ell ;\omega^{''}\ell }{}^{*}K^{\omega^{'''}\ell ^{'''}}_{\omega^{'}\ell ;\omega^{''}\ell}
\right\}\ .
\end{eqnarray}
Taking the summation of $\ell $ and $m$ in (\ref{keisanshitai2}),
 the effect of (\ref{hituyou2}) on (\ref{keisanshitai2}) is 
\begin{eqnarray}
\frac{1}{L}\int d\omega'''
\sum_{\ell }\sum_{-\ell \leq m\leq \ell }\sum_{\ell ^{'''}}\left|C^{\ell ^{'''}2m}_{\ell m;\ell m}\right|^2
\left\{
K^{\omega^{'''}\ell ^{'''}}_{\omega^{'}\ell ;\omega^{''}\ell }{}^{*}K^{\omega^{'''}\ell ^{'''}}_{\omega^{'}\ell ;\omega^{'}\ell }
+K^{\omega^{'''}\ell ^{'''}}_{\omega^{''}\ell ;\omega^{''}\ell }{}^{*}K^{\omega^{'''}\ell ^{'''}}_{\omega^{'}\ell ;\omega^{''}\ell }
\right\}\ .
\label{7_12}
\end{eqnarray}   

Next, we consider  $\psi^{RW}$. Same calculation leads to 
almost the same result as (\ref{7_12})
\begin{eqnarray}
\frac{1}{L} \int d\omega'''
\sum_{\ell }\sum_{-\ell \leq m\leq \ell }\sum_{\ell ^{'''}}\left|D^{\ell ^{'''}2m}_{\ell m;\ell m}\right|^2
\left\{
H^{\omega^{'''}\ell ^{'''}}_{\omega^{'}\ell ;\omega^{''}\ell }{}^{*}H^{\omega^{'''}\ell ^{'''}}_{\omega^{'}\ell ;\omega^{'}\ell }
+H^{\omega^{'''}\ell ^{'''}}_{\omega^{''}\ell ;\omega^{''}\ell }{}^{*}H^{\omega^{'''}\ell ^{'''}}_{\omega^{'}\ell ;\omega^{''}\ell }
\right\}\ .\nonumber
\end{eqnarray}   
However, because of $H^{\omega\ell}_{\omega^{'}\ell^{'};\omega^{''}\ell^{''}}=-H^{\omega\ell}_{\omega^{''}\ell^{''};\omega^{'}\ell^{'}}$, 
this  vanishes.

\subsubsection{ Calculation of (\ref{hituyou1})}

First, we consider  $\psi^{Z}$.
In this case, we must consider $n^{'}_{b\omega^{'}\ell m}=1,2$. 
If $n^{'}_{b\omega^{'}\ell m}=2$, the same calculation as before yields  
\begin{eqnarray}
\frac{1}{L}\int d\omega'''
\sum_{\ell ^{'''}m^{'''}}2\left|\langle 0\ {\rm in}|\left(b^{\phi}_{\omega^{'}\ell m}\right)^2,d^{Z}_{\omega^{'''}\ell ^{'''}m^{'''}}\
 {\rm out}\rangle \right|^2
=\frac{1}{L}\int d\omega'''
\sum_{\ell ^{'''}}\left|C^{\ell ^{'''}2m}_{\ell m;\ell m}\right|^2\left|K^{\omega^{'''}\ell ^{'''}}_{\omega^{'}\ell ;\omega^{'}\ell }\right|^2
      \ .   \label{127}
\end{eqnarray}
Next, we consider $n^{'}_{b\omega^{'}\ell m}=1$. 
In this case, we must consider ``another $\phi$-particle and one graviton". 
Furthermore, if this additional $\phi$-particle is ``b-particle" with
 $({\bar \omega},{\bar \ell },{\bar m})$, we must consider a different mode 
$(\omega^{'},\ell ,m)\neq ({\bar \omega},{\bar \ell },{\bar m})$.   
Then, we have to calculate the followings 
\begin{eqnarray}
\sum_{({\bar \omega},{\bar \ell },{\bar m})\neq (\omega^{'},\ell ,m)}\left|\langle 0\ {\rm in}|b^{\phi}_{\omega^{'}\ell m},b_{{\bar \omega}{\bar \ell }{\bar m}}^{\phi}, \psi^Z{\rm graviton}\times 1\ {\rm out}\rangle \right|^2\ ,\nonumber\\
\sum_{{\bar \omega},{\bar \ell }}\left|\langle 0\ {\rm in}|b^{\phi}_{\omega^{'}\ell m},c_{{\bar \omega}{\bar \ell }}^{\phi}, \psi^Z{\rm graviton}\times 1\ {\rm out}\rangle \right|^2\ ,\nonumber\\
\sum_{{\bar \omega},{\bar \ell },{\bar m}}\left|\langle 0\ {\rm in}|b^{\phi}_{\omega^{'}\ell m},d_{{\bar \omega}{\bar \ell }{\bar m}}^{\phi}, \psi^Z{\rm graviton}\times 1\ {\rm out}\rangle \right|^2\ .
\label{128}
\end{eqnarray} 
Note that the rule $\sum_{{\bar \omega}}=L \int d{\bar \omega}$ should be used
 when we evaluate these quantities. 
Taking look at (\ref{C_expand}), we see that the graviton in the first equation of
 (\ref{128}) must be ``d-particle". The result is given by 
\begin{eqnarray}
&& L \int d\omega''' \sum_{({\bar \omega},{\bar \ell },{\bar m})\neq (\omega^{'},\ell ,m)}
\sum_{\ell ^{'''}m^{'''}}
\left|\langle 0\ {\rm in}|b^{\phi}_{\omega^{'}\ell m},b_{{\bar \omega}{\bar \ell }{\bar m}}^{\phi},d^{Z}_{\omega^{'''}\ell ^{'''}m^{'''}}\ {\rm out}\rangle \right|^2
    \nonumber \\
&& \qquad = L \int d\omega'''\sum_{({\bar \omega},{\bar \ell },{\bar m})\neq (\omega^{'},\ell ,m)}
\sum_{\ell ^{'''}}\left|C^{\ell ^{'''}m+{\bar m}}_{\ell m;{\bar \ell }{\bar m}}\right|^2\left|K^{\omega^{'''}\ell ^{'''}}_{{\bar \omega}{\bar \ell };\omega^{'}\ell }\right|^2\ .\label{129}
\end{eqnarray}
We can also calculate the second equation of (\ref{128}). 
In this case, the graviton must be ``d-particle".
The answer becomes
\begin{eqnarray}
L \int d\omega'''  \sum_{{\bar \omega},{\bar \ell }}
\sum_{\ell ^{'''}}\left|C^{\ell ^{'''}m}_{\ell m;{\bar \ell }0}\right|^2\left|K^{\omega^{'''}\ell ^{'''}}_{{\bar \omega}{\bar \ell };\omega^{'}\ell }\right|^2\ .
\label{130}
\end{eqnarray} 
In the last equation of (\ref{128}), we have to consider all types of gravitons. 
More precisely, the sign of ${\bar m}-m$ determines ``particle" species. 
Thus, we obtain
\begin{eqnarray}
& &L \int d\omega''' \sum_{{\bar \omega},{\bar \ell },{\bar m}}
\sum_{\ell ^{'''}}\Biggl[
\sum_{m^{'''}}\biggl\{
\left|\langle 0\ {\rm in}|b^{\phi}_{\omega^{'}\ell m},d^{\phi}_{{\bar \omega}{\bar \ell }{\bar m}},b^{Z}_{\omega^{'''}\ell ^{'''}m^{'''}}\ {\rm out}\rangle \right|^2
+\left|\langle 0\ {\rm in}|b^{\phi}_{\omega^{'}\ell m},d^{\phi}_{{\bar \omega}{\bar \ell }{\bar m}},d^{Z}_{\omega^{'''}\ell ^{'''}m^{'''}}\ {\rm out}\rangle \right|^2
\biggr\}
\nonumber\\
& &\hspace{7cm}+\left|\langle 0\ {\rm in}|b^{\phi}_{\omega^{'}\ell m},d^{\phi}_{{\bar \omega}{\bar \ell }{\bar m}},c^{Z}_{\omega^{'''}\ell ^{'''}}\ {\rm out}\rangle \right|^2
\Biggr]\nonumber\\
&& \quad = L \int d\omega'''  \sum_{{\bar \omega},{\bar \ell },{\bar m}}
\sum_{\ell ^{'''}}\Biggl[
\sum_{m^{'''}}\biggl\{
\left|C^{\ell ^{'''}m^{'''}}_{{\bar \ell }{\bar m};\ell -m}\right|^2\left|K^{\omega^{'''}\ell ^{'''}}_{\omega^{'}\ell ;{\bar \omega}{\bar \ell }}\right|^2
+\left|C^{\ell ^{'''}m^{'''}}_{{\bar \ell }-{\bar m};\ell m}\right|^2\left|K^{\omega^{'''}\ell ^{'''}}_{\omega^{'}\ell ;{\bar \omega}{\bar \ell }}\right|^2
\biggr\}
+\left|C^{\ell 0}_{{\bar \ell }{\bar m};\ell -m}\right|^2\left|K^{\omega^{'''}\ell ^{'''}}_{\omega^{'}\ell ;{\bar \omega}{\bar \ell }}\right|^2
\Biggr]
\end{eqnarray}
This equation is divided into three parts, but only one has nonzero value for
fixed $m$ and ${\bar m}$ because 
$C^{\ell m}_{ell^{'}m^{'};\ell^{''}m^{''}}\propto \delta_{m,m^{'}+m^{''}}$. 
Then, after taking the summation of $m^{'''}$, we have 
\begin{eqnarray}
L \int d\omega'''  \sum_{{\bar \omega},{\bar \ell },{\bar m}}
\sum_{\ell ^{'''}}
\left|C^{\ell ^{'''}m-{\bar m}}_{{\bar \ell }-{\bar m};\ell m}\right|^2\left|K^{\omega^{'''}\ell ^{'''}}_{\omega^{'}\ell ;{\bar \omega}{\bar \ell }}\right|^2
\ .\label{132}
\end{eqnarray}
Summing up Eqs.(\ref{127}), (\ref{129}), (\ref{130}) and (\ref{132}), 
we see that the contribution from the tree diagrams to (\ref{hituyou1}) becomes 
\begin{eqnarray}
\frac{1}{L} \int d\omega'''
\sum_{{\bar \omega},{\bar \ell }}\sum_{-{\bar \ell }\leq {\bar m}\leq {\bar \ell }}
\sum_{\ell ^{'''}}
\left|C^{\ell ^{'''}m+{\bar m}}_{{\bar \ell }{\bar m};\ell m}\right|^2\left|K^{\omega^{'''}\ell ^{'''}}_{\omega^{'}\ell ;{\bar \omega}{\bar \ell }}\right|^2
\end{eqnarray}

So far, we have only considered the case there are at least one ``b-particle" for $\phi$.
 Calculating  ``c-particle" and ``d-particle" and taking the summation of $\ell$ and $m$
in (\ref{keisanshitai2}), we obtain 
\begin{eqnarray}
\frac{1}{L} \int d\omega''' \sum_{{\bar \omega}}
\sum_{\ell ,\ell ^{'''},{\bar \ell }}
\sum_{-\ell \leq m\leq \ell }
\sum_{-{\bar \ell }\leq {\bar m}\leq {\bar \ell }}
\left|C^{\ell ^{'''}m+{\bar m}}_{{\bar \ell }{\bar m};\ell m}\right|^2\left|K^{\omega^{'''}\ell ^{'''}}_{\omega^{'}\ell ;{\bar \omega}{\bar \ell }}\right|^2\ .
\label{7_20}
\end{eqnarray}

Next, we consider $\psi^{RW}$. With similar calculations with (\ref{hituyou2}),
 we can get the result by replacing C with D and K with H in (\ref{7_20}). 
The resultant expression becomes
\begin{eqnarray}
\frac{1}{L} \int d\omega'''  \sum_{{\bar \omega}}
\sum_{\ell ,\ell ^{'''},{\bar \ell }}
\sum_{-\ell \leq m\leq \ell }
\sum_{-{\bar \ell }\leq {\bar m}\leq {\bar \ell }}
\left|D^{\ell ^{'''}m+{\bar m}}_{{\bar \ell }{\bar m};\ell m}\right|^2\left|H^{\omega^{'''}\ell ^{'''}}_{\omega^{'}\ell ;{\bar \omega}{\bar \ell }}\right|^2\ .
\label{7_21}
\end{eqnarray}

\subsubsection{Summary}

To summarize, substituting these S-matrix into (\ref{keisanshitai2}),
we can write down the effect of interaction
 on Hawking radiation from the tree diagrams as
\begin{eqnarray}
&\ &\frac{1}{L} \int d\omega''' \sum_{\ell}\sum_{-\ell\leq m\leq \ell}\Biggl[
 \sum_{\ell ^{'''}} \int_{\omega^{'}\neq\omega^{''}} d\omega' d\omega''
   \Biggl\{
(\alpha_{\omega\ell,\omega^{''}\ell}\alpha_{\omega\ell,\omega^{'}\ell}^{*}+\beta_{\omega\ell,\omega^{'}\ell}\beta_{\omega\ell,\omega^{''}\ell}^{*})
\nonumber\\
&\ &\hspace{4cm}\times
\left|C^{\ell ^{'''}2m}_{\ell m;\ell m}\right|^2
\left(
K^{\omega^{'''}\ell ^{'''}}_{\omega^{'}\ell ;\omega^{''}\ell }{}^{*}K^{\omega^{'''}\ell ^{'''}}_{\omega^{'}\ell ;\omega^{'}\ell }
+K^{\omega^{'''}\ell ^{'''}}_{\omega^{''}\ell ;\omega^{''}\ell }{}^{*}K^{\omega^{'''}\ell ^{'''}}_{\omega^{'}\ell ;\omega^{''}\ell }
\right)
\Biggr\}\nonumber\\
&\ &\hspace{3.5cm}+\sum_{{\bar \omega},\omega^{'}}
\sum_{\ell ^{'''},{\bar \ell }}
\sum_{-{\bar \ell }\leq {\bar m}\leq {\bar \ell }}
\Biggl\{
(\alpha_{\omega\ell,\omega^{'}\ell}\alpha_{\omega\ell,\omega^{'}\ell}^{*}+\beta_{\omega\ell,\omega^{'}\ell}\beta_{\omega\ell,\omega^{'}\ell}^{*})
\nonumber\\
&\ &\hspace{5cm}
\times 
\left(\left|C^{\ell ^{'''}m+{\bar m}}_{{\bar \ell }{\bar m};\ell m}\right|^2\left|K^{\omega^{'''}\ell ^{'''}}_{\omega^{'}\ell ;{\bar \omega}{\bar \ell }}\right|^2
+\left|D^{\ell ^{'''}m+{\bar m}}_{{\bar \ell }{\bar m};\ell m}\right|^2\left|H^{\omega^{'''}\ell ^{'''}}_{\omega^{'}\ell ;{\bar \omega}{\bar \ell }}\right|^2\right)
\Biggr\}
\Biggr]\ ,
\label{conclusion}
\end{eqnarray}
where $\alpha$ and $\beta$ are calculated in (\ref{Hawking_beta}).  
Recall that we should use the rule
$\sum_{{\bar \omega}}=L\int d{\bar \omega}$  when we evaluate 
the second term in the above formula (\ref{conclusion}) and 
 $\sum_{\omega^{'}}$ in the second term should be
  $\frac{1}{L}\int d\omega^{'}$ as understood from (\ref{effect_b_int}). 

\subsection{Cancelation of Divergences at horizon}

There are coupling function $h_{int}$ which is proportional to $1/f(r)$ in the definition of $K^{\omega\ell}_{\omega^{'}\ell;\omega^{''}\ell^{''}}$, 
so it seems that $K^{\omega\ell}_{\omega^{'}\ell;\omega^{''}\ell^{''}}$ would not converge. However, as you can see, the divergences cancel out each other. 
In the interaction (\ref{even_interaction}), terms proportional to $1/f(r)$ are 
\begin{eqnarray}
&\ &\sqrt{\frac{\lambda+1}{\lambda}}\Biggl[\left(-\frac{2r}{\gamma_sf}\partial_{r^{*}}^2\psi_{\ell m}^{Z*}+\frac{rf^{'}}{\gamma_s f}\partial_{r^{*}}\psi_{\ell m}^{Z*}\right){\dot \phi}_{\ell^{'} m^{'}}{\dot \phi}_{\ell^{''} m^{''}}
+\left(-\frac{2r}{\gamma_s f}\partial_{r^{*}}^{2}\psi_{\ell m}^{Z*}+\frac{rf^{'}}{\gamma_s f}\partial_{r^{*}}\psi_{\ell m}^{Z*}\right) \partial_{r^{*}} \phi_{\ell^{'}m^{'}} \partial_{r^{*}}\phi_{\ell^{''}m^{''}}\nonumber\\
&\ &\hspace{5cm}+\left(\frac{4r}{\gamma_s f}\partial_{r^{*}}{\dot \psi}_{\ell m}^{Z*}-\frac{2rf^{'}}{\gamma_s f}{\dot \psi}_{\ell m}^{Z*}\right){\dot \phi}_{\ell^{'}m^{'}}\partial_{r^{*}}\phi_{\ell^{''}m^{''}}
\Biggr]   \nonumber
\end{eqnarray}
or
\begin{eqnarray}
&\ & \sqrt{\frac{\lambda+1}{\lambda}}\Biggl[
\frac{2r}{\gamma_sf}\biggl(-\partial_{r^{*}}^2\psi_{\ell m}^{*Z}{\dot \phi}_{\ell^{'} m^{'}}{\dot \phi}_{\ell^{''} m^{''}}-\partial_{r^{*}}^2\psi_{\ell m}^{Z*}\partial_{r^{*}}\phi_{\ell^{'}m^{'}} \partial_{r^{*}}\phi_{\ell^{''}m^{''}}
+2\partial_{r^{*}}{\dot \psi}_{\ell m}^{Z*} {\dot \phi}_{\ell^{'}m^{'}} \partial_{r^{*}}\phi_{\ell^{''}m^{''}} \biggr)
\nonumber\\
&\ &\hspace{3cm}+\frac{rf^{'}}{\gamma_s f}\biggr(\partial_{r^{*}}\psi_{\ell m}^{Z*}{\dot \phi}_{\ell^{'} m^{'}}{\dot \phi}_{\ell^{''} m^{''}}+\partial_{r^{*}}\psi_{\ell m}^{Z*}\partial_{r^{*}} \phi_{\ell^{'}m^{'}} \partial_{r^{*}}\phi_{\ell^{''}m^{''}}
-2{\dot \psi}_{\ell m}^{Z*}{\dot \phi}_{\ell^{'}m^{'}}\partial_{r^{*}}\phi_{\ell^{''}m^{''}}
\biggl)
\Biggr]\ .
\label{interaction_f}
\end{eqnarray}
Then, the contribution from these interactions to 
$K^{\omega\ell}_{\omega^{'}\ell^{'};\omega^{''}\ell^{''}}$ becomes
\begin{eqnarray}
&\ &\int dr^*dt\frac{1}{\sqrt{8\omega\omega^{'}\omega^{''}}}\Biggl[
\frac{2r}{\gamma_sf}\biggl(-\partial_{r^{*}}^2u_{\omega \ell}^{Z*}{\dot u}^{\phi*}_{\omega^{'} \ell^{'}}{\dot u}^{\phi*}_{\omega^{''}\ell^{''}}
-\partial_{r^{*}}^2u_{\omega\ell }^{Z*}\partial_{r^{*}}u^{\phi *}_{\omega^{'}\ell^{'}} \partial_{r^{*}}u^{\phi *}_{\omega^{''}\ell^{''}}\nonumber\\
&\ &\hspace{4cm}+2\partial_{r^{*}}{\dot u}^{Z*}_{\omega\ell } {\dot u}^{\phi*}_{\omega^{'}\ell^{'}} \partial_{r^{*}}u^{\phi*}_{\omega^{''}\ell^{''}}+\left\{(\omega^{'},\ell^{'})\leftrightarrow (\omega^{''},\ell^{''})\right\} \biggr)\nonumber\\
& & +\frac{rf^{'}}{\gamma_s f}\biggr(\partial_{r^{*}}u_{\omega \ell}^{Z*}{\dot u}^{\phi*}_{\omega^{'}\ell^{'}}{\dot u}_{\omega^{''}\ell^{''}}^{\phi *}
+\partial_{r^{*}}u^{Z*}_{\omega\ell}\partial_{r^{*}} u^{\phi*}_{\omega^{'}\ell^{'}} \partial_{r^{*}}u^{\phi*}_{\omega^{''}\ell^{''}}
-2{\dot u}_{\omega\ell}^{Z*}{\dot u}^{\phi*}_{\omega^{'}\ell^{'}}\partial_{r^{*}}u^{\phi*}_{\omega^{''}\ell^{''}}
+\left\{(\omega^{'},\ell^{'})\leftrightarrow (\omega^{''},\ell^{''})\right\}\biggl)
\Biggr]\ ,
\label{K_f}
\end{eqnarray} 
where $u^{Z}$ and $u^{\phi}$ satisfy Zerilli equation 
\begin{eqnarray}
\partial_{r^{*}}^2u_{\omega \ell}^{Z}-\partial_{t}^2u_{\omega \ell}^{Z}+V_{Zerilli}u_{\omega\ell}^{Z}=0\ ,\nonumber
\end{eqnarray}
and equation of motion for $\phi$
\begin{eqnarray}
\partial_{r^{*}}^2u_{\omega \ell}^{\phi}-\partial_{r^{*}}^2u_{\omega \ell}^{\phi}+V_{\phi}u_{\omega\ell}^{\phi}=0
\ .\nonumber
\end{eqnarray}
In these equation, $V_{Zerilli}$ and $V_{\phi}$ dump sufficiently first 
as $r^{*}\rightarrow -\infty$ (or near horizon). 
Therefore, $u^{Z}$ and $u^{\phi}$ behave as 
\begin{eqnarray}
\left\{
\begin{array}{l}
u_{\omega \ell}^{Z}\simeq \alpha(t-r^*)+\beta(t+r^*)\\
u_{\omega \ell}^{\phi}\simeq {\bar \alpha}(t-r^*)+{\bar \beta}(t+r^*)
\end{array}
\right.\ .
\label{u_zennkinn}
\end{eqnarray}
Substituting these expressions into (\ref{K_f}), we see 
that the contribution from near horizon vanishes. 
Indeed, the contribution from the pattern $(u^Z,u^{\phi},u^{\phi})
=(\alpha,{\bar \alpha},{\bar \alpha})$ is clearly zero and 
that from the pattern $(\alpha,{\bar \alpha},{\bar \beta})$ is cancelled out
by the pattern $(\beta,{\bar \beta},{\bar \alpha})$. 
Thus, it turns out that $K^{\omega\ell}_{\omega^{'}\ell^{'};\omega^{''}\ell^{''}}$
 converge by taking into account the equations of motion for $\psi^{Z}$ and $\phi$.
  
\subsection{Deviation from Planck Spectrum}

\begin{figure}[htbp]
 \begin{center}
  \includegraphics[width=60mm]{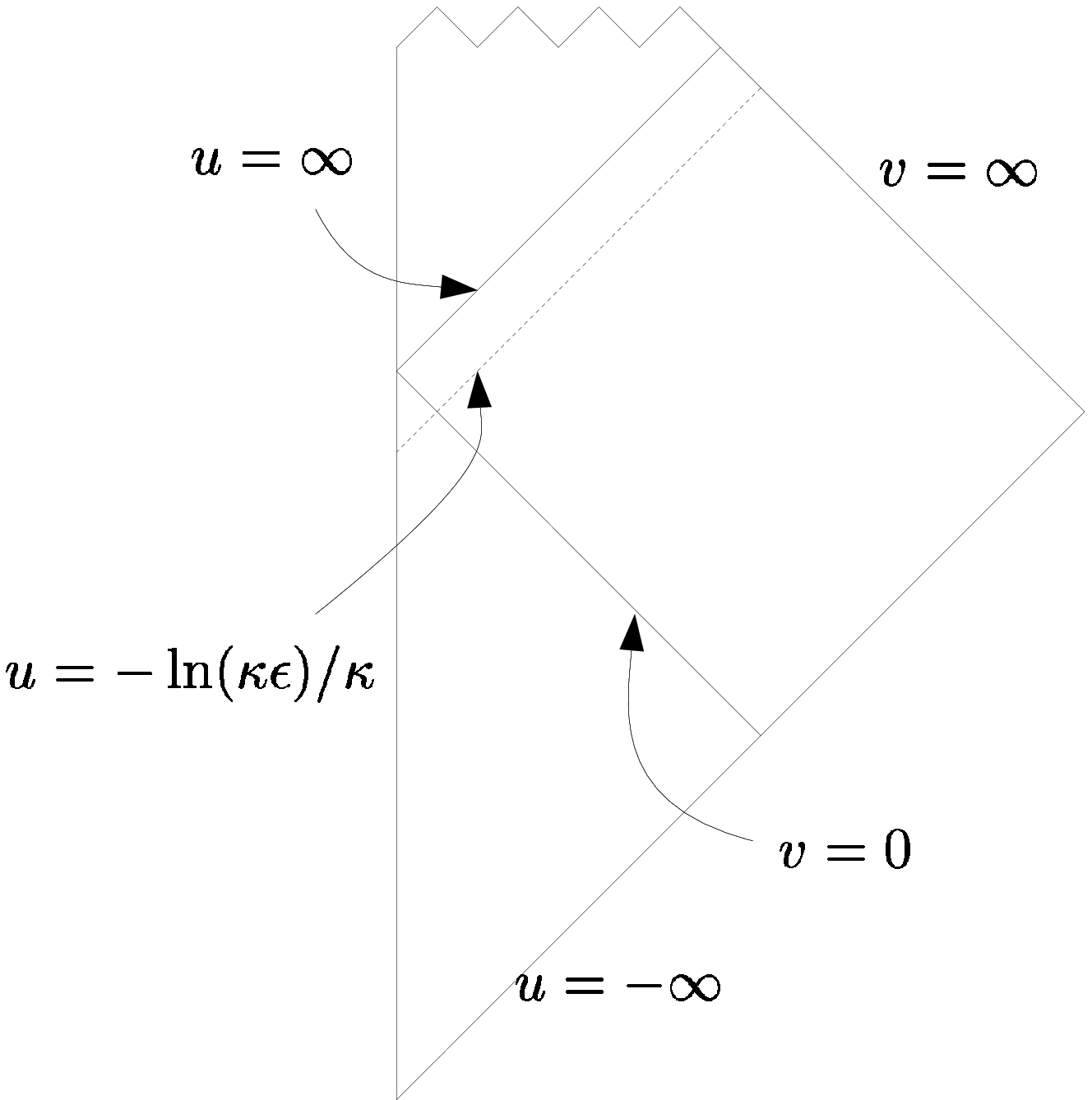}
 \end{center}
 \caption{We separately consider near  $u\sim \infty$ and the other area. We put
  this boundary at $u=-\frac{1}{\kappa}\ln \kappa \epsilon$. }
 \label{fig:estimate}
\end{figure}

Now, we estimate the effect of interaction (\ref{conclusion}). 
To begin with, we estimate $K^{\omega\ell}_{\omega^{'}\ell^{'};\omega^{''}\ell^{''}}$ and  $H^{\omega\ell}_{\omega^{'}\ell^{'};\omega^{''}\ell^{''}}$  
defined by 
\begin{eqnarray}
K^{\omega \ell}_{\omega^{'}\ell^{'};\omega^{''}\ell^{''}}=\sum_{int}\int dr^*dt \frac{h_{int}}{\sqrt{8\omega\omega^{'}\omega^{''}}}
u^{Z*}_{\omega\ell}\left(u^{\phi*}_{\omega^{'}\ell^{'}}{\bar u}^{\phi*}_{\omega^{''}\ell^{''}}+{\bar u}^{\phi*}_{\omega^{'}\ell^{'}}u^{\phi *}_{\omega^{''}\ell^{''}}\right)\ ,\nonumber\\
H^{\omega \ell}_{\omega^{'}\ell^{'};\omega^{''}\ell^{''}}=\sum_{int}\int dr^*dt \frac{w_{int}}{\sqrt{8\omega\omega^{'}\omega^{''}}}
u^{RW*}_{\omega\ell}\left(u^{\phi*}_{\omega^{'}\ell^{'}}{\bar u}^{\phi*}_{\omega^{''}\ell^{''}}-{\bar u}^{\phi*}_{\omega^{'}\ell^{'}}u^{\phi *}_{\omega^{''}\ell^{''}}\right)\ ,\nonumber\label{}
\end{eqnarray}
where $u^{Z,RW,\phi}$ satisfy the boundary condition of ``in region" (\ref{in_condition}). 
Using the geometric optics approximation, we can deduce
the asymptotic solution  
\begin{eqnarray}
	u^{Z,RW,\phi}_{\omega\ell}\simeq\left\{
	\begin{array}{l}
	e^{-i\omega v}-e^{-i\omega u} \quad ({\rm as}\ u\rightarrow -\infty) \\
	e^{-i\omega v}-e^{-i\frac{\omega}{\kappa} e^{-\kappa u}} \quad ({\rm as }\ u\rightarrow \infty)
	\end{array}
	\right. \ ,\label{}
\end{eqnarray}
%
where $u=t-r^*$ and $v=t+r^*$. Then, in order to estimate the effect of interaction,
 we approximate the solution in two ranges as  
\begin{eqnarray}
	u^{Z,RW,\phi}_{\omega\ell}\simeq\left\{
	\begin{array}{l}
	e^{-i\omega v}-e^{-i\omega u} \quad ( {\rm for}\  -\infty <u<-\frac{1}{\kappa}\ln \kappa\epsilon) \\
	e^{-i\omega v}-e^{-i\frac{\omega}{\kappa} e^{-\kappa u}} \quad ({\rm for}\ -\frac{1}{\kappa}\ln \kappa\epsilon<u<\infty) 
	\end{array}
	\right. \ ,\label{u_assume}
\end{eqnarray}
%
and we call the former range  ``outer area" and 
the latter range ``inner area" (see FIG.\ref{fig:estimate}). 

We consider the case that all $u^{Z,RW, \phi}$ in 
$K^{\omega \ell}_{\omega^{'}\ell^{'};\omega^{''}\ell^{''}}$ and 
 $H^{\omega \ell}_{\omega^{'}\ell^{'};\omega^{''}\ell^{''}}$ are $e^{-i\omega v}$. 
For both ``inner area" and ``outer area", this case does not contribute to
 the integral because this pattern is proportional to the delta function
\begin{eqnarray}
	\int dt e^{i(\omega+\omega^{'}+\omega^{''})t}\propto 
      \delta(\omega+\omega^{'}+\omega^{''})=0  \ .
\end{eqnarray}
Since we only consider the positive frequency mode, the delta function gives zero. 
Hence, we need to consider only the mixed cases. 
We can show that the contributions from ``outer area" always vanish. 
 We have already considered the case that all $u^{Z,RW,\phi}$ are $e^{-i\omega v}$. 
 Let us consider the cases that one of $u$ is $e^{-i\omega u}$. 
 If we consider only the ``outer area", this effect is negligible 
 because the integral becomes 
\begin{eqnarray}
\int dr^{*}dt h_{int}(r) e^{i(\omega+\omega^{'})v}e^{i\omega^{''}u} 
	&=&\int dt e^{i(\omega+\omega^{'}+\omega^{''})t}\int dr^{*} h_{int}(r)e^{i(\omega+\omega^{'}-\omega^{''})r^*}\nonumber\\
	&\sim&\delta(\omega+\omega^{'}+\omega^{''})\int dr^{*}h_{int}(r)e^{i(\omega+\omega^{'}-\omega^{''})r^*}
	=0   \ .
\end{eqnarray}
Similarly, we can show that the contributions from ``outer area" for other cases 
are also zero. 
Therefore, we can concentrate on the contribution from ``inner area".

We consider only the dominant terms in interactions (\ref{odd_interaction}) and (\ref{even_interaction}). 
First, we check the interaction terms  proportional to $1/f(r)$. 
However, these terms do not affect the integral. Because we assumed (\ref{u_assume}), 
the same calculation as the last subsection shows that these terms do not affect. 
Secondly, we consider the terms which are proportional to $f\times 1/f$ in (\ref{even_interaction}). These terms are 
\begin{eqnarray}
	&\ & \frac{4}{\gamma_s}\left(\partial_{r^{*}}^2\psi^{Z*}\phi\partial_{r^*}\phi-
	\partial_{r^*}{\dot \psi}^{Z*}{\dot \phi}\phi\right)
      -\frac{2f^{'}}{\gamma_s}\left(\partial_{r^{*}}\psi^{Z*}\phi\partial_{r^*}\phi-
	{\dot \psi}^{Z*}{\dot \phi}\phi\right)
      \ ,
\end{eqnarray}
again these terms do not affect the integral from the same
 consideration as the last subsection. 

There are $O(f^0)$ in (\ref{even_interaction}); 
\begin{eqnarray}
A(r)\partial_{r^*}\psi^{Z*}{\dot \phi}{\dot \phi}
+C(r)\partial_{r^*}\psi^{Z*}\partial_{r^*}\phi\partial_{r^*}\phi
	+E(r){\dot \psi}^{Z*}{\dot \phi}\partial_{r^*}\phi  
+B(r)\psi^{Z*}{\dot \phi}{\dot \phi}+D(r)\psi^{Z*}\partial_{r^*}\phi\partial_{r^*}\phi 
            \ .	\label{even_phi_0}
\end{eqnarray}
And also exist in (\ref{odd_interaction}); 
\begin{eqnarray}
	\frac{1}{r}\psi^{RW*}{\dot \phi}\partial_{r^*}\phi\ . 
	\label{odd_phi_0}
\end{eqnarray}
In (\ref{even_phi_0}), we can show that the first three terms becomes zero
  by using the same method as the last subsection and (\ref{ACE}). 
Furthermore, the other terms in (\ref{even_phi_0}) behave $B(r)\rightarrow 0$ and $D(r)\rightarrow 0$ as $r\rightarrow \infty$; and coupling function in (\ref{odd_phi_0}) 
dumps as $r\rightarrow \infty$. Then, these terms do not affect in the asymptotic region. This means that we can estimate the effect of these terms 
using not coupling functions but their value at the horizon , that is $B(r=2M)\sim 1/M$, $D\sim 1/M$ and $1/r\sim 1/M$. 
Other terms can be neglected because their coupling functions 
are proportional to $f/r$. In fact, these terms do not affect both near the horizon and 
in the asymptotic region. 
Therefore, summarizing these, the dominant interactions in ``inner area" are given by
\begin{eqnarray}
	\frac{1}{M}\left\{\psi^{Z*}{\dot \phi}{\dot \phi}+\psi^{Z*}\partial_{r^*}\phi\partial_{r^*} \phi\right\}
\end{eqnarray}
and
\begin{eqnarray}
	\frac{1}{M}\left\{\psi^{RW*}{\dot \phi}\partial_{r^*}\phi\right\}\ . \label{}
\end{eqnarray}
Thus, $K^{\omega\ell}_{\omega^{'}\ell^{'};\omega^{''}\ell^{''}}$ can be estimated as 
\begin{eqnarray}
	K^{\omega\ell}_{\omega^{'}\ell^{'};\omega^{''}\ell^{''}} \sim
	\int dr^{*}dt\frac{1}{\sqrt{\omega\omega^{'}\omega^{''}}}\frac{1}{M}\left[u^{Z*}_{\omega\ell}{\dot u}^{\phi*}_{\omega^{'}\ell^{''}}{\dot u}^{\phi*}_{\omega^{''}\ell^{''}}
	+u^{Z*}_{\omega\ell}\partial_{r^*}u^{\phi*}_{\omega^{'}\ell^{'}}\partial_{r^*}u^{\phi*}_{\omega^{''}\ell^{''}}\right]
      \ .
\end{eqnarray}
Note that we ignored the constant factor. And $H^{\omega\ell}_{\omega^{'}\ell^{'};\omega^{''}\ell^{''}}$ is 
\begin{eqnarray}
	H^{\omega\ell}_{\omega^{'}\ell^{'};\omega^{''}\ell^{''}} \sim
	\int dr^{*}dt\frac{1}{\sqrt{\omega\omega^{'}\omega^{''}}}\frac{1}{M}\left[u^{RW*}_{\omega\ell}\left({\dot u}^{\phi*}_{\omega^{'}\ell^{'}} \partial_{r^*}u^{\phi*}_{\omega_{''}\ell^{''}}
	-{\dot u}^{\phi*}_{\omega^{''}\ell^{''}}\partial_{r^*}u^{\phi*}_{\omega^{'}\ell^{'}}\right)\right]
      \ .
\end{eqnarray}
However, from (\ref{u_assume}), we see this vanishes. 
Then, we consider $K^{\omega\ell}_{\omega^{'}\ell^{'};\omega^{''}\ell^{''}}$ hereafter. 
We assume $u^{Z,\phi}=e^{-i\omega v}-e^{-i\frac{\omega}{\kappa} e^{-\kappa u}}$. 
In the above, the pattern $(u^Z,u^{\phi},u^{\phi})=(e^{-i\omega v},e^{-i\omega v},e^{-i\omega v})$ has been already considered. Hence,  at least, one of $(u^Z,u^{\phi},u^{\phi})$
should be $e^{-i\frac{\omega}{\kappa} e^{-\kappa u}}$. Among these cases, the most dominant one
 is $(u^Z,u^{\phi},u^{\phi})=(e^{-i\frac{\omega}{\kappa} e^{-\kappa u}},e^{-i\omega v},e^{-i\omega v})$. 
This is because $u\in (-\frac{1}{\kappa} \ln \kappa \epsilon,\infty)$ implies that 
\begin{eqnarray}
	\partial_{r^{*}}e^{-i\omega e^{-\kappa u}}=-i\omega e^{-\kappa u}e^{-i\frac{\omega}{\kappa} e^{-\kappa u}}=O(\epsilon)
      \ . 
\end{eqnarray}
Thus, the case with no derivative of $e^{-i\frac{\omega}{\kappa} e^{-\kappa u}}$ is dominant. 
From these consideration, $K^{\omega\ell}_{\omega^{'}\ell^{'};\omega^{''}\ell^{''}}$ 
can be estimated as follows;
\begin{eqnarray}
	K^{\omega\ell}_{\omega^{'}\ell^{'};\omega^{''}\ell^{''}} &\sim&
	\int dr^{*}dt\frac{1}{\sqrt{\omega\omega^{'}\omega^{''}}}\frac{1}{M}\omega^{'}\omega^{''}e^{i\frac{\omega}{\kappa} e^{-\kappa u}}e^{i(\omega^{'}+\omega^{''})v} \nonumber\\
	&\sim&\sqrt{\frac{\omega^{'}\omega^{''}}{\omega}}\frac{1}{M}\int_0^{\infty}dv e^{i(\omega^{'}+\omega^{''})v}\int_{-\frac{1}{\kappa}\ln \kappa \epsilon}^{\infty}du e^{i\frac{\omega}{\kappa} e^{-\kappa u}}\nonumber\\
	&\sim&\sqrt{\frac{\omega^{'}\omega^{''}}{\omega}}\frac{1}{M}\frac{1}{\omega^{'}+\omega^{''}}\int_0^{\epsilon}dU\frac{1}{\kappa U}e^{i\omega U}
      \ .
\end{eqnarray}
The integral in the above equation diverges. However, this divergence occurs
 because we only considered outside of the horizon. In fact,
since the horizon fluctuates due to metric perturbations, 
we must extend the range of integral to inside the horizon. 
We assume the range can be extended by $\epsilon$. 
Then, the integral converges as 
\begin{eqnarray}
	\int_{-\epsilon}^{\epsilon}dU\frac{1}{\kappa U}e^{i\omega  U}
      \sim \frac{\epsilon \omega}{\kappa} 
      \ .
\end{eqnarray}
So, $K^{\omega\ell}_{\omega^{'}\ell^{'};\omega^{''}\ell^{''}}$ can be estimated as 
\begin{eqnarray}
	K^{\omega\ell}_{\omega^{'}\ell^{'};\omega^{''}\ell^{''}} \sim \frac{\epsilon}{\kappa M}\frac{\sqrt{\omega\omega^{'}\omega^{''}}}{\omega^{'}+\omega^{''}} 
\end{eqnarray}

Finally, using this result, we can evaluate the nonlinear effect. 
Assuming $C^{\ell m}_{\ell^{'}m^{'};\ell^{''}m^{''}}\sim O(1)$, (\ref{7_12}) can be 
deduced as 
\begin{eqnarray}
	\frac{1}{L} \int d\omega'''
      \frac{\epsilon^2 \omega^{'''}}{\kappa^2 M^2}\frac{\sqrt{\omega^{'}\omega^{''}}}{\omega^{'}+\omega^{''}}\sim
	\frac{1}{L}\frac{\epsilon^2\omega_{cut}^2}{\kappa^2 M^2}\frac{\sqrt{\omega^{'}\omega^{''}}}{\omega^{'}
      +\omega^{''}}\ ,
\end{eqnarray}
where $\omega_{cut}$ is a cut-off frequency. 
In the same manner, (\ref{7_20}) can be estimated as 
\begin{eqnarray}
	\frac{1}{L} \int d\omega'''
      \sum_{{\bar \omega}}\frac{\epsilon^2\omega^{'''}}{\kappa^2 M^2}\frac{\omega^{'}{\bar\omega}}{(\omega^{'}+{\bar\omega})^2}\sim
	\frac{1}{L}\sum_{{\bar \omega}}\frac{\epsilon^2\omega_{cut}^2}{\kappa^2M^2}\frac{\omega^{'}{\bar \omega}}{(\omega^{'}+{\bar \omega})^2}
      \ . 
\end{eqnarray}
Here, we note the Bogolubov coefficients are explicitly given by
\begin{eqnarray}
	\alpha_{\omega\ell,\omega^{'}\ell^{'}}\sim \frac{1}{\sqrt{\omega \omega^{'}}}\omega^{'}{}^{-i\omega/\kappa}e^{\pi \omega/(2\kappa)}\Gamma\left(1+\frac{i\omega}{\kappa}\right)
      \ , \quad
	\beta_{\omega\ell,\omega^{'}\ell^{'}}\sim \frac{1}{\sqrt{\omega \omega^{'}}}\omega^{'}{}^{-i\omega/\kappa}e^{-\pi \omega/(2\kappa)}\Gamma\left(1+\frac{i\omega}{\kappa}\right)
      \ , 
\end{eqnarray}
from which we obtain
\begin{eqnarray}
|\alpha_{\omega\ell,\omega^{'}\ell^{'}}|^2=e^{2\pi\omega/\kappa}|\beta_{\omega\ell,\omega^{'}\ell^{'}}|^2  \ ,\qquad 
|\beta_{\omega \omega^{'}}|^2=\frac{1}{4\pi\kappa\omega^{'}}\frac{1}{e^{2\pi\omega/\kappa}-1}
\ .
\end{eqnarray}
Now, we can calculate the contribution from (\ref{7_20}) to (\ref{keisanshitai2})  
 as
\begin{eqnarray}
\sum_{{\bar \omega}}\sum_{\omega^{'}}\left[\left(e^{2\pi\omega/\kappa}+1\right)|\beta_{\omega\ell,\omega^{'}\ell^{'}}|^2\frac{\epsilon^2\omega_{cut}^2}{\kappa^2M^2 L}\frac{\omega^{'}{\bar \omega}}{(\omega^{'}+{\bar \omega})^2}\right]
\sim\frac{\epsilon^2\omega_{cut}^3}{\kappa^3M^2 L}\frac{e^{2\pi\omega/\kappa}+1}
{e^{2\pi\omega/\kappa}-1}
\ .
\end{eqnarray}
And, the contribution from (\ref{7_12}) becomes
\begin{eqnarray}
&\ &\int d\omega' d\omega''  \left[(
	\alpha_{\omega\ell,\omega^{'}\ell^{'}}\alpha_{\omega\ell,\omega^{''}\ell^{''}}^{*}
+\beta_{\omega\ell,\omega^{'}\ell^{'}}\beta_{\omega\ell,\omega^{''}\ell^{''}}^{*})\frac{\epsilon^2\omega_{cut}^2}{\kappa^2M^2L}\frac{\sqrt{\omega^{'}\omega^{''}}}{\omega^{'}+\omega^{''}}\right]\nonumber\\
   &\ &  	\sim
\frac{\epsilon^2\omega_{cut}^3}{\kappa^3M^2L}\frac{e^{\pi\omega/\kappa}+e^{-\pi\omega/\kappa}}{e^{\pi\omega/\kappa}-e^{-\pi\omega/\kappa}}=\frac{\epsilon^2\omega_{cut}^3}{\kappa^3M^2L}\frac{e^{2\pi\omega/\kappa}+1}{e^{2\pi\omega/\kappa}-1}
\ .
\end{eqnarray}
Taking into account that $u^Z$ has dimension of the length, we can recover
  $\ell_p$. Thus, it turns out that
  the cubic interaction modifies Hawking radiation by the term proportional to
\begin{eqnarray}
	 \frac{\epsilon^2\omega_{cut}^3 \ell_p^2}{\kappa^3M^2 L}\coth(\pi\omega/\kappa)
       \ .
\end{eqnarray}
This effect is suppressed by the Planck scale. Therefore,
the deviation from Planck distribution is very small but always exists .


\section{Conclusion}
\label{5}
In this paper, we have studied Hawking radiation from fluctuating black holes.
For this purpose, we constructed the canonical quadratic action for metric perturbations
and also obtained the cubic interaction terms in 4-dimensional Schwarzschild background. 
Taking into account cubic interaction terms in the action for a real scalar field, 
we derived the deviation from the Planck spectrum by considering only tree level diagrams. 
It turned out that the deviation is proportional to $\coth(\pi\omega/\kappa)$ . 
 This result may have 
implications for black holes produced at the LHC and in the early universe.
As is mentioned in the introduction, our analysis is also related to the 
trans-Planckian problem in Hawking radiation.

There are many works to be done. First of all, it is intriguing to investigate 
higher order corrections for metric perturbations,
 i.e. self-interaction terms of $\psi$. As the amplitude of quantum fluctuations 
of the metric is estimated as $\ell_p/r_h$, these higher order corrections must
 be important if black holes become as small as Planck scale. Considering these terms,
  we can investigate ``non-Planckianity" of Hawking radiation from black holes
   like Non-Gaussianity in inflation~\cite{Maldacena:2002vr}.
Furthermore, from the AdS/CFT correspondence point of view, 
these interaction terms will be important for three point correlation functions in CFT. 
It is also interesting to generalize the present analysis 
to higher dimensional Schwarzschild black holes. 
This generalization is important because higher dimensional black holes 
might be created at the LHC if dimension of space-time is more than six. 
The calculations would be more complicated, but the linear equations of motion
 have been already derived in \cite{Kodama:2003jz}.
Therefore, it would be possible to compute the effect of interaction even for
higher dimensional black holes. 
It is also possible to consider gauge fields and fermions instead of a scalar field. 
Finally, since Einstein theory may not be valid near the Planck scale,
it would be important to change the gravitational theory from Einstein theory to other theories motivated by quantum gravity.  In fact, string theory predicts 
higher curvature corrections in addition to Einstein-Hilbert action~\cite{Boulware:1985wk}.  
Again, the linear analysis has been already done in \cite{Dotti:2004sh,Takahashi:2009dz}.
Hence, we can extend our analysis to these cases. 

\begin{acknowledgements}
This work is supported by  the
Grant-in-Aid for  Scientific Research Fund of the Ministry of 
Education, Science and Culture of Japan No.22540274, the Grant-in-Aid
for Scientific Research (A) (No. 22244030), the
Grant-in-Aid for  Scientific Research on Innovative Area No.21111006
and the Grant-in-Aid for the Global COE Program 
``The Next Generation of Physics, Spun from Universality and Emergence". 
\end{acknowledgements}


\appendix
\section{Scalar Harmonics and Vector Harmonics}
\label{harmonics}

We classify metric perturbations into scalar and vector perturbations
in terms of the symmetry in 2-dimensional maximally symmetric space with the 
line element $ds^2=\gamma_{ab}dx^a dx^b$ and the curvature $k = \pm 1, 0$.
They can be expanded by scalar harmonics and vector harmonics. 
In this Appendix, we summarize our conventions. 

Scalar harmonics $Y_k$ are defined by the eigenvalue equations
\begin{eqnarray}
	Y_\rho^{|a}{}_{|a}=-\gamma_s Y_\rho . \label{scalar harmonics}
\end{eqnarray}
where the eigenvalue $\gamma_s$ depends on the index $\rho$.
For example, if the 2-dimensional curvature is positive $k=1$, 
scalar harmonics are nothing but spherical harmonics $Y_{\ell m}$
 and the eigenvalue is given by $\gamma_s=\ell (\ell+1)$. In this case,
 $\rho$ corresponds to $\ell m$. 
We impose  normalization conditions  
\begin{eqnarray}
\int d^2x\sqrt{\gamma}Y_\rho Y_{\rho^{'}}=\delta_{\rho,\rho^{'}}\ .
\end{eqnarray}

Using scalar harmonics, we can define vector harmonics
\begin{eqnarray}
	V_\rho {}^a\equiv\epsilon^{a b}Y_{\rho|b}
      \ ,
\end{eqnarray}
where $\epsilon_{ab}=\sqrt{\gamma}\varepsilon_{ab}$ and $\varepsilon_{23}=-1=-\varepsilon_{32}$. 
The vector harmonics satisfy the transverse conditions $V_{\rho}{}^{a}{}_{|a}=0$ 
and eigenvalue equations 
\begin{eqnarray}
	V_{\rho}{}^{a}{}^{|b}{}_{|b}=-(\gamma_s- k)V_\rho{}^a
      \ . 
\end{eqnarray}
 Note that the vector harmonics satisfy the normalization conditions
\begin{eqnarray}
\int d^2x \sqrt{\gamma}V_{\rho}{}_{a}V_{\rho^{'}}{}^{a}
=\gamma_s\delta_{\rho,\rho^{'}}\ .
\end{eqnarray} 

\section{Quadratic Action in (A)dS-Schwarzschild background}
\label{appendixA}

\subsection{3+1 decomposition}

In this Appendix, we calculate the quadratic  action for metric perturbations
using 3+1 decomposition. 
The background spacetime we consider is a (A)dS-Schwarzschild black hole
\begin{eqnarray}
	ds^2=-f(r)dt^2+1/f(r)dr^2+r^2\gamma_{ab}dx^adx^b\ , \label{}
\end{eqnarray}
where $\gamma_{ab}$ denotes the metric of 2-dimensional maximally symmetric space
with the  curvature $k=\pm 1, 0$. 
Using this metric ansatz, Einstein equations with a cosmological constant $\Lambda$
$R_{\mu\nu}-Rg_{\mu\nu}/2+\Lambda g_{\mu\nu}=0$ read 
\begin{eqnarray}
	f^{'}=(k-f)/r-r\Lambda \ , \qquad
	f^{''}=-2(k-f)/r^2
	\ . 
\end{eqnarray}
This can be easily solved as $f(r)= k-2M/r+\Lambda r^2/3$. 
Here, $M$ is the constant of integration. 

The lapse function $N$ and shift vector $N^i$ of this space-time are given by
\begin{eqnarray}
N=\sqrt{f} \ ,\quad  N^{i}=0
     \ .
\end{eqnarray}
The induced metric is 
\begin{eqnarray}
	g_{ij}=
	\left(
	\begin{array}{c|cc}
	1/f&0&0\\ \hline
	0&\multicolumn{2}{c}{\raisebox{-0.4cm}{$r^2\gamma_{ab}$}}\\
	\raisebox{0.2cm}{0}&
	\end{array}
	\right)\ . 
\end{eqnarray}
Apparently, the extrinsic curvature of a $t={\rm const.}$ surface vanishes
\begin{eqnarray}
	K_{ij}=0 \ .
      \label{back}
\end{eqnarray}
The spatial curvatures are calculated as
\begin{eqnarray}
	{\hat R}_{rr}=-\frac{f^{'}}{rf} \ , \quad
        {\hat R}_{ab}=\left( k -f-\frac{rf^{'}}{2}\right)\gamma_{ab} \ , \quad
        {\hat R}=2\Lambda \ . 
        \label{back2}
\end{eqnarray}
According to 3+1 decomposition, Einstein-Hilbert action becomes 
\begin{eqnarray}
	S=\int dtd^3x N\sqrt{g}\left({\hat R}+K^{ij}K_{ij}-K^2-2\Lambda \right)\ , \label{EHaction}
\end{eqnarray}
where $g$ is the determinant of 3 dimensional metric 
and $K$ is the trace part of extrinsic curvature. 
From now on, we call ${\hat R}-2\Lambda$ ``potential term" 
and $K^{ij}K_{ij}-K^2$ ``kinetic term". 

Let us define ${\bar g}\equiv N\sqrt{g}$ and $L={\hat R}+K^{ij}K_{ij}-K^2-2\Lambda$.
In order to calculate the quadratic action, we must expand ${\bar g}$ and $L$; 
${\bar g}={\bar g}^{(0)}+{\bar g}^{(1)}+{\bar g}^{(2)}+\cdots$ and $L=L^{(0)}+L^{(1)}+L^{(2)}+\cdots$, 
where $(n)$ means the n-th order perturbation. Substituting these expressions into
(\ref{EHaction}), we obtain
\begin{eqnarray}
       S&=&\int dt d^3x({\bar g}^{(0)}+{\bar g}^{(1)}+{\bar g}^{(2)}+\cdots)
       (L^{(0)}+L^{(1)}+L^{(2)}+\cdots)\nonumber\\
       &=&\int dt d^3x\Bigl[{\bar g}^{(0)}L^{(0)}+\left\{{\bar g}^{(1)}L^{(0)}+{\bar g}^{(0)}L^{(1)}\right\}
       +\left\{{\bar g}^{(2)}L^{(0)}+{\bar g}^{(1)}L^{(1)}+{\bar g}^{(0)}L^{(2)}\right\}+\cdots \Bigr]\ .
\label{expand}
\end{eqnarray}
In the above action (\ref{expand}), we notice $L^{(0)}=0$ 
from (\ref{back}) and (\ref{back2}). We also have 
 ${\bar g}^{(1)}L^{(0)}+{\bar g}^{(0)}L^{(1)}=0$ 
because this term is proportional to background equations of motion. 
Furthermore, because the extrinsic curvature of background vanishes, 
$L^{(1)}$ is given by the first order perturbation of ${\hat R}$, ${\hat R}^{(1)}$, 
and $L^{(2)}$ can be calculated as
\begin{eqnarray}
	L^{(2)}={\hat R}^{(2)}+ K^{(1)\ ij} K^{(1)}{}_{ij}-\left(K^{(1)}\right)^2
      \ . 
\end{eqnarray}
Hence, the action becomes
\begin{eqnarray}
	 S&=&\int dtd^3x \left(N\sqrt{g}\right)^{(1)}{\hat R}^{(1)}
       +\int dtd^3xN\sqrt{g}\left({\hat R}^{(2)}+ K^{(1)\ ij} K^{(1)}{}_{ij}-(K^{(1)})^2\right)+\cdots\ . \label{nijiaction}
\end{eqnarray}
This tells us that we only need the first order  lapse function, 
shift vector and extrinsic curvature.  
Hereafter, we omit ``$\hat{\hspace{0.5cm} }$", that is, $R_{ij}^{(1)}$ means first order perturbation of 3-dimensional Ricci tensor, for example.

Finally, we list up formulae for calculating perturbed quantities. 
The 3-dimensional background and perturbed metric are $g_{ij}$
 and $h_{ij}$, respectively. With this notation, 
 the first and second order perturbations of the inverse metric are 
\begin{eqnarray}
	g^{(1)ij}&=&-g^{ik}g^{jl}h_{kl}\ , \nonumber\\
	g^{(2)ij}&=&-g^{(1)ik}g^{jl}h_{kl} =g^{im}g^{kn}g^{jl}h_{mn}h_{kl}
      \ . \label{formula1}
\end{eqnarray}
We can calculate Christoffel symbols from the following formulae
\begin{eqnarray}
	\Gamma^{(1)}{}^{i}_{jk}
      &=&\frac{1}{2}g^{il}\left(h_{lj;k}+h_{lk;j}-h_{jk;l}\right)\ ,\nonumber\\
	\Gamma^{(2)}{}^{i}_{jk}
      &=&\frac{1}{2}g^{(1)il}\left(h_{lj;k}+h_{lk;j}-h_{jk;l}\right)
	                  = g^{im}h_{ml}\Gamma^{(1)}{}^{l}_{jk}\ .
\end{eqnarray}
And, Ricci tensor are given by
\begin{eqnarray}
	R^{(1)}{}_{ij}&=&\Gamma^{(1)}{}^{l}_{ij;l}-\Gamma^{(1)}{}^{l}_{il;j}\ ,\nonumber\\
	R^{(2)}{}_{ij}&=&\Gamma^{(2)}{}^{l}_{ij;l}-\Gamma^{(2)}{}^{l}_{il;j}
	                  +\Gamma^{(1)}{}^{l}_{ij}\Gamma^{(1)}{}^{m}_{lm}-\Gamma^{(1)}{}^{l}_{im}\Gamma^{(1)}{}^{m}_{jl}\ .
	                  \label{formula2}
\end{eqnarray}
Note that $;$ means covariant derivative with respect to 
the 3-dimensional background metric   $g_{ij}$.

\subsection{Scalar Type Perturbations}

We treat scalar perturbations using Zerilli gauge. Later, we will  
change the gauge to a more convenient gauge. 
In the Zerilli gauge, metric perturbations are expressed by
\begin{eqnarray}
h_{\mu\nu}=
\left(
\begin{array}{cc|cc}
f{\bar H}&H_1&0&0\\
H_1&H/f&0&0\\ \hline
0&0&\multicolumn{2}{c}{\raisebox{-0.4cm}{$r^2K\gamma_{ab}$}}\\ 
\raisebox{0.2cm}{0}&\raisebox{0.2cm}{0}&&
\end{array}
\right)
\end{eqnarray}
Then, up to the first order, the lapse function and the shift vector are
given by 
\begin{eqnarray}
	N=\sqrt{f}\left(1-\frac{{\bar H}}{2}\right)\ , \quad
	N^{r}=fH_1 \ ,\quad N^a=0 \ ,\quad N_r=H_1 \ ,\quad N_a=0
      \ , 
\end{eqnarray}
and the induced metric becomes 
\begin{eqnarray}
	h_{ij}=
	\left(
	\begin{array}{c|cc}
	H/f&0&0\\ \hline
	0&\multicolumn{2}{c}{\raisebox{-0.4cm}{$r^2K\gamma_{ab}$}}\\
	\raisebox{0.2cm}{0}&
	\end{array}
	\right)\ . 
\end{eqnarray}
First, we calculate the second order perturbation of ``potential term" 
in the Einstein-Hilbert action. 
Using formulae (\ref{formula1})$\sim$(\ref{formula2}), 
in the Zerilli gauge, perturbed quantities are calculated as follows:
\begin{itemize}
\item inverse metric
\begin{eqnarray}
	g^{(1)rr}=-fH\ ,\ g^{(1)ra}=0\ ,\ g^{(1)ab}=-\frac{1}{r^2}K\gamma^{ab}\ .
\end{eqnarray}
\begin{eqnarray}
	g^{(2)rr}=fH^2\ ,\ g^{(2)ra}=0\ ,\ g^{(2)ab}=\frac{1}{r^2}K^2\gamma^{ab}\ .
\end{eqnarray}

\item Christoffel symbols
\begin{eqnarray}
	\Gamma^{(1)}{}^{r}_{rr}&=&\frac{H^{'}}{2}\ ,\ 
      \Gamma^{(1)}{}^{r}_{ra}=\frac{H_{|a}}{2}\ ,\ 
      \Gamma^{(1)}{}^{r}_{ab}=f\left(rH-\frac{1}{2}(r^2K)^{'}\right)\gamma_{ab}\ ,
      \nonumber\\
	\Gamma^{(1)}{}^{a}_{rr}&=&-\frac{1}{2r^{2} f} H^{|a}\ ,\ 
      \Gamma^{(1)}{}^{a}_{rb}=\frac{1}{2}K^{'}\delta^{a}_{b}\ ,\
	\Gamma^{(1)}{}^{a}_{bc}
      =\frac{1}{2}\left(K_{|c}\delta^{a}_{b}+K_{|b}\delta^{a}_{c}-K^{|a}\gamma_{cb}\right)
      \ . \label{B19}
\end{eqnarray}
\begin{eqnarray}
	\Gamma^{(2)}{}^{r}_{rr}&=&-\frac{HH^{'}}{2}\ ,\ 
      \Gamma^{(2)}{}^{r}_{ra}=-\frac{HH_{|a}}{2}\ ,\ 
      \Gamma^{(2)}{}^{r}_{ab}=-fH\left(rH-\frac{1}{2}(r^2K)^{'}\right)\gamma_{ab}
      \ ,\nonumber\\
	\Gamma^{(2)}{}^{a}_{rr}&=&\frac{1}{2r^{2} f} KH^{|a}\ ,\ 
      \Gamma^{(2)}{}^{a}_{rb}=-\frac{1}{2}KK^{'}\delta^{a}_{b}\ ,\
	\Gamma^{(2)}{}^{a}_{bc}=-\frac{K}{2}\left(K_{|c}\delta^{a}_{b}+K_{|b}\delta^{a}_{c}-K^{|a}\gamma_{cb}\right)\ .
\end{eqnarray}
\item Ricci tensor
\begin{eqnarray}
	R^{(1)}{}_{rr}&=&-K^{''}-\frac{2}{r}K^{'}-\frac{f^{'}}{2f}K^{'}+\frac{H^{'}}{r}-\frac{1}{2r^2f}H^{|a}{}_{|a}\ ,\nonumber\\
	R^{(1)}{}_{ra}&=&\left[\frac{1}{2r}H-\frac{1}{2}K^{'}\right]_{|a}\ ,\nonumber\\
	R^{(1)}{}_{ab}&=&-\frac{1}{2}H_{|ab}+\Bigl[f\left(rH-\frac{1}{2}(r^2K)^{'}\right)^{'}+\frac{f^{'}}{2}\left(rH-\frac{1}{2}(r^2K)^{'}\right)
      -\frac{1}{2}rfH^{'}-\frac{1}{2}K^{|a}{}_{|a}\Bigr]\gamma_{ab}\ .
\end{eqnarray}
\item Scalar curvature
\begin{eqnarray}
	 R^{(1)}&=&g^{(1)ij}R_{ij}+g^{ij}R^{(1)}{}_{ij}\nonumber\\
	       &=&-\frac{1}{r^2}\left(H+K\right)^{|a}{}_{|a}
 +\frac{1}{r^2}\left(-2 k K+2rHf^{'}-r^2f^{'}K^{'}+2fH+2fr(H^{'}-3K^{'}-rK^{''})\right)
\end{eqnarray}
\item $N\sqrt{g}$
\begin{eqnarray}
	(N\sqrt{g})^{(1)}=r^2\sqrt{\gamma}\left(K+\frac{H}{2}-\frac{{\bar H}}{2}\right)
\end{eqnarray}
\end{itemize} 
Using these results, we can obtain ``potential terms" of the action . 
The second order scalar curvature perturbation can be calculated as 
$R^{(2)}=g^{(2)ij}R_{ij}+g^{(1)ij}R^{(1)}{}_{ij}+g^{ij}R^{(2)}_{ij}$. 
The first two term can be easily calculated by using the previous results.  
Using Eq.(\ref{formula2}), the last term can be written as 
\begin{eqnarray}
g^{ij}\left(\Gamma^{(2)}{}^{l}_{ij;l}-\Gamma^{(2)}{}^{l}_{il;j}
	                  +\Gamma^{(1)}{}^{l}_{ij}\Gamma^{(1)}{}^{m}_{lm}-\Gamma^{(1)}{}^{l}_{im}\Gamma^{(1)}{}^{m}_{jl}\right)
\label{R2last}
\end{eqnarray}
In the above Eq.(\ref{R2last}), the last two terms must be calculated using the formulae
(\ref{B19}), while the first two terms can be calculated by using the integration by parts
\begin{eqnarray}
	\int \sqrt{g}N\left(g^{ij}\Gamma^{(2)}{}^{l}_{ij;l}-g^{ij}\Gamma^{(2)}{}^{l}_{il;j}\right)&=&
	\int \sqrt{g}(-\partial_l N)\left(g^{ij}\Gamma^{(2)}{}^{l}_{ij}-g^{il}\Gamma^{(2)}{}^{m}_{im}\right)\nonumber\\
	&=&\int \sqrt{g}(-\partial_r N)\left(\frac{\gamma^{ab}}{r^2}\Gamma^{(2)}{}^{r}_{ab}-f\Gamma^{(2)}{}^{a}_{ra}\right)\ .
\label{tric}
\end{eqnarray}
Thus, we can show that the second order  ``potential terms" become
\begin{eqnarray}
\int\sqrt{\gamma}\Bigl[\frac{1}{2}HH^{|a}{}_{|a}-r^2f^{'}HK^{'}+\frac{1}{2}r^2fK^{'2}-r^2fK^{'}H^{'}
+\frac{1}{2}\left(H-{\bar H}\right)r^2R^{(1)}\Bigr]\ .
\end{eqnarray}

Next, we calculate the ``kinetic terms" of the Einstein-Hilbert action. 
Here, we define $E_{ij}=NK_{ij}$. At the first order, we have
\begin{eqnarray}
	E_{rr}=\frac{1}{2f}\left(\dot{H}-2fH_1^{'}-f^{'}H_1\right)\ ,\ E_{ra}=-\frac{1}{2}H_{1|a}\ ,\ E_{ab}=-\frac{1}{2}\left(r^2\dot{K}-2rfH_1\right)\gamma_{ab}\ .
\end{eqnarray}
Then, the trace part of the extrinsic curvature is given by
\begin{eqnarray}
E=g^{ij}E_{ij}=\frac{1}{2}\left(\dot{H}-2fH_1^{'}-f^{'}H_1\right)+\left(\dot{K}-\frac{2f}{r}H_1\right)
\end{eqnarray}
and the quadratic part reads
\begin{eqnarray}
	E_{ij}E^{ij}=\frac{1}{4}\left(\dot{H}-2fH_1^{'}-f^{'}H_1\right)^2+\frac{f}{2r^2}H_{1|a}H_{1}{}^{|a}+\frac{1}{2r^4}\left(r^2\dot{K}-2rfH_1\right)^2
      \ .
\end{eqnarray}
Therefore, the second order perturbation of ``kinetic terms" can be deduced as
\begin{eqnarray}
\int \sqrt{\gamma}\Bigl[-\frac{1}{2}H_1H_{1}{}^{|a}{}_{|a}-\frac{r^2}{2f}{\dot K}^2-\frac{r^2}{f}\dot{H}\dot{K}+2r\dot{H}H_1
      +r^2\dot{K}H_1^{'}+\frac{\dot{K}}{f}\left(r^2fH_1\right)^{'}\Bigr]
       \ .
\end{eqnarray}
Then, we obtain the total quadratic action for perturbed metric 
\begin{eqnarray}
	 &\ &\int\sqrt{\gamma}\Bigl[\frac{1}{2}HH^{|a}{}_{|a}-r^2f^{'}HK^{'}+\frac{1}{2}r^2fK^{'2}-r^2fK^{'}H^{'}
       +\frac{1}{2}\left(H-{\bar H}\right)r^2R^{(1)}\nonumber\\
	&\ &\hspace{2cm}-\frac{1}{2}H_1H_{1}{}^{|a}{}_{|a}-\frac{r^2}{2f}{\dot K}^2-\frac{r^2}{f}\dot{H}\dot{K}+2r\dot{H}H_1
      +r^2\dot{K}H_1^{'}+\frac{\dot{K}}{f}\left(r^2fH_1\right)^{'}\Bigr]\ .
	\label{Zerilli_second_order}
\end{eqnarray}

We can derive the constraint equations taking the variation of this second order action 
with respect to $H_1$ and ${\bar H}$. However, we can not easily solve resultant
 constraint equations. Therefore, we need to change the gauge from Zerilli gauge to the convenient gauge 
defined by (\ref{fs_gauge}). This gauge transformation can be obtained from
 (\ref{even_gauge_transport}).  The gauge transformation we use is 
\begin{eqnarray}
	\left(
	\begin{array}{cc|cc}
	f{\bar H}&H_1&0&0\\
	sym&H/f&\multicolumn{2}{c}{\ \ w_{|a}}\\ \hline
	\bf{sym}&\bf{sym}&0&0\\
	\bf{sym}&\bf{sym}&0&0\\
	\end{array}
	\right)
	\rightarrow
	\left(
\begin{array}{cc|cc}
f{\bar H}^Z&H_1^Z&0&0\\
H_1^Z&H^Z/f&0&0\\ \hline
0&0&\multicolumn{2}{c}{\raisebox{-0.4cm}{$r^2K^Z\gamma_{ab}$}}\\ 
\raisebox{0.2cm}{0}&\raisebox{0.2cm}{0}&&
\end{array}
\right),
\end{eqnarray}
where the index $Z$ means Zerilli gauge. 
This gauge transformation can be obtained by setting $\xi_{\mu}=(0,-w,0,0)$. 
Then, from (\ref{even_gauge_transport}), 
the relation between these variables can be read off as 
\begin{eqnarray}
	{\bar H^{z}}={\bar H}+f^{'}w \ ,\quad 
	H_1^{z}= H_1-{\dot w} \ ,\quad
	H^{z}= H-2fw^{'}-f^{'}w \ ,\quad
	K^{z}= -\frac{2fw}{r}   \ .
	\label{eq:Zerilli2fs}
\end{eqnarray}
Substituting the transformation rules (\ref{eq:Zerilli2fs}) into the action
(\ref{Zerilli_second_order}), we get the quadratic action for metric perturbations
 in the convenient gauge as
\begin{eqnarray}
	&\ &\int\sqrt{\gamma}\Biggl[-\frac{1}{2}H_{1}H_{1}^{|a}{}_{|a}+{\dot w}^{|a}{}_{|a}H_{1}-\frac{1}{2}{\dot w}{\dot w}^{|a}{}_{|a}+2r\dot{H}H_1 
      +\frac{2f+rf^{'}}{2r}wH^{|a}{}_{|a}+\frac{(fr)^{'}}{2}H^2-\frac{k f}{r^2}ww^{|a}{}_{|a}\nonumber\\
	&\ &\hspace{1cm}+\frac{1}{2}{\bar H}\left\{\left(H-2fw^{'}-f^{'}w-\frac{2f}{r}w\right)^{|a}{}_{|a}-\left(2rfH\right)^{'}\right\}\Biggr] 
	\ .
\end{eqnarray}
Now, we expand metric perturbations by scalar harmonics and only consider the real 
mode (for example, if $k=1$, $m=0$ modes are real). 
Of course, other modes can be calculated and we can get almost the 
same result as real modes. After the expansion, the quadratic action becomes
\begin{eqnarray}
\int dr dt\Biggl[\frac{\gamma_s}{2}H_{1}^2
-\gamma_s{\dot w}H_{1}+\frac{\gamma_s}{2}{\dot w}^2+2r\dot{H}H_1 
      -\gamma_s\frac{2f+rf^{'}}{2r}wH+\frac{(fr)^{'}}{2}H^2+\gamma_s\frac{k f}{r^2}w^2
      -\frac{1}{2}{\bar H}q_2\Biggr] \ ,\label{even_fs_action}
\end{eqnarray}
where 
\begin{eqnarray}
	q_1=2rfH-2\gamma_s fw \ , \quad
	q_2=\gamma_s\left(H+f^{'}w-\frac{2fw}{r}\right)+q_1^{'}
	 \label{app_q}
\end{eqnarray}
or
\begin{eqnarray}
	w&=&\frac{r}{\gamma_s T(r)}\left(q_2-q_1^{'}-\frac{\gamma_s}{2rf}q_1\right)\\
	H&=&\frac{1}{T(r)}\left(q_2-q_1^{'}+\frac{rf^{'}-2f}{2rf}q_1\right) \ , \quad
	T(r)\equiv rf^{'}-2f+\gamma_s
 \ .\label{app_wH}
\end{eqnarray}
Therefore, the Hamiltonian constraint equation obtained by varying the action
with respect to ${\bar H}$ gives
\begin{eqnarray}
q_2=0
\label{app_q2}
\end{eqnarray}
and the momentum constraint equation obtained by varying $H_1$ is 
\begin{eqnarray}
	H_1=\dot{w}-\frac{2r\dot{H}}{\gamma_s}\ . \label{app_H1}
\end{eqnarray}
Substituting (\ref{app_wH}), (\ref{app_q2}) and (\ref{app_H1}) into (\ref{even_fs_action}), 
we get the quadratic action which is a functional of $q_1$. 
Using a new variable $\phi\equiv q_1/T(r)$ instead of $q_1$, 
we obtain the quadratic action 
\begin{eqnarray}
	\int dr dt \Biggl[\frac{1}{2f}\frac{\lambda}{\lambda+k}(\partial_t \phi)^2
	                      -\frac{f}{2}\frac{\lambda}{\lambda+k}(\partial_r \phi)^2
	-\frac{1}{2f}\frac{\lambda}{\lambda+k}V_{Z}(r)\phi^2\Biggr] \ ,\label{}
\end{eqnarray}
where we defined $\lambda=(\gamma_s-2 k)/2$ and $V_{Z}$ is given by
\begin{eqnarray}
	V_{Z}&=&-\frac{f}{4\left(r^2f^{'}-2rf+2r\lambda+2r k \right)^2}
      \Bigl[8\lambda( k +\lambda)^2+4(k+\lambda)^2rf^{'}+2(2 k+\lambda)r^2f^{'2}+r^3f^{'3}
      \nonumber\\
	&\ &   \hspace{0.3cm}             +4f^2(2\lambda+rf^{'})
      -4f\left\{2\lambda(2 k+\lambda)+2( k +\lambda)rf^{'}
      +r^2f^{'2}\right\}\Bigr]\nonumber\\
	&=&\frac{2f}{r^3}\frac{\lambda^2(\lambda+ k )r^3+3\lambda^2Mr^2
      +(9\lambda r-3r^3\Lambda)M^2+9M^3}{(r\lambda+3M)^2}
      \ . 
\end{eqnarray} 
Finally, defining  
\begin{eqnarray}
	\psi^Z\equiv \sqrt{\frac{\lambda}{\lambda+ k }}\phi \label{}
\end{eqnarray}
and using the tortoise coordinate, we obtain the canonical quadratic action 
for scalar type perturbations 
\begin{eqnarray}
	 \int dr^{*} dt \left[\frac{1}{2}(\partial_t \psi^Z)^2-\frac{1}{2}(\partial_{r^{*}} \psi^Z)^2-\frac{1}{2}V_{Z}(r) (\psi^Z)^2\right]\ .\label{app_even}
\end{eqnarray}

\subsection{Vector Type Perturbations}

Next, we derive the quadratic action for vector perturbations 
using Regge-Wheeler gauge  
\begin{eqnarray}
h_{\mu \nu}=
	\left(
	\begin{array}{cc|cc}
	0&0&\multicolumn{2}{c}{v_a}\\
	0&0&\multicolumn{2}{c}{w_a}\\ \hline
	sym&sym&0&0\\
	sym&sym&0&0
	\end{array}
	\right)\ ,
\end{eqnarray}
where $v_a$ and $w_a$ satisfy transverse conditions $v_{a}{}^{|a}=0 , w_{a}{}^{|a}=0$. 
In this gauge, the zeroth and first order lapse function
 and shift vector are given by
\begin{eqnarray}
	N=\sqrt{f} \ ,\quad N^{r}=0 \ ,\quad N^{a}=\frac{v^{a}}{r^2} \ ,
\end{eqnarray}
and induced metric becomes 
\begin{eqnarray}
	h_{ij}=
	\left(
	\begin{array}{c|cc}
	0&\multicolumn{2}{c}{w_a}\\ \hline
	sym&0&0\\
	sym&0&0
	\end{array}
	\right) \ .
\end{eqnarray}

First, we consider the second order ``potential term" in the Einstein-Hilbert action. 
In our gauge, perturbed quantities are calculated as follows;
\begin{itemize}
\item inverse metric
\begin{eqnarray}
	g^{(1)rr}=0\ ,\quad 
      g^{(1)ra}=-\frac{f}{r^2}w^{a} \ ,\quad 
      g^{(1)ab}=0 \ .
\end{eqnarray}
\begin{eqnarray}
	g^{(2)rr}=\frac{f^2}{r^2}w^aw_a\ ,\quad 
      g^{(2)ra}=0 \ ,\quad 
      g^{(2)ab}=\frac{f}{r^4}w^aw^b\ .
\end{eqnarray}
\item Christoffel symbols
\begin{eqnarray}
	\Gamma^{(1)}{}^{r}_{rr}&=&0 \ , \quad
      \Gamma^{(1)}{}^{r}_{ra}=-\frac{f}{r}w^a \ ,\quad 
      \Gamma^{(1)}{}^{r}_{ab}=\frac{f}{2}\left(w_{a|b}+w_{b|a}\right)
                                 \ , \nonumber\\
	\Gamma^{(1)}{}^{a}_{rr}&=&\frac{1}{r^2}\left(w^{a'}+\frac{f^{'}}{2f}w^a\right)\ ,\quad
      \Gamma^{(1)}{}^{a}_{rb}=\frac{1}{2r^2}\left(w^a_{|b}-w_b^{|a}\right) \ ,\quad
	\Gamma^{(1)}{}^{a}_{bc} = \frac{f}{r}\gamma_{bc}w^a \ .
\end{eqnarray}
\begin{eqnarray}
	\Gamma^{(2)}{}^{r}_{rr}&=&
      -\frac{f}{r^2}w_a\left(w^{a'}+\frac{f^{'}}{2f}w^a\right)\ ,\quad 
      \Gamma^{(2)}{}^{r}_{ra}=-\frac{f}{2r^2}w_b\left(w^b_{|a}-w_a^{|b}\right)\ ,\quad
	\Gamma^{(2)}{}^{r}_{ab} = -\frac{f^2}{r}w^cw_c\gamma_{ab}\ ,\nonumber\\
	\Gamma^{(2)}{}^{a}_{rr}&=&0\ ,\quad 
      \Gamma^{(2)}{}^{a}_{rb}=\frac{f}{r^3}w^aw_b\ ,\quad 
      \Gamma^{(2)}{}^{a}_{bc}=-\frac{f}{2r^2}w^a\left(w_{b|c}+w_{c|b}\right)\ .
\end{eqnarray}
\item Ricci tensor
\begin{eqnarray}
	R^{(1)}{}_{rr}&=&0\ ,\ R^{(1)}{}_{ra}=\left(-\frac{f}{r^2}-\frac{f^{'}}{2r}\right)w_a+\frac{1}{2r^2}\left(k w_a-w_{a|d}{}^{|d}\right)\ ,\nonumber\\
	R^{(1)}{}_{ab}&=&\frac{f^{'}}{4}\left(w_{a|b}+w_{b|a}\right)+\frac{f}{2}\left(w_{a|b}+w_{b|a}\right)^{'}\ .
\end{eqnarray}
\end{itemize}
This proves that the first order scalar curvature $R^{(1)}$ vanishes.
Therefore, in (\ref{nijiaction}), we only calculate $\int d^4x N\sqrt{g} R^{(2)}$ 
for the ``potential term". 
This term is easily calculated using the technique (\ref{tric}) and the result is 
\begin{eqnarray}
	\int\sqrt{\gamma}\frac{f}{2r^2}\left( k w^aw_a+w_aw^{a}{}^{|b}{}_{|b}\right)
\end{eqnarray}
Next, we calculate the ``kinetic term". The first order $E_{ij}=NK_{ij}$ are
\begin{eqnarray}
	E_{rr}=0\ ,\ E_{ra}=\frac{1}{2}\left(\dot{w_a}+\frac{2}{r}v_a-v_a^{'}\right)\ ,\ E_{ab}=-\frac{1}{2}\left(v_{a|b}+v_{b|a}\right)\ .
\end{eqnarray}
These lead $E=g^{ij}E_{ij}=0$ and 
\begin{eqnarray}
	E_{ij}E^{ij}
  =\frac{f}{2r^2}\left(\dot{w_a}+\frac{2}{r}v_a-v_a^{'}\right)
  \left(\dot{w^a}+\frac{2}{r}v^a-v^{a'}\right)
      +\frac{1}{4r^4}\left(v_{a|b}+v_{b|a}\right)\left(v^{a|b}+v^{b|a}\right)\ . 
\end{eqnarray}
Then, the second order ``kinetic term" is given by
\begin{eqnarray}
	\int \sqrt{\gamma}\left[\frac{1}{2}\left(\dot{w_a}+\frac{2}{r}v_a-v_a^{'}\right)\left(\dot{w^a}+\frac{2}{r}v^a-v^{a'}\right)
		-\frac{1}{2fr^2}\left(k v_av^a+v^{b|a}{}_{|a}v_b\right)\right] 
            \ .
\end{eqnarray}
Note that we ignored total derivative terms. 
Thus, we have obtained the total quadratic action for vector type perturbations 
\begin{eqnarray}
\int\sqrt{\gamma}\Biggl[\frac{1}{2}\left(\dot{w_a}+\frac{2}{r}v_a-v_a^{'}\right)\left(\dot{w^a}+\frac{2}{r}v^a-v^{a'}\right)
		-\frac{1}{2fr^2}\left(k v_av^a+v^{b|a}{}_{|a}v_b\right) 
            +\frac{f}{2r^2}\left(k w^aw_a+w_aw^{a}{}^{|b}{}_{|b}\right)\Biggr] 
            \ .
\end{eqnarray}

Let us expand metric perturbations by vector harmonics $V_k^a$ and only consider 
the real modes. The quadratic action becomes 
\begin{eqnarray}
	2(\lambda+k)\int dr dt\Biggl[\frac{1}{2}\left(\dot{w}+\frac{2}{r}v
      -v^{'}\right)\left(\dot{w}+\frac{2}{r}v-v^{'}\right)
		-\frac{\lambda}{2fr^2}v^2 -\frac{\lambda f}{2r^2}w^2\Biggr] 
            \ ,
\end{eqnarray}
where $\lambda=(\gamma_s-2 k)/2$ and the overall factor comes from 
$\int \sqrt{\gamma}V_\rho^{a*}V_{\rho a}$.
Taking the variation of this action with respect to $v$, 
we obtain the constraint equation which cannot be solved easily. 
However, this difficulty can be circumvented by working in phase space. 

From the  Lagrangian  
\begin{eqnarray}
	 L=(2\lambda+2 k)\int dr \Bigl[\frac{1}{2}\left(\dot{w}+\frac{2}{r}v
       -v^{'}\right)^2+\frac{\lambda}{fr^2}v^2-\frac{\lambda f}{r^2}w^2\Bigr]
       \ ,\nonumber
\end{eqnarray}
we can define conjugate momentum $p$ as
\begin{eqnarray}
	p=\frac{\delta L}{\delta w}=(2\lambda+2 k)(\dot{w}+\frac{2}{r}v-v^{'}) 
      \nonumber
\end{eqnarray}
which leads to 
\begin{eqnarray}
\dot{w}=\frac{p}{2\lambda+2 k}-\frac{2}{r}v+v^{'}\nonumber\ .
\end{eqnarray}
The Hamiltonian is 
\begin{eqnarray}
	H&=&\int dr p\dot{w}-L\nonumber\\
	 &=&(2\lambda+2 k)\int dr \left[\frac{1}{2}\left(\frac{p}{2\lambda+2 k}\right)^2
       + \frac{\lambda f}{r^2}w^2-\left(\frac{2}{r}\frac{p}{2\lambda+2 k}
       +\left(\frac{p}{2\lambda+2 k}\right)^{'}+\frac{\lambda v}{fr^2}\right)v\right]
            \ .   \nonumber
\end{eqnarray}
In phase space, constraint equation can be written as 
\begin{eqnarray}
\frac{2}{r}\frac{p}{2\lambda+2 k}+\left(\frac{p}{2\lambda+2 k}\right)^{'}
+\frac{2\lambda v}{fr^2}=0 \ .
\end{eqnarray}
It is easy to solve the above constraint as
\begin{eqnarray}
v=-\frac{fr^2}{4\lambda(\lambda+ k)}\left(\frac{2}{r}p+p^{'}\right)
\end{eqnarray}
Substituting this result into the Hamiltonian, we have reduced Hamiltonian
\begin{eqnarray}
	H=(2\lambda+2 k)\int dr\left[\left(\frac{1}{2}-\frac{(fr)^{'}}{2\lambda}+\frac{f}{\lambda}\right)\left(\frac{p}{2\lambda+2 k}\right)^2+\frac{fr^2}{4\lambda}
\left(\frac{p}{2\lambda+2 k}\right)^{'}{}^{2}+\frac{\lambda f}{r^2}w^2\right]
 \ ,\label{hamiltonian}
\end{eqnarray}
which is a functional of $w$ and $p$.  
Let us make the following simple canonical transformation 
\begin{eqnarray}
  Q=-p \ ,\quad	P=w \ ,\quad       H(w,p)=K(Q,P) 
\ ,
\end{eqnarray}
where $K(Q,P)$ is a new Hamiltonian. Then, we have 
\begin{eqnarray}
	K(Q,P)&=&H(P,-Q)\nonumber\\
	      &=&(2\lambda+2 k)\int dr\left[\left(\frac{1}{2}
       -\frac{(fr)^{'}}{2\lambda}+\frac{f}{\lambda}\right)
       \left(\frac{Q}{2\lambda+2 k}\right)^2+
\frac{fr^2}{4\lambda}\left(\frac{Q}{2\lambda+2 k}\right)^{'}{}^{2}
+\frac{\lambda f}{r^2}P^2\right] 
          \ .  
\end{eqnarray}
 Now, we define
\begin{eqnarray}
	\dot{Q}=\frac{\delta K}{\delta P}=\frac{4\lambda(\lambda+ k) f}{r^2}P
     \ . 
\end{eqnarray}
Thus, the action becomes 
\begin{eqnarray}
   \int dr dt P\dot{Q}-\int dt K(Q,P) 
   =\frac{1}{2\lambda+2 k}\int dr dt\Biggl[\frac{r^2}{4\lambda f}\dot{Q}^2-\left(\frac{1}{2}-\frac{(fr)^{'}}{2\lambda}+\frac{f}{\lambda}\right)Q^2-\frac{fr^2}{4\lambda}Q^{'}{}^{2}\Biggr] \ .
\end{eqnarray}
Finally, changing normalization by 
\begin{eqnarray}
	\psi^{RW}\equiv \frac{r}{2\sqrt{\lambda(\lambda+ k)}}Q 
\end{eqnarray}
and using the tortoise coordinate,
 we obtain the canonical quadratic action for vector perturbation
\begin{eqnarray}
	\int dr^{*}dt\left[\frac{1}{2}\left(\partial_t \psi^{RW}\right)^2-\frac{1}{2}\left(\partial_{r^*} \psi^{RW}\right)^2-\frac{1}{2}V_{RW}(r)\psi^{RW}{}^2\right] 
      \ ,\label{app_odd}
\end{eqnarray}
where we have defined
\begin{eqnarray}
	V_{RW}=\frac{f}{r^2}\left(-rf^{'}+2f+2\lambda\right)
      \ . 
\end{eqnarray}

\section{Symmetry in Quadratic Action}

In this Appendix, we identify the symmetry in the quadratic action. 
As shown in \cite{Bakas:2008gz}, we can rewrite $V_{Z}$ and $V_{RW}$ as 
\begin{eqnarray}
	\left\{
	\begin{array}{l}
	V_{RW}=W^2-\frac{dW}{dr^{*}}+\omega_s^2\\
	V_{Z}=W^2+\frac{dW}{dr^{*}}+\omega_s^2
	\end{array}
	\right. \ ,
\end{eqnarray}
where 
\begin{eqnarray}
	W(r)=\frac{3Mf}{r(\lambda r+3M)}+i\omega_s \ , \quad
	\omega_s=-\frac{i}{3M}\lambda(\lambda+ k)
\ . 
\end{eqnarray}
Now we consider the quadratic action (\ref{app_even}) and (\ref{app_odd}).
 Using $W(r)$, this action can be rewritten as 
\begin{eqnarray}
	\frac{1}{2}\int dr^{*}dt\left[{\bf \psi}^{\dagger}{\hat H}{\bf \psi}\right]
      \ ,\label{double}
\end{eqnarray}
where ${\bf \psi}$ represents $(\psi^Z,\psi^{RW})^T$.
 Here, the operator ${\hat H}$ is defined by
\begin{eqnarray}
	{\hat H}\equiv -\partial_t^2+\partial_{r^{*}}^2
      -\left(\omega_s^2+W^2+\sigma_3\frac{dW}{dr^{*}}\right) \ , 
\end{eqnarray}
where $\sigma_3$ is a Pauli matrix. 
Then, if there exist operators which commutate with ${\hat H}$, 
the canonical action (\ref{double}) has symmetry. 
In fact, there are two such operators defined by 
\begin{eqnarray}
	\hat{Q}_1 =\sigma_1\frac{1}{i}\partial_{r^{*}}+\sigma_2W \ ,\quad
	\hat{Q}_2 =\sigma_2\frac{1}{i}\partial_{r^{*}}-\sigma_1W
      \nonumber\ ,
\end{eqnarray}
where $\sigma_1$ and $\sigma_2$ are also Pauli matrices.
These operators satisfy anti-commutation relation $\{{\hat Q}_1,{\hat Q}_2\}$=0. 
Then we can conclude that the quadratic action 
in the (A)dS-Schwarzschild background has ``N=2 supersymmetry".

\end{document}